\documentclass[12pt]{iopart}

\usepackage{aabib}
\usepackage{epsfig}

\newcommand{\Jupiter}{{\rm J}}   

\newcommand{\ddeg}{\hbox{.\hskip-3pt $^\circ$}}

\newcommand{\la}{\mathrel{\mathchoice {\vcenter{\offinterlineskip\halign{\hfil
  $\displaystyle##$\hfil\cr<\cr\sim\cr}}}
  {\vcenter{\offinterlineskip\halign{\hfil$\textstyle##$\hfil\cr
  <\cr\sim\cr}}}
  {\vcenter{\offinterlineskip\halign{\hfil$\scriptstyle##$\hfil\cr
  <\cr\sim\cr}}}
  {\vcenter{\offinterlineskip\halign{\hfil$\scriptscriptstyle##$\hfil\cr
  <\cr\sim\cr}}}}}
 
\newcommand{\ga}{\mathrel{\mathchoice {\vcenter{\offinterlineskip\halign{\hfil
  $\displaystyle##$\hfil\cr>\cr\sim\cr}}}
  {\vcenter{\offinterlineskip\halign{\hfil$\textstyle##$\hfil\cr
  >\cr\sim\cr}}}
  {\vcenter{\offinterlineskip\halign{\hfil$\scriptstyle##$\hfil\cr
  >\cr\sim\cr}}}
  {\vcenter{\offinterlineskip\halign{\hfil$\scriptscriptstyle##$\hfil\cr
  >\cr\sim\cr}}}}}

\begin{document}
\jl{19}

\review{Extra-solar planets}

\author{M A C Perryman}

\address{Astrophysics Division, European Space Agency, ESTEC,
Noordwijk 2200AG, The Netherlands; and Leiden Observatory,
University of Leiden, The Netherlands}

\begin{abstract}
The discovery of the first extra-solar planet surrounding a
main-sequence star was announced in 1995, based on very precise radial
velocity (Doppler) measurements.  A total of 34 such planets were
known by the end of March 2000, and their numbers are growing steadily. 
The newly-discovered systems confirm some of the features predicted by
standard theories of star and planet formation, but systems with
massive planets having very small orbital radii and large eccentricities
are common and were generally unexpected.

Other techniques being used to search for planetary signatures include
accurate measurement of positional (astrometric) displacements,
gravitational microlensing, and pulsar timing, the latter resulting in the
detection of the first planetary mass bodies beyond our Solar System
in 1992.  The transit of a planet across the face of the host star
provides significant physical diagnostics, and the first such detection
was announced in 1999.  Protoplanetary disks, which represent an
important evolutionary stage for understanding planet formation, are
being imaged from space.  In contrast, direct imaging of extra-solar
planets represents an enormous challenge.  Long-term efforts are
directed towards infrared space interferometry, the detection of 
Earth-mass planets, and measurement of their spectral characteristics.

Theoretical atmospheric models provide predictions of planetary
temperatures, radii, albedos, chemical condensates, and spectral
features as a function of mass, composition and distance from the 
host star.  Efforts to characterise planets occupying the `habitable
zone', in which liquid water may be present, and indicators of the
presence of life, are advancing quantitatively.

\end{abstract}


\submitted

\maketitle

\section{Introduction}
\label{sec:introduction}

There are hundreds of billions of galaxies in the observable Universe,
with each galaxy such as our own containing some $10^{11}$~stars.
Surrounded by this seemingly limitless ocean of stars, mankind has
long speculated about the existence of planetary systems other than
our own, and the possibility of the development of life elsewhere in
the Universe.   Only recently has evidence become available to begin
to distinguish the extremes of thinking that has pervaded for more
than 2000 years, typified by opinions ranging from {\it `There are
infinite worlds both like and unlike this world of ours'\/} (Epicurus,
341--270~BC) to {\it `There cannot be more worlds than one'\/}
(Aristotle, 384--322~BC).  The last 10--20 years has seen rapid advances
in theoretical understanding of planetary formation, the development
of a variety of conceptual methods for extra-solar planet detection,
the implementation of observational programmes to carry out targeted
searches and, within the last five years, the detection of a number
of planets beyond our own Solar System.

Shining only by reflected starlight, extra-solar planets comparable to
bodies in our own Solar System should be typically billions of times
fainter than their host stars and, depending on their distances from
us, at angular separations from their accompanying star of, at most, a
few seconds of arc. This combination makes direct detection
extraordinarily demanding, particularly at optical wavelengths where
the star/planet intensity ratio is large, and especially from the
ground given the perturbing effect of the Earth's atmosphere.
Alternative detection methods, based on the dynamical perturbation of
the star by the orbiting planet, on planetary transits, and on gravitational
lensing, have therefore been developed, although these effects are
also extremely subtle.  

Radio pulsar timing achieved the first
detection of planetary mass bodies beyond our Solar System in 1992. 
High-precision radial velocity (Doppler) measurements resulted in the
detection of the first extra-solar planetary systems surrounding
main-sequence stars similar to our own in 1995, and in 1999
gravitational microlensing provided (contested) evidence for a planet
near the centre of our Galaxy nearly 30\,000 light-years away.
Observational progress in extra-solar planet detection and
characterization is now moving rapidly on a number of fronts.

The extra-solar planets detected from radial velocity measurements, of
which 34 were known by the end of March 2000, have masses in the range
0.2--11\,$M_\Jupiter$, orbital periods in the range 3--1700~days, and
semi-major axes in the range 0.04--2.8~AU.  Planets significantly less
massive than Jupiter cannot be detected with current radial velocity
techniques, while systems with large orbital radii will have escaped
detection so far due to their long-periods and hence small deviation
from rectilinear photocentric motion over measurement periods of a few
years. Nevertheless, the newly-discovered systems do not typify
planetary properties that had been expected.  More than one third have
significantly elliptical orbits, with $e>0.3$, compared with the
largest eccentricities in our Solar System, of about~0.2 for Mercury
and Pluto, and 0.05 for Jupiter.  Even more significantly, about two
thirds are orbiting their host star much closer than Mercury orbits
the Sun (0.39~AU).  While the Doppler technique preferentially selects
systems in tight orbits, giant planets so close to the parent star
were generally unexpected. Theoretical progress in understanding their
formation and their properties has been rapid, but it remains far from
complete.
%
%
\footnote{The text employs a number of standard astronomical terms and
units.  Planetary systems are conveniently characterized in terms of
corresponding Solar System quantities 
($\odot$ = Sun; $\oplus$ = Earth; \Jupiter\ = Jupiter): 
$M_\odot\simeq2.0\times10^{30}$~kg; 
$M_\Jupiter\simeq9.5\times10^{-4}\,M_\odot$; 
$M_{\rm Saturn}\simeq2.9\times10^{-4}\,M_\odot$; 
$M_{\rm Uranus}\simeq4.4\times10^{-5}\,M_\odot$; 
$M_\oplus\simeq3.0\times10^{-6}\,M_\odot\simeq 3\times10^{-3}\,M_\Jupiter$. 
1~AU~=~1~astronomical unit (mean Sun-Earth distance) 
$\simeq1.5\times10^{11}$~m. 
For Jupiter, $a=5.2$~AU and $P=11.9$~yr. 
Stellar distances are conveniently given in parsec (pc), defined as the 
distance at which 1~AU subtends an angle of 1~second of arc (or arcsec);
1~pc~$\simeq3.1\times10^{16}$~m $\simeq$~3.26 light-years. 
For reference, distances to the nearest stars are of order 1~pc; 
there are about 2000 known stars within a radius of 25~pc of our Sun, 
and the distance to the Galactic centre is about 8.5~kpc. 
Stellar masses range from around $0.1-30\,M_\odot$, with spectral
types providing convenient spectral classification related to the 
primary stellar properties of temperature and luminosity. Our Sun is 
of spectral type G2V: cooler stars (types K, M) are of lower mass and 
have longer lifetimes; hotter stars (types F, A, etc.) are of higher 
mass and have shorter lifetimes.
Object names such as 70~Vir (for 70~Virginis) reflect standard 
astronomical (constellation-based) nomenclature, while other designations 
reflect discovery catalogues or techniques variously labelled with catalogue 
running numbers (e.g.\ HD~114762) or according to celestial coordinates (e.g.\ 
PSR~1257+12).
The International Astronomical Union is in the process of
formulating recommendations for the nomenclature of extra-solar
planets (cf.\ \cite{wd98}), meanwhile the {\it de facto\/} custom
denotes (multiple) planets around star~X as X~b, c,... according to 
discovery sequence.
}

Based on present knowledge from the radial velocity surveys, about 5\%
of solar-type stars may harbour massive planets, and
an even higher percentage may have planets of lower mass
or with larger orbital radii.  If these numbers can be extrapolated,
the number of planets in our Galaxy alone would be of order 1~billion.
Although only one main-sequence extra-solar planetary system is known 
which contains more than one planet, present understanding of planet
formation could imply that many of these massive planets are 
accompanied by other objects in the same orbiting system.  For the
future, experiments capable of detecting tens of thousands of
extra-solar planetary systems, lower mass planets down to around
$1\,M_\oplus$, and spectral signatures which may indicate the presence
of life, are now underway or are planned.  The most substantial
advances may come from space observatories over the next 10--20~years.

This review covers published literature to March 2000. 
It provides a summary of the theoretical understanding of
planet formation, a review of detection methods, reporting on major
relevant experiments both ongoing and planned, and outlines the
properties of the extra-solar planetary systems detected to date.
Amongst others it addresses the following questions: What defines a
planet?  What prospects are there for the various search programmes
underway? How common are planetary systems?  What is their formation
process?  What information can be deduced from them?  What fraction
are likely to lie in the `habitable zone'? What are the prospects for
detecting the presence of life?

Section~\ref{sec:stars-planets} provides a background to the processes
believed to underpin star formation and planetary formation.
Section~\ref{sec:methods} reviews detection methods, including
instruments and programmes under development, and their prospects for
planet detection in the future. The detection of protoplanetary disks, 
from which planets are formed, is also covered.
Section~\ref{sec:properties} presents the observed and inferred
properties of the known systems and their host stars, explaining how 
these characteristics may fit into a revised picture of the formation and
evolutionary of extra-solar planets.
Section~\ref{sec:life} touches upon some of the issues relevant for
the development of the field from planet discovery to the detection of
life.

\section{Stars, brown dwarfs and planets}
\label{sec:stars-planets}

\subsection{Star formation}
\label{sec:star-formation}

The existence and formation of planets should first be placed in a broader
astronomical context.
Over the last two decades a standard paradigm has emerged for the
origin and evolution of structure in the Universe (e.g.\ \cite{ws00}).
In this picture, the Universe expanded from a hot, dense and smooth
state, the Hot Big Bang (the Planck form of the cosmic microwave
background provides evidence for this back to a Universe age of a few
months, while the light element abundances extend this to an age of a
few minutes). The observed near-homogeneity and isotropy were produced
by an early phase of inflationary expansion, with all observed
structure originating from quantum zero-point fluctuations during the
inflationary period, and growing by gravitational amplification.
Galaxies were formed by cooling and condensation of gas in the cores
of heavy halos produced by non-linear hierarchical clustering of dark
matter -- the unseen and unknown form of non-baryonic, weakly
interacting matter considered to dominate gravitating mass at the
present time.

In the standard model of star formation, stars form from gravitational
instabilities in interstellar clouds of gas and `dust' grains, leading
to collapse and fragmentation (e.g.\ \cite{sal87}; \cite{bos88}).  In 
particular, a density perturbation (e.g.\ due to a shock wave) may  
cause the gravitational binding energy  of a certain region of a cloud 
to exceed its thermal energy, in which case the region begins to  
contract (the Jeans instability criterion).  The
details, including effects such as stellar rotation and magnetic
fields, are complex and incompletely known, and the early phases are
considered particularly uncertain (\cite{af96}; \cite{elm99}). The
most massive stars evolved rapidly, creating new elements by
nucleosynthesis, and dispersing them through gaseous outflows or
supernova explosions. Part of the chemically enriched material
remained in the gas phase, while part condensed into solid dust
grains (e.g.\ silicates), together providing material for subsequent 
generations of star formation.

For a cloud with some initial rotation gravitational collapse will
lead, from conservation of angular momentum, to a flattened system,
with a high proportion of double and multiple stars arising from the
fragmentation process (\cite{ab92}; \cite{bon99}).  Single star formation 
then involves three fairly distinct stages, within which flattened disks 
appear to be a natural by-product:
(i) collapse, under self-gravity, of an extended cloud of gas and dust
grains assembled from the debris of processed stars and remnants of
the early Universe (the gas consists primarily of H$_2$ molecules
along with H and He atoms and simple molecules such as CO, CO$_2$,
N$_2$, CH$_4$ and H$_2$O, while the dust grains, of order
10~$\mu$m in size, each contain typically $10^6$ atoms of C, Si, O
etc.\ with outer coatings of H$_2$O or CO$_2$).  The material
accumulates quickly towards the central proto-star, but with enough
residual angular momentum to prevent total collapse onto the central object.
The average angular momentum of the collapsing region defines the
rotation axis of the resulting disk, whose thickness is much smaller than
its radius, and which therefore forms a thin plane or circumstellar 
disk extending out to about 100~AU (in the case of our Solar System,
well beyond the orbit of Pluto). The formation of the relatively
stable disk probably occurs over about $10^5-10^6$~years after the
onset of free-fall collapse;
(ii) inflow of gas and dust from the disk onto the central object by
self-gravitation, heating the centrally condensed gas by compression
until nuclear fusion occurs in the central parts and the star is formed
on time scales of $10^5-10^7$~years.  Material in the disk is
replenished by infall from the surrounding molecular cloud;
(iii) through redistribution of mass and angular momentum in
the disk, a centrifugally-supported residual `solar nebula' is formed
containing the material that accumulates to form planets
(Section~\ref{sec:planet-formation}). The process ends when all the
residual nebula gas has been lost, either by escape to interstellar
space, or to the central star (\cite{war81}).

The resulting stars shine by thermonuclear fusion, with stable hydrogen 
burning occurring for masses above about $0.08\,M_\odot$ (about
$80\,M_\Jupiter$), when the central temperature triggers 
nucleosynthesis, this mass limit being slightly sensitive to initial
chemical composition. Objects having marginally smaller masses
(0.075--0.070\,$M_\odot$) are not stars, in the sense that they are
never fully supported by nuclear energy generation, but their thermal
energy output means that they nevertheless spend about $10^{10}$~years
(roughly the present age of the Universe) at luminosities
$L\sim10^{-4}-10^{-5}L_\odot$ (\cite{dm85}). Brown dwarfs (\cite{tar86}; 
\cite{bl93}; \cite{bas00}) are objects which occupy the mass range of about
12--80\,$M_\Jupiter$ ($0.01-0.08M_\odot$), and are also considered to
have formed by gravitational instability in a gas.  These objects are
not massive enough to ignite stable hydrogen burning although
deuterium fusion in their cores, contributing to their luminosity, is
possible down to masses of about $12\,M_\Jupiter$.  Below this limit,
which is again sensitive to chemical composition, and which also
depends on structural uniformity, nuclear processes, and the role of
dust (\cite{tin99}), objects should retain essentially their entire
deuterium complement of around the protosolar value,
D/H~$\sim2\times10^{-5}$, and derive no luminosity from thermonuclear
fusion at any stage in their evolutionary lifetime (\cite{bhs+93};
\cite{shb+96}).

The minimum mass for star formation via fragmentation is determined by
the minimum Jeans mass along the track of a collapsing cloud in the
($\log\rho, \log T$) plane, which occurs when the cloud or fragment
first becomes optically thick. The minimum mass that can be obtained 
in this way appears to be around 7--20\,$M_\Jupiter$ (\cite{bos86};
\cite{bos88}; \cite{bod98}).  Consequently, a mass limit of
$\simeq7-12\,M_\Jupiter$ represents a useful boundary, from both
formation and nuclear reaction perspectives, to provisionally distinguish 
(giant) planets from the more massive brown dwarfs.  The extent to 
which this distinction is confirmed by the recent discoveries will 
be considered in Section~\ref{sec:properties}. The recent detection of 
`free-floating' objects of brown dwarf and planetary mass, which appear 
to have formed by this fragmentation process in young clusters and 
star-forming regions (e.g.\ \cite{bzr99}; \cite{ots99}; \cite{lr00}) 
is not considered here in further detail.

\subsection{Planet formation and our Solar System}
\label{sec:planet-formation}

A number of theories of the origin of our Solar System have been
advanced, starting with the scientific theory of \cite*{lap1796} (see,
e.g., \cite{woo00}). In the most widely considered `solar nebula theory'
planet formation in our Solar System (and, by inference, planetary
formation in general) follows on from the process of star formation
described previously, through the agglomeration of residual
protoplanetary disk material.

In this paradigm, planet formation proceeds through several sub-stages
characterised by differences in the respective particle interactions
(\cite{saf69}; \cite{gw73}; \cite{lis93}; \cite{lis95}; \cite{wet96b};
\cite{rud99}).
First, the dust grains settle into a dense layer in the mid-plane of
the disk, and begin to stick together as they collide, forming
macroscopic objects with sizes of order 0.01--10\,m, all orbiting 
the protostar in the same direction and in the same plane, analogous 
to the few hundred metre thick rings around Saturn. 
In a second stage, over the next $10^4-10^5$~years, further collisions
lead to the formation of `planetesimals', objects up to a km or so in
size, driven by gravitational interactions, and leading to the
concentration of objects into particular orbits, with nearly empty
gaps between them. 
In the third phase, the mutual gravitational interaction between
planetesimals leads to small changes in their Keplerian orbits,
resulting in subsequent collisions, some of which would shatter the
planetesimals, but most of which occur at velocities producing a single
larger object, or embryo. These are of mass $\sim10^{23}$~kg in the
terrestrial planet region, and of larger but uncertain mass in the
outer Solar System.

In the presence of an adequate mass of planetesimals, runaway growth of
embryos occurs as a result of dynamical friction, three-body effects,
and fragmentation (\cite{ki96}; \cite{ki00}).  Runaway growth of these
lunar- to Mercury-sized bodies requires as little as $10^5$~years at
1~AU, but of the order of a few million years in the outer part of the
asteroid belt.
When the gravitational pull of the largest planetesimals is
sufficient, they grow rapidly to the size of small planets by mutual
collisions and mergers, with the terrestrial planets growing over time
scales of between $10^7$ to a few times $10^8$~years (\cite{wet90}),
primarily limited by the time required to sweep up bodies accelerated
to high velocities by close encounters and outer planet resonances.
Mergers proceed by pairwise accretion until the spacing of planetary
orbits becomes large enough that the configuration is stable for the
lifetime of the system (\cite{saf69}; \cite{lis95}; \cite{wsd+97};
\cite{cw98}).

In the case of our Solar System, a few solid cores in the outer parts
of the proto-Solar disk became large enough ($\sim10-30\,M_\oplus$) to
accrete residual gas (H and He) slowly, giving rise to the giant
planets -- Jupiter, Saturn, Uranus and Neptune (\cite{pol84}; \cite{bp86};
\cite{phb+96}). In the generalised picture of gas giant formation
by core accretion (\cite{miz80}; \cite{pol84}; \cite{bhl00}) these objects 
formed far out from the central proto-star --- closer in, temperatures
were too high to allow ice formation and gas accretion (`ice' 
generally refers to a combination of volatiles, such as water, 
methane, and ammonia, in either a solid or liquid phase), and the total
mass of disk material close in was anyhow smaller (\cite{bos95}).
Alternative theories of giant planet formation through
gravitational instability of a massive cool solar nebula, leading to
fragmentation on a dynamical time scale (\cite{kui51}; \cite{ab92}) and
thus circumventing the lengthier formation time scale demanded by core 
accretion, is discussed by, e.g., \cite*{pol84} and \cite*{bos98c}.

Termination of the planet's accretion process, by clearing a gap
around its orbit, depends on the (unknown) viscosity of the
protoplanetary disk (\cite{lp79}).  Factors determining the ultimate 
size and spacing of gas giant planets are complex and poorly
understood. Modern theories of planetary growth do not yield
deterministic `Bode's Law' formulae for planetary orbits
(\cite{nie72}), but characteristic orbital spacings do exist,
suggesting that gaps develop roughly in proportion to the distance
from the central star (\cite{ht98}; \cite{las00}).

Long-term (Hill) stability of the resulting orbits depends on the
separation of the semi-major axes (e.g.\ \cite{gla93}; \cite{lis93};
\cite{cwb96}; \cite{lis97b}). Orbital resonances can either increase
orbital stability (e.g.\ in the case of Neptune--Pluto) or lead to
instabilities.  Larger eccentricities tend to destabilise systems
because bodies can approach each other more closely, while large
inclinations usually increase stability since the bodies remain
farther apart.  
Modelling of the Solar System stability shows that Mercury, for 
example, could be chaotically ejected following close encounters with 
Venus within 3.5~Gyr (\cite{las94p}).
Planetesimal formation in binary systems has been considered by
various authors (\cite{har77}; \cite{hep78b}; \cite{wmc+98}).

In the Solar System, planetesimals which grew to modest size without
joining larger objects, as well as collisional debris, are represented
by meteoroids and comets. The former comprise objects made of `rock' 
(a combination of iron- and magnesium-bearing silicates and metallic iron)
with random orbits and ranging up to a few hundred metres in
size, while the latter comprise `dirty snowballs' of frozen H$_2$O,
CO$_2$, and dust grains up to a few km in size (\cite{hm99x}). Comets
comprise two families: the Edgeworth-Kuiper Belt, extending from 50~AU
(beyond Pluto) to hundreds or a few thousand~AU, forming a vast system
possibly identified with the remnant of the Sun's protoplanetary disk
(\cite{wil97}; \cite{jew99}).  The Oort Cloud consists of some
$10^{12}$ comets of all orbital orientations, extending out to tens of
thousands of~AU, but with a total mass of only a few $M_\oplus$.  It
originated from the gravitational scattering of planetesimals early in
the history of the Solar System by Uranus and Neptune, objects sent
outwards at close to the Solar System escape velocity, and perturbed
into long-lived orbits by nearby stars or the Galactic tidal field.
Bodies scattered from Jupiter and Saturn, deeper within the Sun's
gravitational potential well, would have been lost from the Solar
System.

The smallest objects in the Solar System, swarms of sand to small
meteoroids, are presumably debris left over from the earliest phase. 
The final formation stage would be accompanied by the impact of the
last fragments of matter, evidence of which is seen in the impact
craters of the Moon (whose origin now tends to be attributed to a giant
impact event with the Earth during the later stages of planetary
formation, rather than to capture, e.g.\ \cite{cls99}). Even today at
intervals of several tens of millions of years, a small planetesimal
or comet in an Earth-crossing orbit strikes our planet, these impacts
being considered responsible for mass extinctions such as at the
Cretaceous/Tertiary boundary some 65~million years ago (e.g.\
\cite{sto98}).

In the case of our Solar System formation theories can be confronted with
a wealth of diverse observational constraints: orbital motions,
stability, and spacings of the nine planets; planetary masses,
rotations, and the existence of planetary satellites and rings; angular momentum
distribution; bulk and isotopic composition; radio-isotope ages;
cratering records; and the occurrence of comets, asteroids and
meteorites including the presence of the Oort Cloud and the
Edgeworth-Kuiper Belt (\cite{pol84}; \cite{lis93}; \cite{woo00}). 

While the present paradigm of planet formation offers many attractive 
features, various problems remain: for example, the details of the
redistribution of angular momentum, the requirements for and dominant
source of turbulence during the early nebular phase, the size and
`stickiness' of the dust grains required for the first phase of large
particle formation from micron-sized dust to km-sized planetesimals
(\cite{bw00}), the formation time scale of the giant planets and, in
the case of our Solar System, issues including the $7^\circ$ tilt of
the solar spin axis with respect to the mean orbital plane, the
detailed distribution of the planetary satellites, the isotopic
anomalies in meteorites, and the formation of chondrites (e.g.\ \cite{ssl96};
\cite{mh99}; \cite{dc00}).

Alternative theories exist, most notably developments of the `capture
theory' (\cite{woo64}) which built upon early ideas by \cite*{jea17},
and which involves the tidal interaction between the Sun and a
lower-mass, diffuse cool protostar. The subsequent developments of
this model, and its merits, are reviewed by \cite*{woo00}.

\section{Detection methods}
\label{sec:methods}

\begin{figure}[tbh]
\begin{center}
\centerline{\epsfig{file=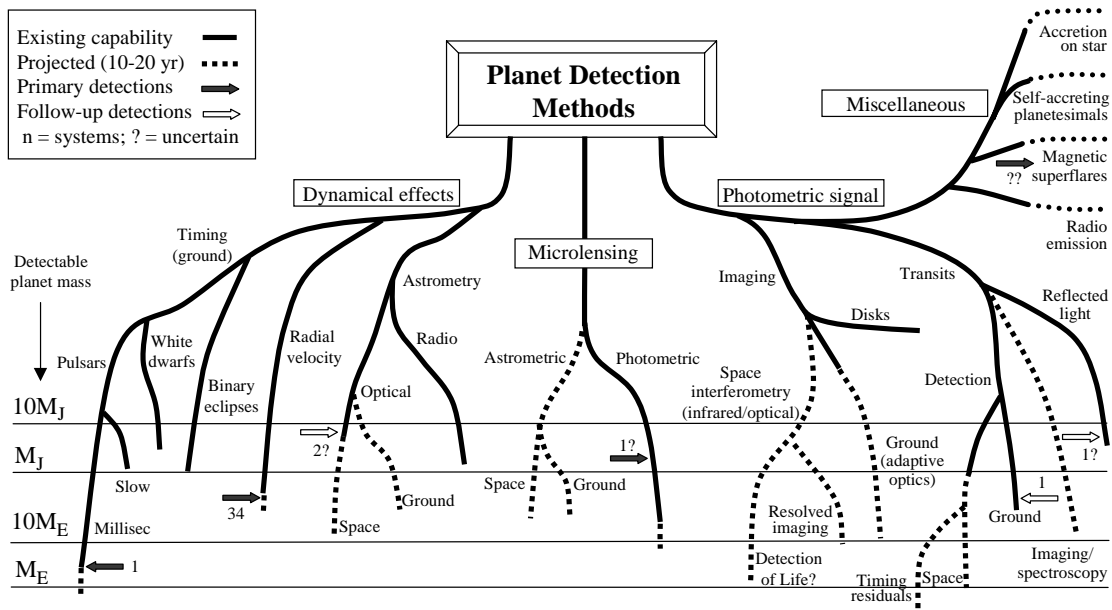,height=16.0cm,angle=270,	
	bbllx=200,bblly=20,bburx=570,bbury=720, clip=}}
\end{center}
\vspace{-20pt}
\caption{Detection methods for extra-solar planets described in the
text.  The lower extent of the lines indicates, roughly, the
detectable masses that are in principle within reach of present 
measurements (solid lines), and those that might be expected within 
the next 10--20~years (dashed). The (logarithmic) mass scale is 
shown at left.  The miscellaneous signatures to the upper right are 
less well quantified in mass terms. Solid arrows indicate 
(original) detections according to approximate mass, while open arrows
indicate further measurements of previously-detected systems. 
`?'~indicates uncertain or unconfirmed detections. The figure takes no 
account of the numbers of planets that may be detectable by each method.
}
\label{fig:methods}
\end{figure}

This review concentrates on the detection and characterisation of
discrete extra-solar planetary mass bodies, although observational
evidence for disks is also summarised.  In considering detection
methods, the order followed in this section progresses roughly from
the most to the least intuitive, although this is somewhat 
uncorrelated with technical feasibility and progress to date.  Although
almost all systems currently known (loosely classified as giant
planets with masses of order $1\,M_\Jupiter$) have been detected by
high-precision radial velocity measurements, the prospects for
the detection of planets with masses significantly below that of
Jupiter appear somewhat limited by this method.  The development of
other methods will be required to lower planetary mass detection
limits towards the `habitable zone' (Section~\ref{sec:life}), to
enlarge the sample sizes to provide better constraints on formation
theories, and to enhance the knowledge of the physical properties of
the detected systems.

Parameters used are mass $M$, radius $R$, and luminosity $L$, with
subscripts $*$ and $p$ referring to star and planet respectively.
Systems are characterised by their orbital period $P$, semi-major axis
$a$, eccentricity $e$, orbital inclination with respect to the plane 
of the sky $i$ ($i=0^\circ$ face-on, $i=90^\circ$ edge-on), and
distance from the Solar System $d$.  Unless explicitly noted, it is
assumed that a single dominant planet is in orbit around the star, a
simplification in the case of our own Solar System, and perhaps the
majority of others, but one which appears to be an acceptable starting
point for describing the systems detected to date.
Figure~\ref{fig:methods} summarises the detection possibilities
referred to in this section.

\subsection{Imaging}
\label{sec:imaging}

Imaging of an extra-solar planet generally refers to the detection of
a point source image of the object seen in the reflected light from
the parent star. This is to be distinguished from resolution of the
planet surface, prospects for which are discussed briefly at the end
of this section.

The ratio of the planet to stellar brightness depends on the 
planet's size and proximity to the star, and on the scattering 
properties of the planet's atmosphere. For reflected light of wavelength 
$\lambda$:
\begin{equation}
\label{equ:imaging}
\frac{L_p}{L_*}= p(\lambda, \alpha) \left ( \frac{R_p}{a}\right )^2 
\end{equation}
where $p(\lambda, \alpha)$ is a phase-dependent function, including
the effects of orbital inclination, depending on the amplitude and
angular dependence of the various sources of scattering in the
planetary atmosphere, integrated over the surface of the sphere.
$\alpha$ is the angle between the star and observer as seen from
the planet. The scattering properties are characterised by the
geometric albedo, $p$, being the flux from the planet at $\alpha=0$ 
compared to the equivalent flux from a Lambert law (perfectly 
diffusing) disk of the same diameter (e.g.\ \cite{cjn98}). The 
formula, including the phase dependence, is modified if thermal emission 
from the planet is significant. 

$L_p/L_*$ is very small, of order $10^{-9}$ for a Jupiter-type object
at maximum elongation. Viewed from a ground-based telescope, with a
star-planet separation of 1~arcsec (Jupiter viewed from 5~pc) the
planet signal is immersed in the photon noise of the telescope's
diffraction profile ($\lambda/D\simeq0.02$~arcsec at 500~nm for a 5-m
telescope) and more problematically within the `seeing' profile (of
order 1~arcsec) arising from turbulent atmospheric refraction.  Under
these conditions elementary signal-to-noise calculations imply that
obtaining a direct image of the planet is not feasible.

Imaging efforts are directed at ways of reducing the angular size of
the stellar image, suppressing scattered light (including use of 
coronographic masks), minimising the effects of atmospheric turbulence 
(including eliminating them altogether using space observations), and 
enhancing the contrast between the planet and the star by observing at 
longer wavelengths (favourable for detection of thermal emission).  None 
of these imaging techniques has been successfully applied to extra-solar
planetary detection so far. But they represent important long-term efforts 
since they provide the basis for attempts to measure the
spectral features of planets, and therefore the possibility
of detecting signatures of life in their atmospheres. Specifically,
direct imaging will in future be capable of providing broad-band
colours and eventually spectral energy distributions, giving
constraints on temperature and chemical composition, and ultimately
insights into the planet's chemical and biological processes.

\begin{figure}[tbh]
\begin{center}
\centerline{
\epsfig{file=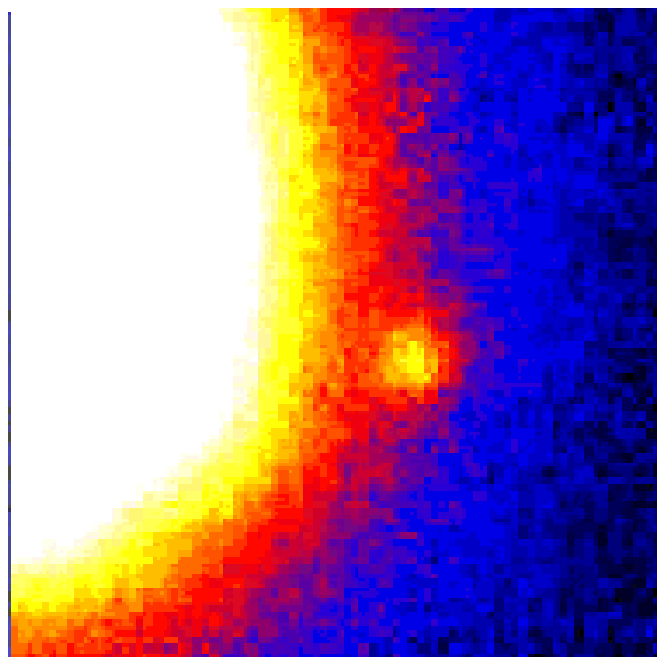,width=6.5cm,angle=0}
\hfil
\epsfig{file=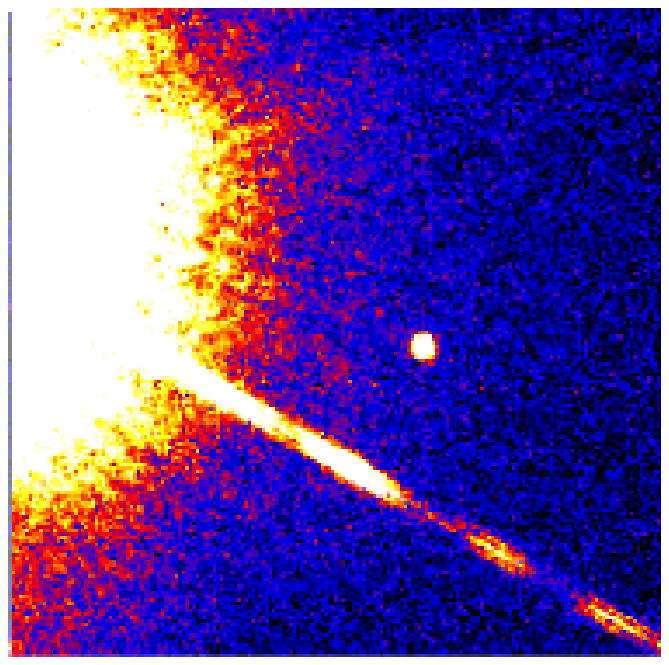,width=6.5cm,angle=0}
}
\end{center}
\vspace{-20pt}
\caption{ 
  Image of the brown dwarf Gliese 229~b, obtained with an adaptive 
  optics system at the Palomar Observatory 60-inch telescope (left)
  and with the Hubble Space Telescope (right).  The brown dwarf is 
  7~arcsec to the lower right of the companion star, Gliese~229.  
  The star/brown dwarf brightness ratio is $\sim$5000, and the 
  distance between the two objects corresponds roughly to the 
  Sun-Pluto separation. A Jupiter mass
  planet at a distance of 10~pc would be 14~times closer to its parent
  star, and roughly 200\,000 times dimmer than Gliese~229~b 
  (courtesy of Tadashi Nakajima).}
\label{fig:gliese229b}
\end{figure}

A promising route to the detection and characterisation of extra-solar
terrestrial planets, and especially for detecting signatures of life, would
be an infrared space interferometer, with baselines of order 50~m.  
Before detailing these ideas, ground-based efforts will be described, 
since a necessary prerequisite for space experiments will be the building 
and testing of ground-based optical interferometers, a task complicated 
considerably by atmospheric turbulence. 

Prospects for ground-based planetary imaging have concentrated on the
use of adaptive optics (e.g.\ \cite{ang94}; \cite{ss95p}; cf.\ the
related but simpler use of active optics which compensates for, e.g.,
gravitational flexure at much lower temporal and spatial frequencies).
Adaptive optics is under intensive development for the latest
generation of large (8--10~m class) astronomical telescopes, aiming to
compensate for atmospheric phase fluctuations across the telescope pupil in
order to achieve diffraction-limited resolution.  The method relies on
continuous measurement of the wavefront from a reference star, then
applying an equal but opposite correction using a deformable mirror
containing actuators distributed across its surface, at frequencies of
order 1~kHz. Adaptive systems typically rely on a nearby bright reference
star to measure these phase fluctuations. Measurements must be made within 
a narrow coherent region, the isoplanatic patch, and over pupil
sub-apertures of size $\simeq r_0$, where $r_0$ is the atmospheric
coherence length (0.15--0.2~m at a good site at visible wavelengths,
increasing to $\sim$1~m at 2~$\mu$m). The use of artificial laser
guide stars from resonant scattering in the mesospheric sodium layer
at 95~km is expected to extend the applicability of the technique to
arbitrary locations on the sky (e.g.\ \cite{lag+98}).

If the position of the planet is known, adjustments can be made so
that the stellar light from the different parts of the mirror arrives
destructively at the planet location.  In the `dark speckle' technique
(\cite{lab95}), rapid random changes in optical path length due to the
atmospheric turbulence are exploited, with the goal of detecting the
planet in very short exposures ($\sim1$~ms) when, by chance, the star
light interferes destructively at the planet location.  The challenges 
posed by such measurements are made clearer by reference to
Figure~\ref{fig:gliese229b}, which shows the brown dwarf Gliese 229~b
imaged from the ground (with adaptive optics) and from space.

Adaptive optics programmes underway at the world's largest
ground-based telescopes being developed for interferometry (the
European Southern Observatory's Very Large Telescope, the ESO VLT at
Cerro Paranal, Chile; and the US Keck telescopes, Hawaii) may
ultimately have the sensitivity to detect giant planets around several
nearby bright stars, taking into account the effects of photon noise
and speckle noise (caused by residual wavefront errors), diffraction
and scattering from the telescope mirrors, the `dark speckle'
technique, and eventual interferometric combination of the beams from
the individual telescopes.  The Keck~II telescope has a 349-actuator
system (\cite{was+00}), the main effect of which will be to achieve a
concentration in light from the planet and hence improve the
planet/star intensity ratio by $\sim100$.  The reduction
of scattered light is not greatly affected beyond field angles
$\lambda/d$ where $d$ is the actuator spacing, corresponding to
0.5~arcsec at 1.25~$\mu$m for a 0.56-m actuator spacing.  Higher-order
systems, e.g.\ the $10^4$ actuator system proposed by \cite*{ang94},
would bring into reach the detection of giant planets around some 100
stars. Massive optical apertures such as the 100-m OWL telescope
(\cite{gdd+98}) are under consideration, and with a planet detection
sensitivity $\propto D^4$, such a telescope equipped with $10^6$
actuators might lead to the detection of Earth-like planets around 100
stars.

The most promising imaging opportunities, however, exist from space.
\cite*{bjl75} considered Lyot filtering to decrease the brightness of 
the Airy rings, while \cite*{ken77} suggested an analogue of phase-contrast 
microscopy to attenuate scattered light arising from the imperfect figure of 
a 2-m space telescope.  \cite*{ell78} proposed a space telescope 
in an orbit yielding a stationary lunar occultation of any star
lasting two hours, using the black limb of the moon as an occulting
edge to reduce the background light from the planet's star, a concept
which has been extended recently to the idea of a large occulting
satellite (\cite{cs00}).

\cite*{bra78} and \cite*{bm79} noted that with Sun/Jupiter
temperatures of 6000\,K and 128\,K, detection of thermal emission in
the Rayleigh-Jeans regime longward of $\sim20\,\mu$m (where the
thermal infrared from the planet is strongest) would result in a
factor of $10^5$ improvement in the intensity contrast.  They also
introduced the principle of nulling interferometry to enhance the
planet/star signal. In this technique, light from two or more small
apertures, typically 20--50~m apart, are combined out of phase, the
baseline being adjusted such that the stellar light interferes
destructively over a broad wavelength range, while the planet signal
interferes constructively (the baseline can be adjusted such that the
radius of constructive interference corresponds to the `habitable
zone').  Rotating the interferometer about an axis through the star
would allow detecting the faint planet from the signal modulated at
harmonics of the spin frequency. Ideas for improved space missions
(\cite{acw86}; \cite{kdt+94}) or balloon experiments above altitudes
of 30~km (\cite{tf97}) have subsequently been developed, with similar
principles being tested on the ground (\cite{hah+00}).  Multi-element
arrays can provide a deep central null with high-resolution fringes
that can be used for mapping.  These should yield full constructive
interference for a close-in planet even in the presence of a resolved
stellar disk, and should allow planets to be resolved from a dust
cloud in the external system (\cite{aw96}; \cite{wa97}; \cite{aw97a}).

\begin{figure}[t]
\begin{center}
\centerline{\epsfig{file=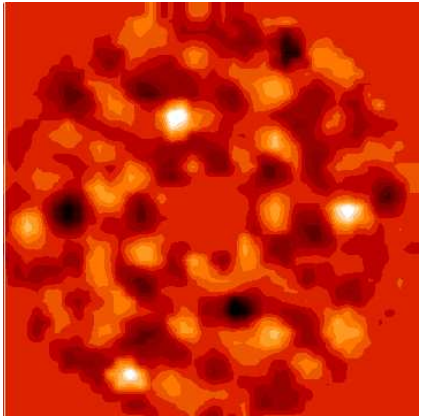,height=5.5cm,angle=0}}
\end{center}
\vspace{-30pt}
\caption{ Simulation of a 60-hour exposure of the proposed ESA Darwin space 
  interferometer (six 1.5-m~telescopes in a 1~AU orbit and 50-m
  baseline).  It simulates a solar-type star at 10~pc, with
  solar-level zodiacal dust, and three inner planets yielding terrestrial-level
  flux at locations corresponding to the positions of Venus, the Earth and
  Mars.  
  The reconstruction method is based on a simple cross-correlation
  algorithm (\protect\cite{aw97a}).  The light from the star, at the
  centre of the image, has been nulled and is therefore invisible. 
  The three brightest regions (top, right, and lower left)
  are the detected planets, with the other features 
  corresponding to artifacts of the image reconstruction
  (courtesy of Bertrand Mennesson).  }
\label{fig:mennesson}
\end{figure}

In addition to the important issue of improving the planet/star
contrast, an infrared space interferometer would provide access to the
spectral region in which molecular species considered as indicators of
life, in particular O$_3$ and H$_2$O, are present (see
Section~\ref{sec:life}).  The European Space Agency is considering the
Darwin Infrared Space Interferometer as a high-priority but longer-term 
programme (\cite{leg+93}; \cite{lmm+96b}; \cite{mlm+97}; \cite{mm97}; 
\cite{lmo+98}; \cite{fri00}).  It would employ nulling interferometry in the 
infrared to detect and obtain spectra of Earth-like planets around 100--200
stars out to distances of 15--20~pc. It would comprise
4--6~free-flying 1-m class telescopes, passively-cooled to 40~K,
separated by up to 50~m, and operating between 6--17\,$\mu$m to cover
spectral lines including H$_2$O, CH$_4$, O$_3$ and CO$_2$
(Figure~\ref{fig:mennesson}).  Observations in the infrared bring one
specific complication, notably the background radiation from the Solar
System zodiacal light, i.e.\ emission from Solar System dust (from
comets and asteroids) which itself peaks around 10--20~$\mu$m --- if 
the instrument were placed at 4--5~AU from the Sun the corresponding 
background contribution would be reduced by
about 100 (\cite{lmo+98}).  NASA is considering a 75--1000~m baseline
infrared interferometer TPF (Terrestrial Planet Finder, \cite{bei96};
\cite{bei98}) as part of its Origins Program (\cite{thr97}).
Space Technology-3 is a NASA experiment to be flown in 2005 to test
interferometry between free-flying satellites using two 12-cm mirrors
up to 1~km apart.


Both Darwin and TPF are targeted for launch some time after 2010.  The 
common programme goals may or may not result in a collaborative venture
between ESA and NASA.  Issues of Solar System zodiacal emission, and
the effects of extra-solar zodiacal emission on detection
probabilities (\cite{ang98}), mean that precursor space missions with
less-ambitious goals have also been proposed (\cite{hwa99});
telescopes with high finesse active optics to reduce rms aberrations
to a few nm could detect Jupiters around a few stars with 3-m
apertures, and Earth-like planets around 100 stars with 30-m apertures
(\cite{mys95}). Meanwhile the capabilities of the Hubble
Space Telescope (\cite{sg96}; see also Section~\ref{sec:disks} and
Figure~\ref{fig:hst-disk}), NASA's SIRTF (a 0.85~m
cryogenically-cooled infrared observatory, to be launched in 2002,
\cite{cw97}), and the planned Next Generation Space Telescope (NGST, to be 
launched 2009, \cite{mbs97}) have been considered for their brown dwarf
and planet imaging capabilities.

Technology precursors for these ambitious space interferometers include
the IOTA interferometer, two 0.45-m telescopes with baselines up to
38~m, in operation since 1995 (\cite{tra98});
the Palomar Testbed Interferometer (\cite{col97}; \cite{pkc+98}), a 
110-m baseline operating at K-band (2--2.4~$\mu$m) with baselines 
stabilised to a few nm against large rapid changes
imposed by atmospheric turbulence and Earth rotation;
and the CHARA array on Mt.~Wilson, six 1-m telescopes along three 200-m 
long arms, operating in the visible and K-band and 
expected to be operational in early 2001 (\cite{mca99}).
Future milestones will be interferometric operation of the two 85-m baseline
10-m Keck telescopes (combined with four auxiliary 2-m telescopes), and
the interferometric combination over a 200-m baseline of ESO's 
four 8-m and (presently) three auxiliary 1.8-m telescopes. 
%
 
NASA's Space Interferometry Mission, SIM, due for launch in 2005, is
primarily designed for optical astrometry at microarcsec level with maximum
baselines of about 10~m. Itself conceived partly as a technological
precursor for TPF, it will be capable of imaging and nulling
(\cite{ba99}) albeit at much lower angular resolution than planned for
TPF. Systems with similar disk size and more than 0.1 of the dust
content of $\beta$~Pic (about 1000 times the dust content and surface
brightness of our Solar System, see Section~\ref{sec:disks}) will be
detected out to a few kpc if nulling efficiencies of $10^{-4}$ are
achieved; inner clear regions indicative of the presence of massive
planets should be detected and imaged for such systems.

Many other ground-based instruments bear some relevance to planetary
search programmes.  After completion in 2003, the Large Binocular
Telescope (two 8.4~m telescopes on a common mount with a 14.4-m baseline) 
will be able to assess the expected sensitivity of TPF to
extra-solar zodiacal emission (\cite{aw97c}; \cite{haw+99}).  At
11~$\mu$m it will be sensitive to solar-level zodiacal emission
at 0.8~AU from a star at 10~pc.  At 4~$\mu$m a Jupiter-size planet,
1~Gyr old, could be detected as close in as~0.3~AU.  Studies have been
made of infrared/sub-mm observations from the high, dry Antarctic
plateau (\cite{bal94}), while further in the future, the Atacama Large 
Millimeter Array (ALMA) is a mm and sub-mm wave interferometer consisting 
of $64\times12$-m antennas to be built in the Atacama desert in Northern Chile.

In summary, imaging of Earth-mass extra-solar planets from large
ground-based telescopes equipped with adaptive optics and operating in
interferometric combination, and observations in the infrared using
space interferometers, are receiving considerable attention.  While
the commitment is impressive, dedicated space missions are probably
10--15~years or more away.

At the start of this section it was noted that extra-solar planetary
imaging generally refers to the detection of a reflection point-source
image of the planet, rather than to resolution of the extra-solar
planet surface.  Ground- or space-based (or lunar) interferometric 
arrays of 10--100~km baseline could start to tackle resolved planetary  
imaging (\cite{lab96}). \cite*{bs96} undertook a partial design of a separated
spacecraft interferometer which could achieve visible light images
with $10\times10$ resolution elements across an Earth-like planet at
10~pc.  This called for 15--25 telescopes of 10-m aperture, spread
over 200~km baselines, with the dominant problem being that of
suppressing starlight to the necessary levels.  Reaching $100\times100$
resolution elements would require 150--200 spacecraft distributed over
2000~km baselines, and an observation time of 10~years per planet.
Maintaining the necessary telescope separation geometry, using laser
metrology and 1\,mN field-emission electric propulsion thrusters
(FEEPs), seems more tractable, and indeed comparable requirements are
necessary for ESA's proposed gravitational wave detection
interferometer, LISA.  \cite*{bs96} noted that the resources they
identified would dwarf those of the Apollo Program or the Space
Station, concluding that it was {\it `difficult to see how such a
program could be justified'}.  In the approach of \cite*{lab99a}
a 30-min exposure using a hyper-telescope comprising 150 3-m diameter 
mirrors in space with separations up to 150~km, would be sufficient 
to detect green spots similar to the Earth's Amazon basin on a planet 
at a distance of 10~light-years.

\subsection{Dynamical perturbation of the star}
\label{sec:perturbation}

The motion of a single planet in a circular orbit around a star causes
the star to undergo a reflex circular motion about the star-planet
barycentre, with orbital radius $a_*=a\cdot(M_p/M_*)$ and period $P$.
This results in the periodic perturbation of three observables, all
of which have been detected (albeit in different systems): in radial
velocity, in angular (or astrometric) position, and in time of arrival
of some periodic reference signal.

\begin{figure}[tbh]
\begin{center}
\centerline{
\epsfig{file=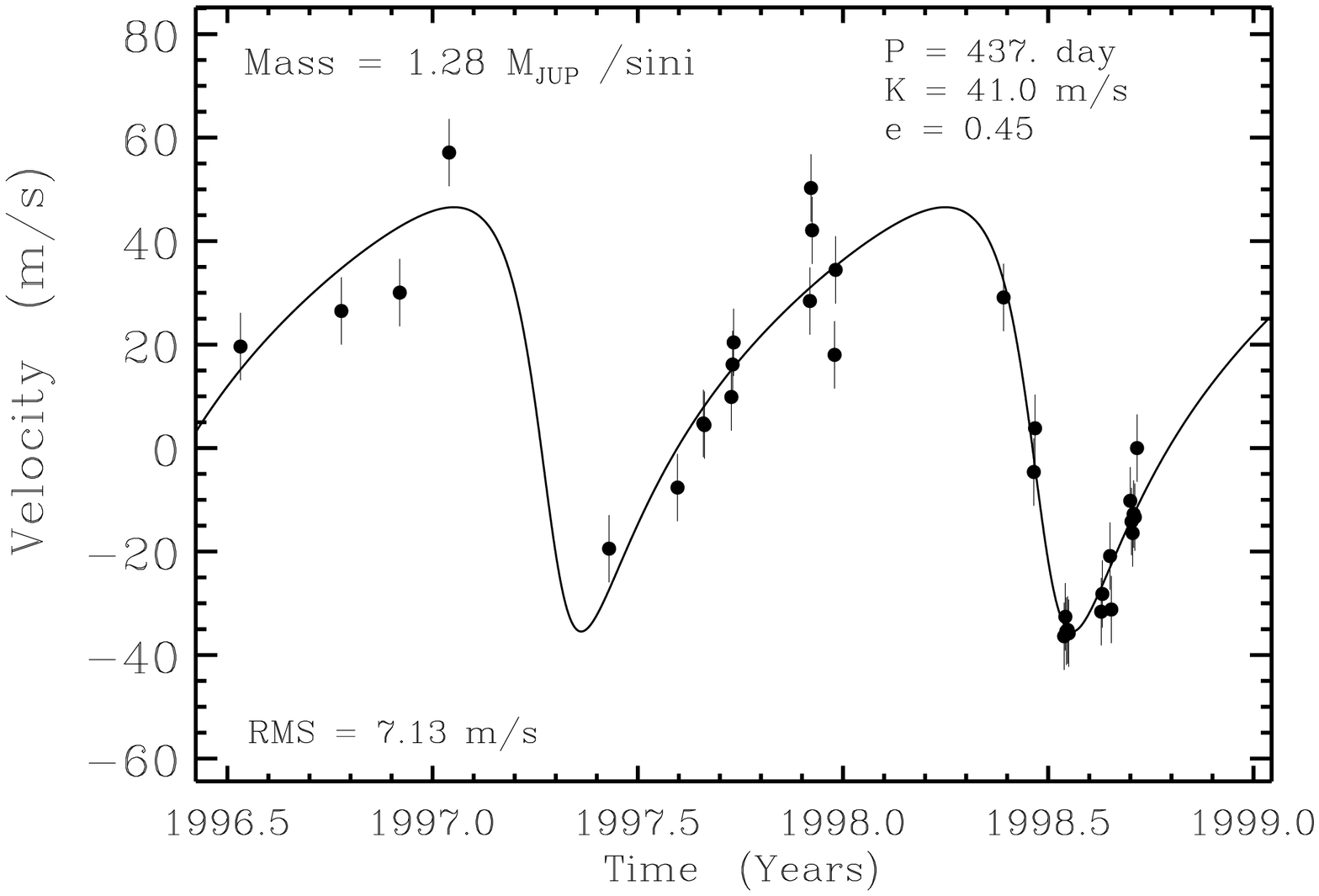,width=6.0cm,angle=90}
\hfil
\epsfig{file=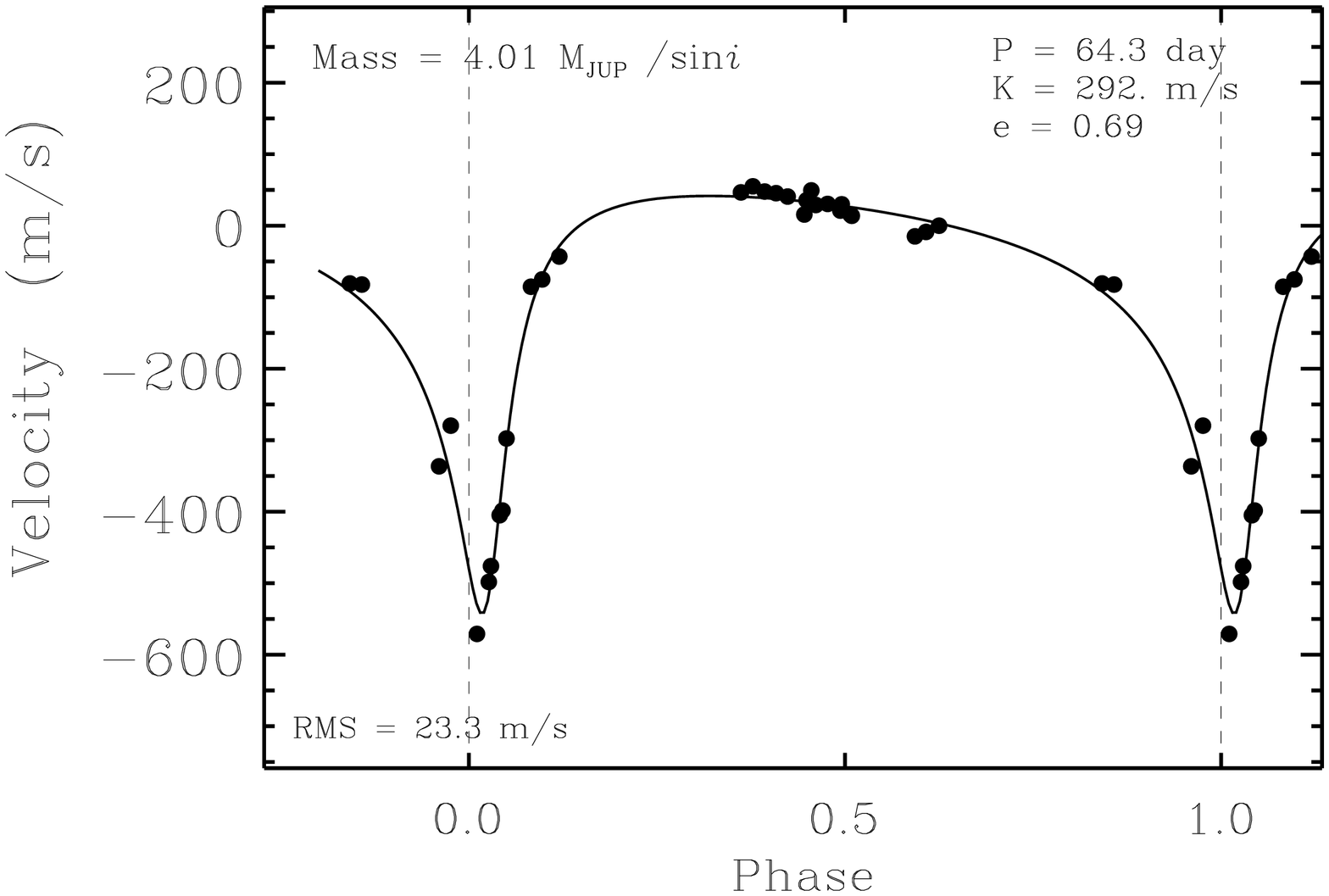,width=6.0cm,angle=90}
}
\end{center}
\vspace{-30pt}
\caption{Examples of radial velocity measurements: 
  HD~210277 (left) and HD~168443 (right), from
  \protect\cite*{mbv+99a}, obtained with the HIRES spectrometer on the
  Keck telescope. The solid lines show the best-fit Keplerian models.
  The non-sinusoidal variations result from the eccentric orbits, and
  the derived $M\sin i$ values are 1.28 and 4.01~$M_\Jupiter$
  respectively. The fit for HD~168443 is improved further by a linear
  velocity trend, suggestive of an additional nearby, long-period
  stellar or brown dwarf companion (courtesy of Geoffrey Marcy).  }
\label{fig:marcy}
\end{figure}

\subsubsection{Radial velocity}
\label{sec:radvel}

The velocity amplitude $K$ of a star of mass $M_*$ due to a companion
with mass $M_p\sin i$ with orbital period $P$ and eccentricity $e$ is
(e.g.\ \cite{cmb99}):
\begin{equation}
\label{equ:radvel1}
K=\left ( \frac{2\pi G}{P} \right )^{1/3}\ 
	\frac{M_p \sin i}{(M_p + M_*)^{2/3}}\ \frac{1}{(1-e^2)^{1/2}}
\end{equation}
In a circular orbit the velocity variations are sinusoidal, and 
for $M_p \ll M_*$ the amplitude reduces to:
\begin{equation}
\label{equ:radvel2}
K=28.4\,  \left ( \frac{P}{1\ {\rm year}} \right )^{-1/3}\ 
         \left ( \frac{M_p \sin i}{M_\Jupiter} \right ) \ 
         \left ( \frac{M_*}{M_\odot} \right )^{-2/3}\ \ {\rm m\,s}^{-1}
\end{equation}
where $P$ and $a$ are related by Kepler's Third Law:
\begin{equation}
\label{equ:radvel3}
P=  \left ( \frac{a}{1~{\rm AU}} \right )^{3/2} \ 
         \left ( \frac{M_*}{M_\odot} \right )^{-1/2}\ \ {\rm year}
\end{equation}

The effect is about $K=12.5$~m~s$^{-1}$ with a period of 11.9~yr in
the case of Jupiter orbiting the Sun, and about 0.1~m~s$^{-1}$ for the
Earth. The $\sin i$ dependence means that orbital systems seen face on
($i=0$) result in no measurable radial velocity perturbation and that,
conversely, radial velocity measurements can determine only $M_p \sin
i$ rather than $M_p$, and hence provide only a lower limit to the
planet mass since the orbital inclination is generally unknown.
Although the radial velocity amplitude is independent of the distance
to the star, signal-to-noise considerations limit observations to the
brighter stars (typically $V<8$~mag). Equation~\ref{equ:radvel1}
indicates that radial velocity measurements favour the detection of
systems with massive planets, and with small $a$ (and hence small~$P$).

All of the known extra-solar planets around normal main-sequence
stars have been discovered, starting with the first in 1995, using
radial velocity techniques.  A compilation of these systems, through
to the end of March 2000, is given in Table~\ref{tab:known-planets}, which
also includes some derived parameters discussed in
Section~\ref{sec:properties}. In order to measure these systems,
accuracies of around 15~m\,s$^{-1}$ or better have been needed. This
is extremely challenging: stars are only faint sources of light, so
that large telescopes and long integration times are required for high
signal-to-noise and sub-pixel accuracies.  In addition, gravitational
flexure affects the telescope which must follow the star's apparent
motion as a result of Earth rotation; and very accurate instrumental
and wavelength calibration is demanded. Typically, spectrographs with
resolution of around $\lambda/\Delta\lambda\sim60\,000$ are operated
in the optical region (450--700~nm), using gas absorption cells (HF or
I) to provide numerous accurate wavelength reference features
superimposed on the stellar spectral lines (\cite{gg73}; \cite{cw79};
\cite{mb92p}).  An instantaneous measurement of the stellar velocity
is given by the small, systematic change in wavelength of the many
absorption lines that make up the star's spectrum.  A precise
ephemeris accounts for all known motions of the Earth, including
gravitational perturbations by other planets in the Solar System.
Current state-of-the-art measurements reach around 3~m\,s$^{-1}$
(\cite{bmw+96}), corresponding to an accuracy of about 1~part in
$10^8$ in wavelength, with developments to about 1--2~m\,s$^{-1}$
planned (\cite{khc+99c}).  A projected precision of 1~m~s$^{-1}$ using
absolute accelerometry is under development (\cite{con94}; \cite{bcb99}). 
Intrinsic accuracy limits of around $\pm1$~m~s$^{-1}$
may arise from the effects of star spots and convective
inhomogeneities on the stellar surface, even in older less-active
stars (\cite{sd97}; \cite{sbm98}; \cite{bm98}).

Early radial velocity surveys on a few (10--30) stars aimed to
characterise the sub-stellar/brown dwarf mass function by searching
for binary companions of main-sequence stars with masses below
1\,$M_\odot$ (\cite{cwy88}; \cite{mm89}; \cite{mb89}; \cite{msp+90};
\cite{dm91}; \cite{tok92}). As accuracies improved towards expected
planetary signals, existing groups intensified their efforts, and
others started new observing programmes, leading to the monitoring of
many more stars over a number of years, at accuracies of typically
15~m\,s$^{-1}$ or better. An incomplete list of these programmes,
which now typically make use of several tens of nights on each of many
telescopes throughout the world, includes:
University of British Columbia (from 1983: \cite{wwi95}); 
Arizona (from 1987, using Fabry-Perot techniques: \cite{mmp+94});
McDonald Observatory Planetary Search, Texas 
	(MPOS, from 1988: \cite{ch94}; \cite{hc98c}); 
Lick Observatory (from 1992: \cite{mb92p}; \cite{bm97}; \cite{cmb99});
Advanced Fibre-Optic Echelle (AFOE) at the Whipple Telescope in
        Arizona (\cite{bnn+94}; \cite{njk+97b});
ESO Planet Search at the European Southern Observatory 
        (\cite{hkc+96}; \cite{khc+99c});
and the Elodie spectrometer at Observatoire de Haute Provence 
        (\cite{mqu+97}; \cite{qms+98}).
Other major programmes started on the Keck~I 10-m telescope with the HIRES
spectrometer in 1996
(monitoring 500 main sequence stars from F7-M5, 
 with a precision of 3~m~s$^{-1}$
  and recently yielding 6 new candidates, \cite{vmb+00}); 
on the Anglo-Australian Telescope in 1997;
on the Swiss 1.2-m Euler telescope at La Silla with the Coralie 
        spectrometer in 1998 (\cite{qmw+00}); 
and on the 10-m Texas Hobby-Eberly Telescope.
These programmes are now each monitoring typically 100--300 stars in a
systematic manner, with Coralie surveying about 1600. 
Most late-type main sequence stars brighter than $V\sim7.5$~mag are currently 
being surveyed.  Reviews of the progress of these radial velocity 
searches have been given by
\cite*{lat97};
\cite*{bm98};
\cite*{lsm+98};
\cite*{mb98b};
\cite*{na98};
and \cite*{mb00}.

\cite*{lms+89} first reported a low-mass companion to a star,
HD~114762, inferring a mass of about 10\,$M_\Jupiter$, which was
further constrained by observations by \cite*{chh91}, and is
consistent with a massive planet or low-mass brown dwarf.  However, by
1995, none of the ongoing programmes had measured any systems with
smaller masses expected to be representative of a postulated planetary
mass population, and there seemed to be a deficiency of sub-solar mass
stars which could represent the tail of the stellar mass function.

The first report of a planetary candidate of significantly lower mass,
surrounding the star 51~Peg, was announced by \cite*{mq95}.  This
discovery was rapidly confirmed by the Lick Observatory group, who
were also quickly able to report two new planets that they had been
monitoring: 70~Vir (\cite{mb96}) and 47~UMa (\cite{bm96}).

With $P=4.2$~days and $a=0.05$~AU, 51~Peg provoked early controversy,
partly in view of its unexpectedly short orbital period and hence
close proximity to the parent star), but also since an alternative
explanation -- that the radial velocity shifts arose from non-radial
oscillations -- was put forward to explain possible distortions in the
absorption line profile bisector (the locus of midpoints from the core
up to the continuum). 
Studies that followed (\cite{gra97}; \cite{hcj97}; \cite{mbw+97};
\cite{gh97}; \cite{whs+97}; \cite{bkh+98a}; \cite{bkh+98b}; 
\cite{gra98}; \cite{hcb98a};
\cite{hcb98b}) finally resulted in a consensus that the planet
hypothesis was the most reasonable possibility.  Other planets have 
typically not proven to be controversial, and work since the discovery 
papers is devoted to studies of their properties or those of their 
parent star.

Additional detections by a number of different groups have followed rapidly.
The present list of stars with inferred planetary mass companions
numbers 34 (see Table~\ref{tab:known-planets}), with further
detections expected to follow steadily as new surveys, more objects,
higher measurement accuracies, and longer temporal baselines
(permitting the discovery of longer period systems) take effect. The
observed and derived properties of these systems are proving fertile
ground for theories of planetary formation
(Section~\ref{sec:properties}).

The dearth of brown dwarf candidates has continued, with the precision
Doppler surveys, along with lower precision surveys of several
thousand stars, having revealed only 11 orbiting brown dwarf
candidates in the mass range $M\sin i = 8-80\, M_\Jupiter$, despite
the fact that there is no bias against brown dwarf companions, 
and the fact that they are much easier to detect than
companions of Jupiter mass.  Most of these objects in fact appear to
be hydrogen-burning stars with low orbital inclination (\cite{mqu+97};
\cite{ham+00}; \cite{umn+00}); this paucity of brown dwarf companions
presently renders the planets distinguishable by their high occurrence
at low masses.

\subsubsection{Astrometric position}
\label{sec:astrometry}

The path of a star orbiting the star-planet barycentre appears
projected on the plane of the sky as an ellipse with angular
semi-major axis $\alpha$ given by:
\begin{equation}
\label{equ:astrom}
\alpha = \frac{M_p}{M_*} \cdot \frac{a}{d}
\end{equation}
where $\alpha$ is in arcsec when $a$ is in~AU and $d$ is in~pc (and
$M_p$ and $M_*$ are in common units).  This `astrometric signature' is
therefore proportional to both the planet mass and the orbital radius,
and inversely proportional to the distance to the star.  Astrometric
techniques aim to measure this transverse component of the
photocentric displacement.  Jupiter orbiting the Sun viewed from a
distance of 10~pc would result in an astrometric amplitude of
500~microarcsec ($\mu$as), while the effect of the Earth at 10~pc is a
one-year period with 0.3~$\mu$as amplitude (the motion of the Sun
over, say 50~years, is complex due to the combined gravitational effect 
of all the planets).  The astrometric accuracy required to detect planets
through this reflex motion is therefore typically in the
sub-milliarcsec range, although it would reach a few milliarcsec for
$M_p=1\,M_\Jupiter$ for very nearby solar-mass stars. 

The astrometric technique is particularly sensitive to relatively long
orbital periods ($P>1$~yr), and hence complements radial velocity
measurements. The method is also applicable to hot or rapidly rotating stars
for which radial velocity techniques are limited. One important 
feature is that if $a$ is known from spectroscopic measurements, $d$ 
from the star's parallax motion, and if $M_*$ can be estimated from its 
spectral type or from evolutionary models, then the astrometric displacement
yields $M_p$ directly rather than $M_p\sin i$ given by radial velocity
measurements; a single measurement of the angular separation at one
epoch from ground-based interferometric astrometry would also provide
orbital constraints on $\sin i$ (\cite{tor99}). For multi-planet systems
astrometric measurements can determine their relative orbital
inclinations (i.e., whether the planets are co-planar), an important
ingredient for formation theories and dynamical stability analyses.
Possible mimicking of astrometric planetary perturbations by a nearby
binary star has been considered by \cite*{sch00b}.

Astrometric detection demands very accurate positional measurements
within a well-defined reference system at a number of epochs, and is
very challenging for optical measurements from the ground, again
because of the atmospheric phase fluctuations. Across the
electromagnetic spectrum accessible to astronomical observations,
milliarcsec positional measurements or better are presently only
achieved at radio and optical wavelengths, and these are considered in
turn.

\begin{figure}[tbh]
\begin{center}
\leavevmode
\centerline{
\epsfig{file=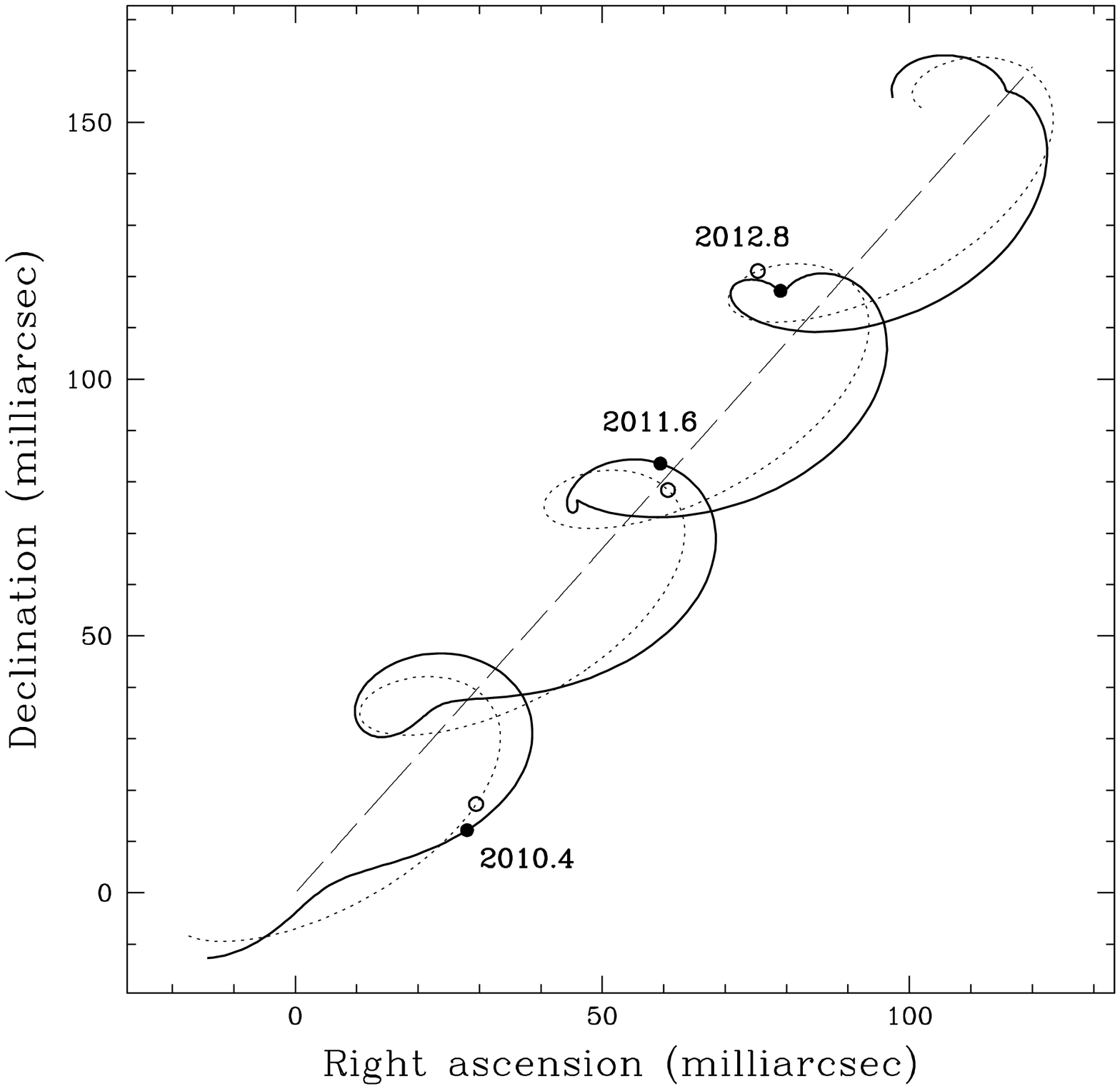,width=7.5cm,angle=0}
\hfil
\epsfig{file=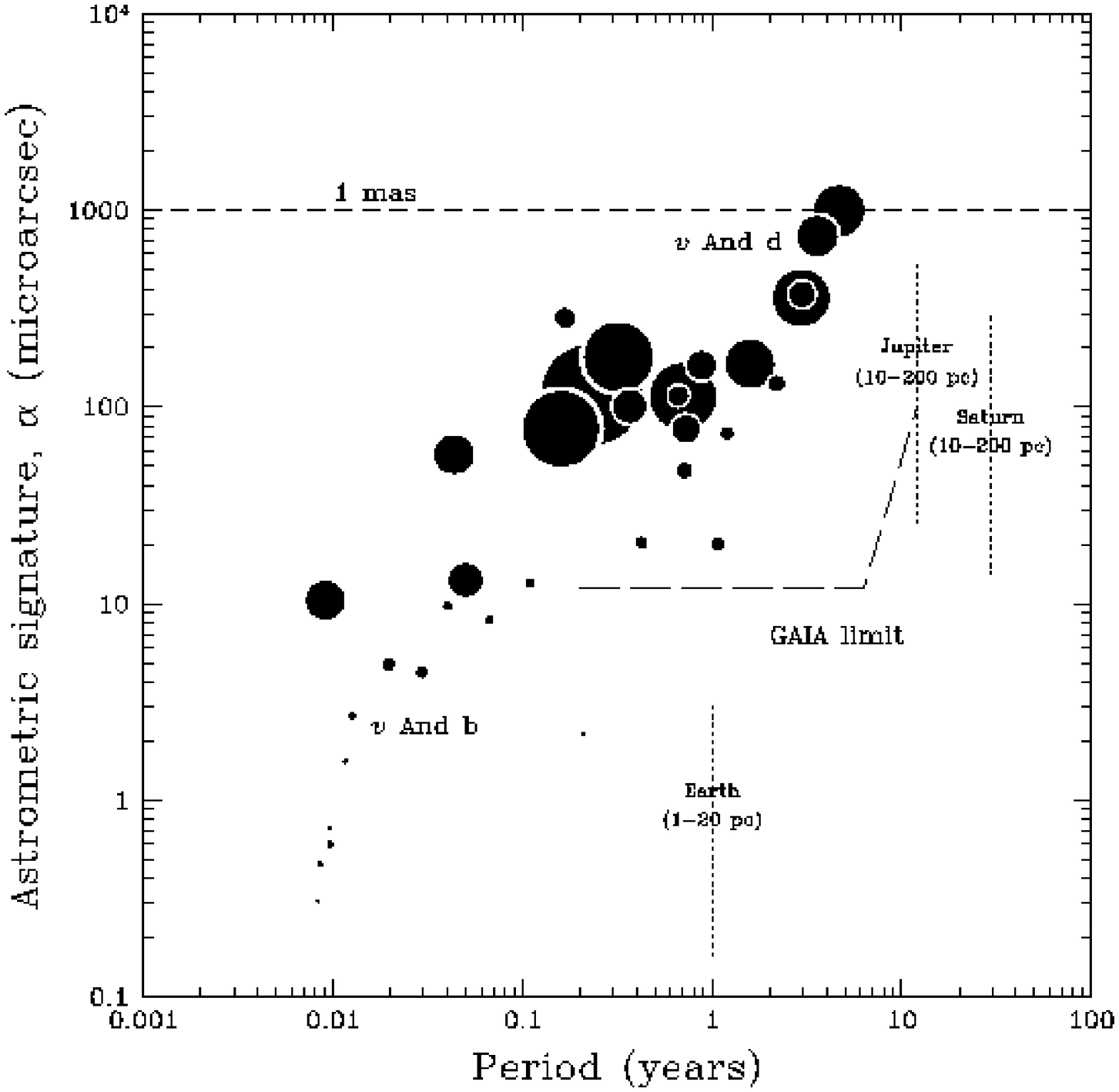,width=7.5cm,angle=0}
}
\end{center}
\vspace{-10pt}
\caption{Left: the modeled path on the sky of a star at a distance of
  50~pc, with a proper motion of 50~mas yr$^{-1}$, and orbited by a
  planet of $M_p=15\,M_\Jupiter$, $e=0.2$, and $a=0.6$~AU. The straight
  dashed line shows the path of the system's barycentric motion viewed
  from the Solar System barycentre. The dotted line shows the effect
  of parallax (the Earth's orbital motion around the Sun, with a
  period of 1~year). The solid line shows the apparent motion of the
  star as a result of the planet, the additional perturbation being
  magnified by $\times30$ for visibility.  Labels indicate times in years
  (relevant for the planned GAIA mission launch date of 2009).
  Right: astrometric signature, $\alpha$ (equation~\ref{equ:astrom}),
  induced on the parent star for the known planetary systems listed in
  Table~\ref{tab:known-planets} as a function of orbital period
  (adapted from \protect\cite{lss+00}).
  Circles are shown with a radius proportional to $M_p\sin i$.
  Astrometry at the milliarcsec level has negligible power in
  detecting these systems, while the situation changes dramatically
  for microarcsec measurements.  The positions of the innermost
  short-period planet (b) and outermost longest period planet (d) in
  the $\upsilon$~And triple system are indicated: short-period systems to
  which radial velocity measurements are sensitive are difficult to
  detect astrometrically, while the longest period systems will be
  straightforward for microarcsec positional measurements.  Effects
  of Earth, Jupiter, and Saturn are shown at the distances indicated. }
\label{fig:astrom}
\vspace{-10pt}
\end{figure}

In the radio, milliarcsec positional accuracy is now reached through
high-precision very long baseline interferometry (VLBI), using
baselines of order 3000~km, with phase referencing techniques
providing sufficient sensitivity to detect a number of nearby radio
emitting stars. The highest accuracies have been reported for the
close binary system $\sigma^2$~CrB (\cite{lpj+96}; \cite{lpj+99}).  In
these systems, named after their prototype RS~CVn, quiescent and
flaring gyro-synchrotron radio emission is generated from MeV
electrons in magnetic structures related to the intra-stellar region
and stellar photosphere respectively. Observations of $\sigma^2$~CrB since 
1987 yielded post-fit rms residuals of 0.20~milliarcsec (\cite{lpj+96}) which
would correspond, at a distance $d=21$\,pc, to the displacement which
would be expected for a Jupiter-like planet around the binary system.
Only about five nearby radio stars are suitable for monitoring as part
of a planetary search programme: Jupiter mass companions would produce
perturbations above 1~milliarcsec, while an Earth-like planet around
the nearest radio-emitting star UV~Ceti would result in a displacement
of about 8~microarcsec.  No candidate planets have so far been
reported using this technique.

Discussions of ground-based optical observations related to planet
detection are given by \cite*{bs82} and \cite*{gat87}. Unconfirmed 
reports of small, long-period astrometric displacements consistent 
with planetary bodies have been made for Barnard's Star, based on 
observations over many years, yielding two proposed planetary mass 
bodies (0.7 and 0.5~$M_\Jupiter$) with periods of 12 and 20 years 
respectively (\cite{kam82}), and for Lalande 21185 with a mass of 
0.9~$M_\Jupiter$ and $P=5.8$~years (\cite{gat96}). 

Measurement of sub-milliarcsec displacements has been impossible to
date because of atmospheric effects.  Accuracies for narrow-angle
ground-based measurements are improving rapidly, with prospects for
measurements at the tens of microarcsec or better, building on the
success of the Mk~III optical interferometer (\cite{hum+94}), e.g.\
from the Palomar Testbed Interferometer (\cite{cs94}), from the Keck
Interferometer (\cite{cbc+98}), and from ESO's VLTI (\cite{mdd+98}). 
These instruments should ultimately provide very high relative
astrometric accuracy over small fields (for example, on bright double
or multiple systems) at the 10~microarcsec level (\cite{sc92};
\cite{sha95}).  Within ESO's VLTI programme, PRIMA (Phase-Referenced
Imaging and Microarcsecond Astrometry) aims to use the four 8.2-m 
telescopes to achieve 10--50~microarcsec astrometry for various
narrow-field applications.

Astrometric measurements can be made more accurately from above the Earth's
atmosphere.  Only a single astrometric space mission has been carried
out to date, Hipparcos (\cite{esa97}; \cite{plk+97}), which provided
$\sim1$~milliarcsec accuracy for about 120\,000 stars.  Specific
methods have been applied to the individual Hipparcos measurements for
the detection of brown dwarfs (\cite{bb95}; \cite{ber97}).  
For known planetary systems, the Hipparcos data have provided some
constraints on planetary masses: \cite*{pla+96} derived weak upper 
limits on $M_p$ in a few cases; \cite*{mzt+99} derived a mass for the 
outer companion of $\upsilon$~And of $10.1^{+4.7}_{-4.6}\,M_\Jupiter$, 
compared with $M_p \sin i=4.1\,M_\Jupiter$; and \cite*{zm00} have 
concluded that the companion of HD~10697 is probably a brown dwarf.

Although some inferences about the properties of systems discovered by
radial velocity measurements have been possible from the Hipparcos
results, it is evident from equation~\ref{equ:astrom} (and
Figure~\ref{fig:astrom}) that milliarcsec astrometry can contribute
only marginally to extra-solar planet detection.  In contrast,
large-scale acquisition of microarcsec astrometric measurements in the
future promises at least three important developments. First, measurements
significantly below 0.1~milliarcsec offer planet detection
possibilities well below the Jupiter mass limit, out to 50--200~pc. 
Second, in combination with spectroscopic measurements they provide
direct determination of the planet mass (in terms of the star mass),
independent of the orbital inclination. Third, the relative orbital 
inclination of multi-planet systems can be determined.

Future space astrometry experiments include those targeting
measurements accuracies of below a milli-arcsecond, specifically DIVA,
a German national project at the proposal stage (\cite{rbb+97}) and
FAME, an approved NASA project (\cite{gum+97}).  At even higher
accuracies are the microarcsec-class SIM mission, a NASA approved
project (\cite{bus97}; \cite{du99}) and GAIA, an ESA project at the
proposal stage (\cite{mig99}; \cite{gbf+00}).  SIM is a pointed
mission which will be especially powerful for detailed orbit
determinations for planetary systems detected from ground-based
experiments.  GAIA will survey approximately a billion stars to
$V\sim20$~mag as part of a census of the Galactic stellar population. 
On the assumption that 4--5~per cent of solar-type stars have
Jupiter-mass companions (\cite{mcm00}) it will detect (and provide
orbital parameters in many cases) for upwards of 10\,000 planetary
systems of mass $\sim1\,M_\Jupiter$ and periods around 1--10~years
(\cite{lss+00}), for stars as faint as about 15~mag. With proposed
launch dates between 2004 (FAME) to 2009 (GAIA) astrometric space
missions will contribute to the detection and orbital determination of
large numbers of planetary systems, providing target lists for further
observations by other techniques (Figure~\ref{fig:astrom}).

While sub-microarcsec astrometry may be feasible in the more distant
future, the ultimate limit to the astrometric detection of Earth-like
planets from space may be the non-uniformity of illumination over the
disk of a star. The Earth causes the Sun's centre of mass to move with
a semi-amplitude of about 500~km (0.03~per cent of the stellar
diameter), while sunspots with up to 1~per cent of the Sun's area will
cause the apparent centre of light of the Sun to move by up to 0.5~per
cent (\cite{wa98}).

\subsubsection{Timing and the pulsar planets}
\label{sec:timing}

Although all orbital systems are affected by changes in light travel
time across the orbit, in general there is no timing reference on
which to base such measurements.  A notable exception are radio
pulsars, rapidly spinning highly-magnetised neutron stars, formed
during the core collapse of massive stars (8--20\,$M_\odot$) in a
supernova explosion.  Pulsars emit narrow beams of radio emission
parallel to their magnetic dipole axis, seen as intense pulses at the
object's spin frequency due to a misalignment of the magnetic and spin
axes. There are two broad classes: `normal' pulsars, with spin periods
around 1~s, and of which several hundred are known; and the
millisecond pulsars, `recycled' old ($\sim10^9$~yr) neutron stars that
have been spun-up to very short spin periods during mass and angular
momentum transfer from a binary companion, with most of the 30~known
objects still having (non-accreting) binary companions, either white
dwarfs or neutron stars.  The latter are extremely accurate frequency
standards, with periods changing only through a tiny spin-down at a
rate $\sim10^{-19}$\,s\,s$^{-1}$ presumed due to their low magnetic
field strength (\cite{bai96}).

For a circular, edge-on orbit, and a canonical pulsar mass of
$1.35\,M_\odot$, the amplitude of timing residuals due to planetary
motion is (\cite{wol97}):
\begin{equation}
\label{equ:wol}
\tau_p = 1.2\,  \left ( \frac{M_p}{M_\oplus} \right )\, 
           \left ( \frac{P}{1\ {\rm yr}} \right )^{2/3}\ \ {\rm ms}
\end{equation}
The extremely high accuracy of pulsar timing allows the detection of
lower mass bodies orbiting the pulsar to be inferred
from changes in pulse arrival times due to orbital motion.  Jovian or
terrestrial planets are expected to be detectable around
`normal' slow pulsars, while substantially lower masses, down to that
of our Moon and largest asteroids, could be recognised in millisec
pulsar timing residuals.

The first planetary system discovered around an object other than our
Sun was found around the 6.2-ms pulsar PSR~1257+12 ($d\sim500$~pc),
with at least two plausible terrestrial-mass companions (\cite{wf92})
having masses of 2.8 and $3.4\,M_\oplus$, and almost circular orbits
with $a=0.47$ and 0.36~AU, and $P=98.22$ and 66.54~days respectively,
close to a 3:2 orbital resonance.  Although a number of alternative
ways of producing the observed timing residuals were examined
(\cite{pt94}), the planet hypothesis was rigorously verifiable: the
semi-major axes of the orbits are sufficiently similar that the two
planets would perturb one another significantly during individual
close encounters, with resulting three-body effects leading to
departures from a simple non-interacting Keplerian model growing
rapidly with time (\cite{mbe+92}; \cite{rns+92}; \cite{mal93};
\cite{pea94}).

Continued monitoring of PSR~1257+12 provided evidence for a third
companion (\cite{wol94a}; \cite{wol94b}) with $P=25.34$~days, and
possibly a fourth, with $P\sim170$~yr (\cite{wol97}), as well as
confirmation of the predicted mutual gravitational perturbations
(\cite{wol94b}; \cite{kmw99}).  The third planet may be an artifact of
temporal changes of heliospheric electron density at the solar
rotation rate at the relevant heliospheric latitude (\cite{sfa+97}),
although this unconfirmed explanation would not invalidate the
existence of the other two planets; further observations will
establish if the 25.3~day periodicity is frequency independent, and
thus unrelated to solar interference.  Dynamical simulations indicate
that the system would be stable over some hundreds of thousands of
years (\cite{gla93}).

Evidence for a long-period planet around the slow pulsar PSR~B0329+54
(spin period 0.71~s) was reported by \cite*{dp79}, based on large
second time derivatives of the spin frequency. Although alternative
explanations were also given, the planetary interpretation was
supported by \cite*{sha95p}, who gave $P=16.9$~yr, $M_p>2\,M_\oplus$,
$e=0.23$ and $a=7.3$~AU. Both groups reported an additional 3~yr
periodicity in pulse arrival times. The planetary hypothesis has 
recently been questioned by \cite*{klw+99} based on further
observations, with variations in the timing residuals for this
relatively young neutron star attributed to spin irregularities or
precession of the pulsar spin axis.

\cite*{ajr+96} used similar timing techniques to detect a
10--30\,$M_\Jupiter$ object orbiting the binary pulsar PSR~B1620--26
in the globular cluster~M4. The pulsar is tightly orbited by a white
dwarf, and the third object is the lightest and most distant member of
a hierarchical triple system, with $a\sim$~35--60\,AU and
$P\sim100$\,yr.  A possible formation scenario for this triple system
involves a dynamical exchange interaction between the binary pulsar
and a primordial star-planet system, in which the planet was captured
by the binary (\cite{fjr+00}). Whether the outer body is a planet or a
brown dwarf remains to be established, but further examples and
detailed evaluation of capture probabilities may indicate whether
significant numbers of extra-solar planetary systems also exist in
old star clusters (\cite{fjr+00}; \cite{tac+99}).

Unconfirmed evidence for a companion ($a=3.3$~AU and $M_p\sin i =
1.7\,M_\odot$) to the radio quiet pulsar Geminga was reported from
$\gamma$~ray observations (\cite{mhc98}), although this may have been
an artifact of the spin period. A
possible companion to PSR~1829--10 (\cite{bls91}) was subsequently
retracted (\cite{lb92}). No other planetary systems around millisec
pulsars have been discovered, suggesting that planetary systems around
old neutron stars are probably rare.

Whilst considered as somewhat distinct from the planetary systems
surrounding main sequence stars, the pulsar planets provide
constraints on plausable formation processes. Models for planet
production around pulsars divide into two broad classes (\cite{ph93};
\cite{pod93}; \cite{brs+93}; \cite{pt94}).  In the first, the planet
formed around either a normal star (possibly the pulsar progenitor,
its continued existence implying that it had survived the supernova
explosion), or around another star and was subsequently captured by
the pulsar. Alternatively, the planet formed after the neutron star
was created, a process requiring the capture of material, perhaps from
a pre-existing stellar companion, into an accretion disk around the
pulsar. Subsequent fragmentation into planets would then follow the
more standard model of planet formation (Section~\ref{sec:planet-formation}).
Difficulties in
modeling multiple planets which survive the supernova explosion make
the accretion disk model more favoured, especially for the PSR~1257+12
system, with residual disk material governing the transport of angular
momentum and possibly providing the means to circularise the orbits
and bring them close to the observed 3:2 resonance (\cite{rud93}).

White dwarfs, the end point of stellar evolution for most stars,
undergo a less violent birth than pulsars, and planets would likely be
surviving members of original systems, more closely related to our own
Solar System, although they may also be created around white dwarfs as
part of the pulsar formation process (\cite{lps92}).  Timing detection
methods similar to those used for radio pulsars can be applied but using 
a very different timing signature (\cite{pro97}). As the white dwarf cools
through certain temperature ranges C/O (at $10^5$~K), He (at
$2.5\times10^4$~K) and H (at $10^4$~K) in its photosphere
progressively become partially ionized, driving multi-periodic
non-radial $g$-mode pulsations in the period range 100--1000~s, with some
objects amongst the most stable pulsators known (\cite{kwn+91}). Data
going back 20~years, originally acquired to probe white dwarf
interiors, have been used to place upper limits on planetary companion
masses for two objects: G117--B15A (\cite{kwn+91}), where the rate of
change of period for the 215~s pulsation is
$\dot{P}=12.0\pm3.5\times10^{-15}$~s\,s$^{-1}$, and for G29--38
(\cite{knw+94}), with present searches being sensitive to planetary
companions down to 0.5\,$M_\Jupiter$ if seen edge on, with
$P\sim0.1-30$~yr (\cite{pro97}).
Formation conditions and detection possibilities for 
surviving planets around stars in the planetary nebula phase have been 
discussed by \cite*{sok99}. 

\begin{figure}[tbh]
\begin{center}
\centerline{\epsfig{file=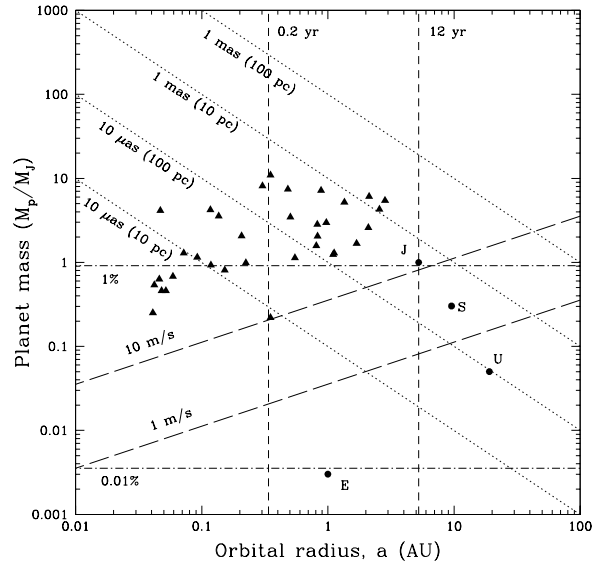,width=8.0cm,angle=0}}
\end{center}
\vspace{-30pt}
\caption{Detection domains for methods exploiting planet 
  orbital motion, as a function of planet mass and orbital radius,
  assuming $M_*=M_\odot$.  Lines from top left to bottom right show
  the locus of astrometric signatures of 1~mas and 10~$\mu$arcsec at
  distances of 10 and 100~pc (equation~\ref{equ:astrom}; a measurement
  accuracy 3--4 times better would be needed to detect a given value
  of $\alpha$).  Very short and very long period planets cannot be
  detected by planned astrometric space missions: vertical lines show
  limits corresponding to orbital periods of 0.2 and 12~years.
  Lines from top right to bottom left show radial velocities
  corresponding to $K=10$ and $K=1$~m~s$^{-1}$
  (equation~\ref{equ:radvel2} and~\ref{equ:radvel3}; a measurement
  accuracy 3--4 times better would be needed to detect a given value
  of~$K$).
  Horizontal lines indicate photometric detection thresholds for
  planetary transits, of 1\% and 0.01\%, corresponding roughly to
  Jupiter and Earth radius planets respectively (neglecting the
  effects of orbital inclination, which will diminish the probability
  of observing a transit as $a$ increases).
  The positions of Earth (E), Jupiter (J), Saturn (S) and Uranus (U)
  are shown, as are the lower limits on the masses of known planetary
  systems (triangles) from Table~\ref{tab:known-planets}.  }
\label{fig:domains}
\end{figure}

Figure~\ref{fig:domains} illustrates the parameter regions probed by
radial velocity, astrometric, and transit measurements at current and 
projected accuracy levels.

\subsection{Periodic photometry: transits and reflections}
\label{sec:photometry}

Detection of extra-solar planets by measuring the photometric
signature of the eclipse of the star by a planet was first considered
by \cite*{str52}. The method is conceptually simple: given a suitable
alignment geometry, star light is attenuated by the transit of the 
orbiting planet
across its disk, with the effect repeating at the orbital period of
the planet.  For a Sun/Jupiter system at 10~pc, the resulting
luminosity change is of order 2\%, or 0.02~mag.  In the early 1970s
this was considered as observationally more feasible than the
prospects of detecting an astrometric shift of 0.0005~arcsec, or a
radial velocity perturbation of around 10\,m\,s$^{-1}$. However, the
probability of observing an eclipse, seen from a random direction and
at a random time, is extremely small.  

The idea was developed by \cite*{ros71}, who proposed detecting the
eclipse colour signature, which could be measured as a change in ratio
due to limb darkening rather than in absolute intensity, and who
considered in detail the effects of stellar noise sources (intrinsic
stellar variations, flares, coronal effects, sunspots, etc.) and Earth
atmospheric effects (air mass, absorption bands, seeing, and
scintillation).  The method is presently considered as one of the most
promising means of detecting planets with masses significantly below
that of Jupiter, with the detection of Earth-class (and hence
habitable) planets within its capabilities.  Extrapolation of the
method down to masses of planetary satellite may even be feasible.
Further refinements have therefore continued (\cite{bs84};
\cite{bsh85}; \cite{sc90}; \cite{hd94}; \cite{sch94}; \cite{hea96};
\cite{jan96}; \cite{sch96}; \cite{dee98}; \cite{ss99}), and recent
reviews are given by \cite*{sac99} and \cite*{sch00a}.

Detection probabilities depend on the transit geometry and on the 
luminosity drop produced by an object on the line of sight to the 
star, which approximates to:
\begin{equation}
\label{equ:photom}
\frac{\Delta L}{L_*} \simeq  \left ( \frac{R_p}{R_*} \right )^2
\end{equation}
under the assumption of a uniform surface brightness of the star.  
Strictly the effect includes a dependence on the
local surface brightness of the stellar disk, which varies with radius
due to `limb darkening': moving radially outwards towards the stellar limb
the line of sight passes through increasing atmospheric depth (cf.\
equation~1 of \cite{ss99}; see also \cite{sac99}). For
a central transit across a star with a limb-darkening coefficient of
0.6, the maximum brightness drop is 25\% larger than given by
equation~\ref{equ:photom}.  Since limb-darkening is wavelength dependent, a
planetary event will also cause a small colour change
(\cite{ros71}). 

Values of $\Delta L/L_*$ for the Earth, Mars, and
Jupiter transiting the Sun are $8.4\times10^{-5}$,
$3\times10^{-5}$, and $1.1\times10^{-2}$ respectively. If the radius
of the star can be estimated from, say, spectral classification,
then $R_p$ can be estimated from equation~\ref{equ:photom}.  With knowledge
of $P$ and an estimate of $M_*$ (also from spectral classification
or via evolutionary models), $a$ can be derived from Kepler's law. 
Other observational parameters are given, to first order, by simple geometry
(\cite{dee98}). Thus the duration of the transit is:
\begin{equation}
\label{equ:duration}
\tau = \frac{P}{\pi}  \left ( \frac{R_* \cos\delta +R_p}{a} \right )
 \simeq 13\ \left ( \frac{M_*}{M_\odot} \right )^{-1/2}\,
	\left ( \frac{1}{1\ {\rm AU}} \right )^{1/2}\,
	\left ( \frac{R_*}{R_\odot} \right) \ \ {\rm hours}
\end{equation}
where $\delta$ is the latitude of the transit on the stellar disk,
giving a transit period of about 25~hr for a Jupiter-type planet 
and 13~hr for an Earth-type system, results rather
insensitive to $a$ (\cite{sab+98}). With the other parameters
estimated as above, $\delta$ can be derived from
equation~\ref{equ:duration}, and hence the orbital inclination from
$\cos i=(R_*\sin\delta)/a$. The minimum inclination where transits can
occur is given by $i_{\rm min}=\cos^{-1}(R_*/a)$, with the probability
of observing transits for a randomly oriented system of $p=R_*/a=\cos
i_{\rm min}$. Evaluation of $i$ and $p$ for realistic cases
demonstrates that $i$ must be very close to $90^\circ$, while $p$ is
very small.

The principal disadvantage of the method is that it requires
configurations in which the viewing direction (to the Earth) lies in
the orbital plane of the planet.  This time-independent geometrical
alignment therefore occurs with only very low probability, such that
surveys without {\it a priori\/} knowledge of system geometry or
orbital period and orbital phase are characterised by very low
detection probabilities. Prospects can be improved by pre-selecting
stars whose planetary companions are likely to have their orbital
planes perpendicular to the plane of the sky, for example stars whose
rotation axis can be inferred to lie in this plane (\cite{doy88}).
Another possibility is to focus attention on eclipsing binary systems
whose geometry implies $i\sim90^\circ$, although this in turn relies
on the further assumption that the planetary and binary orbital planes
are co-aligned (\cite{doy88}; \cite{sc90}; \cite{hal94}).

Configurations in which the planet orbits one component of a
widely-separated binary, or is in orbit around a close binary, are
possible, with both resulting in large domains of dynamical stability
(\cite{har77}; \cite{hep78b}; \cite{ben98b}). \cite*{sc90}
demonstrated that when applied to eclipsing binaries the method leads
to the possibility of detecting planets with radii around 10\,000\,km
($\sim1.5\,R_\oplus$) with a rather large probability and a fairly
good precision in the orbital period. Thus, observations of the
smallest known eclipsing binary, CM~Dra with a period of 1.28~days,
should allow the detection of terrestrial-sized inner planets
(1.5--2.78\,$R_\oplus$), if they exist around this system, using
ground-based 1-m class telescopes over several months. This programme
has been pursued by the TEP (Transits of Extra-Solar Planets) network
since 1994, using six~telescopes located around the world
(\cite{ddd+96}; \cite{dee98}). Candidate (but unconfirmed) transit
events, and current confidence limits versus planetary radius and
period, are given for this specific system by \cite*{ddk+98}. Extension
to other eclipsing binaries and densely populated fields in Baade's Window 
and NGC~6752 is described by \cite*{drb+99} and \cite*{rdb99}. Timing 
of the binary eclipse minima provides a planet detection sensitivity 
down to about $1\,M_\Jupiter$ (\cite{ddk+00}).

Ground-based photometry to better than about 0.1\% accuracy is
complicated by variable atmospheric extinction, while scintillation,
the rapidly varying turbulent refocusing of rays passing through the
atmosphere, imposes limits at about 0.01\%.  Extension of the transit
method to space experiments, where very long uninterrupted
observations can be made above the Earth's atmosphere, therefore holds
particular promise.  STARS, a space telescope of 0.5~m$^2$ collecting
area and 1\ddeg5 field of view, was studied but not accepted by the
European Space Agency (\cite{sch96}).  A revised concept, Eddington,
was selected for study by ESA in early 2000 (\cite{rcf00}).  It has an
enlarged telescope of 1~m$^2$ collecting area and 6~deg$^2$ field of
view, and a CCD focal plane detector array.  The first 2--3~years is
to be devoted to stellar seismology, aiming to detect solar-type
acoustic oscillations in stars in the nearest open clusters.  A
further 2--3~years would be dedicated mainly to planetary transit
detection in up to 700\,000 stars in about 20~fields observed for one
month each.  Reaching a photometric precision of about $10^{-6}$, it
is expected to yield three or more transits of Earth-mass extra-solar
planets in each of about 50 main-sequence stars.

A similar US space mission, Kepler (\cite{bkd+97}), which evolved from
an earlier proposal FRESIP (\cite{kbc+96}), was unsuccessful
in its bid for selection as part of NASA's Discovery Program. With a
1-m aperture and $12^\circ$ field of view, Kepler would have
continuously monitored some 80\,000 main-sequence stars brighter than
14~mag in a specific sky area, at a photometric precision of
$10^{-5}$. It was specifically designed to detect and characterise
Earth-class planets in and near the habitable zones of a wide variety
of stellar types. Expected results included 480~Earth-class planet
detections; 160~inner-orbit giant planets and 24~outer-orbit planets;
and 1400~planet detections with periods below 1~week from the phase
modulation of the reflected light.

%
%
COROT (\cite{dbl+97}; \cite{sab+98}) is a small mission led by the French
space agency CNES, scheduled for launch in 2004.  It has a 27~cm 
aperture, and is also primarily designed for the study of stellar
oscillations. In its planetary detection mode, about 50\,000 stars
will be monitored, with photometric precision between $7\times10^{-4}$
and $5\times10^{-3}$ for $V=11-15.5$~mag and 1~hour integration times.
These values imply a low probability of discovering transits of
Earth-class planets in the habitable zone, but give an increased
probability for planets which are slightly larger, or closer to their
central star, such that several detection of planets with
2\,$R_\oplus$ are expected.   MONS and MOST are related projects of
the Danish Small Satellite Program and Canadian Space Agency
respectively, primarily devoted to studies of stellar oscillations but
with similar applicability to parallel planetary transit studies.

The sensitivity of the transit method is such that the detection of
comets appears feasible with photometric accuracies of around
$10^{-4}$ (\cite{lvf99}).  Simulations based on a geometrical model of
the dust distribution, and optical properties of the cometary grains,
suggest largely achromatic transits, with a `rounded triangular'
shape as typical temporal signature. Photometric variations for
$\beta$~Pictoris have been attributed to either a planetary passage or
a dust cloud, the latter explanation probably requiring a cometary
tail (\cite{lvb+97}; \cite{lvf99}).  The importance of these studies
is evident in the context of comet models for our own Solar System
(\cite{pa98}).
 
Even satellites of extra-solar planets, and planetary rings appear 
detectable by this method (\cite{sab+98}). \cite*{ss99} have
derived detection probabilities, and give illustrative examples of
transit light curves, evaluating the combination of geometric
conditions (orbital inclination angle such that it will transit the
star) and orbital conditions (determined by the location of the
planet), with and without an accompanying transit of the parent
planet. The planet-satellite transits also constrain the
period and orbital radius of the satellite.  Even if the satellite is
not extended enough to produce a detectable signal in the stellar
light curve, it may still be detected indirectly through the rotation
of the planet around the barycentre of the planet-satellite system,
although this necessarily requires that a planetary transit be observed at
least 3~times.  \cite*{ss99} showed that COROT could detect or exclude
satellites much smaller than those photometrically detectable, for
example with radius $\sim0.3\,R_\oplus$ around a Jupiter-like planet.
The prospects for transit spectroscopy and imaging are discussed by 
\cite*{sch00a}.

\begin{figure}[t]
\begin{center}
\centerline{\epsfig{file=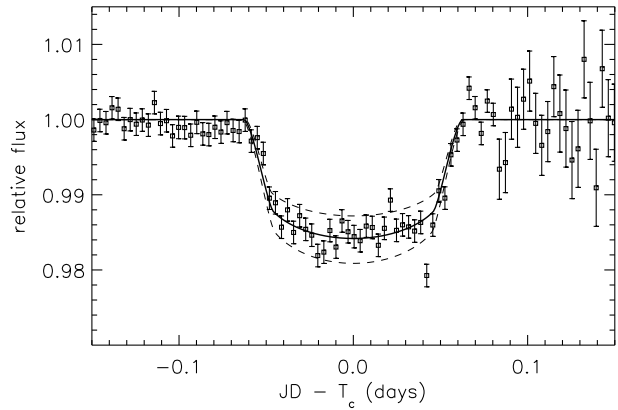,height=6.0cm,angle=0}}
\end{center}
\vspace{-30pt}
\caption{ The first detected transit of an extra-solar planet,
HD~209458 (from \protect\cite{cbl+00}). The figure shows the measured
relative intensity versus time. Measurement noise increases to the
right due to increasing atmospheric air mass.  From the detailed shape
of the transit, some of the physical characteristics of the planet can
be inferred (courtesy of David Charbonneau).}
\label{fig:charbonneau}
\end{figure}

An important recent result has been the detection of the first
extra-solar planetary transit event, observed by \cite*{cbl+00} and
independently by \cite*{hmb+00} for HD~209458, a planet with small~$a$, 
and one of the most recent of the radial velocity detections.
\cite*{cbl+00} observed two well-defined transits
(Figure~\ref{fig:charbonneau}), of duration about 2.5~hours, providing
an orbital period consistent with the radial velocity determination,
and confirming beyond any doubt that its radial velocity variations
arise from an orbiting planet. The importance of this result and the
relative strength of the transit signal have resulted in a number of
independent confirmations of transits for HD~209458, including the
detection of the transit signals in the Hipparcos astrometry
satellite's photometric data from 1990--93 (\cite{sra99};
\cite{ra00}).  The physical information about this planet that can be
inferred from this measurement are described in
Section~\ref{sec:hd209458}.

%
The realisation that stars may be accompanied by close-in planets
yielding measurable transit signatures has led to the development of
dedicated systems to monitor specific fields, which can be based on
small-aperture optics: STARE, a 10~cm instrument used for the
first detection of the HD~209458 transits by \cite*{cbl+00}, is
monitoring some 24\,000 stars in a 5.7~degree square field in the
constellation of Auriga; ASP (Arizona Search for Planets) uses a 20~cm
aperture in a similar manner; and ASAS (All-Sky Automated Survey) has
as its goal the photometric monitoring of $\sim10^7$ stars brighter
than 14~mag over the entire sky, making more than 100 3-min
exposures per night.  Such searches should soon extend the
detection of transits to later spectral types (cooler, less massive K
and M stars) than the Sun-like (F and G-type) stars favoured in the
radial velocity surveys, in which the transit effect should be more
pronounced due to the smaller stellar size. Observations of more than 
$34\,000$ stars in the globular cluster 47~Tucanae, uniformally 
sampled over 9~days by the Hubble Space Telescope in July 1999,
may result in several tens of transit detections if such planets exist
in globular clusters (\cite{gil99}), although preliminary analysis for 
$27\,000$ stars has revealed no convincing planet candidates (\cite{bcg+00}).
15--20 detections would have been expected if the occurrence rate in 
47~Tuc were the same as indicated by radial velocity searches in the 
solar neighbourhood; the discrepancy suggests that at least one of the 
processes of formation, migration, or survival of close-in planets may 
be significantly altered in the cluster environment. 

Equation~\ref{equ:imaging} indicates that $L_p/L_* \propto a^{-2}$.  For 
planets very close to the central star, a modulation due to the 
(Doppler shifted) reflected light intensity could therefore be
expected, even if the planet cannot be imaged as such (\cite{bro92};
\cite{cjn98}; \cite{ss98}; \cite{sws00}).  This may still occur even if 
the orbital plane is somewhat inclined to the line of site, with a signature
dependent upon inclination (no modulation would be observed
for face-on systems).  The close-in planetary system $\tau$~Bootis has
$a=0.046$~AU, $M_p\sin i=3.89\,M_\Jupiter$, and hence $L_p/L_* \sim
10^{-4}$, some $10^4-10^5$ times higher than for the case of a Jupiter
system, although $\tau$~Boo and its planet are separated by, at most,
0.003~arcsec.  The system has been studied by \cite*{cnk+99}.  The
spectrum should contain a secondary component which varies in amplitude
depending on the adopted functional form for the phase variation and
scattering law, with the planet's radial velocity relative to the star
reaching a maximum amplitude of $\pm152$\,km\,s$^{-1}$. The system
appears to be tidally locked (cf.\ equation~1 of \cite{gbh+96}), in
which case there is no relative motion between the planet and star
surfaces. The planet should therefore reflect a non-rotationally
broadened stellar spectrum, yielding relatively narrow planetary lines
dominated only by the stellar photospheric convective motions,
superimposed on much broader stellar lines.  No evidence for a highly
reflective planet was found, yielding a geometric albedo $p\la0.3$
(assuming $R_p\sim 1.2\,R_\Jupiter$, and with a weak dependence on the
assumed orbital inclination), compared to values of 0.39--0.60 for the
Solar System giant planets.  But many plausible model planetary
atmospheres remain consistent with this upper limit,
with predictions for the albedos of the close-in extra-solar giant
planets varying by orders of magnitude, being highly sensitive to the
presence of condensates such as Fe and MgSiO$_3$ in the planetary
atmospheres.  Observations for the same system by \cite*{chp+99}
suggest that detection of the reflected light at the appropriate
orbital period may have been achieved, providing a value for $i$, and
hence an estimate of $M_p\sim8\,M_\Jupiter$, twice the minimum value
for an edge-on orbit. Assuming a Jupiter-like albedo $p=0.55$ yields
$R_p\sim1.6-1.8\,R_\Jupiter$, larger than the structural and evolutionary 
predictions of \cite*{gsb+97} of $\sim1.4-1.1\,R_\Jupiter$ at ages of 2--3~Gyr.
This discrepancy between measured and predicted radius may cast doubt 
on the claimed detection, or may simply imply inadequacy in present
atmospheric modelling (\cite{bgh+00}; \cite{sws00}). A search for a methane
signature in the infrared spectrum also yields suggestive but
presently inconclusive results (\cite{wdb00}).

\subsection{Gravitational microlensing}
\label{sec:lensing}

Gravitational lensing is the focusing and hence amplification of light
rays from a distant source by an intervening object, first considered
by \cite*{ein36} and \cite*{lin36}. For the early history of photometric 
lensing, see \cite*{lin69}, p267.
%
%
%
\cite*{wcw79} discovered the first case of a double image created by
gravitational lensing of a distant source, the quasar 0957+561.
Arc-like images of extended galaxies were first reliably reported by
\cite*{lp89}. Relative motion between the background source, the
intervening lens, and the observer will lead to apparent brightening
and subsequent dimming of the resulting image, which may occur over
time scale of hours and upwards.

The term microlensing was introduced by \cite*{pac86a} to describe
gravitational lensing that can be detected by measuring the intensity
variation of a macro-image made of any number of micro-images, which
are generally unresolved by the observer.  The lensing phenomenon
offers a powerful but experimentally challenging route
to the detection and characterisation of planetary systems.

Many of the essential formulae used to analyze gravitational lensing
were derived by \cite*{ref64}, and are found in numerous forms
throughout the microlensing literature, with the following given by
\cite*{wam97}; see also \cite*{sac99}; \cite*{rs94}. Events are
characterised in terms of the Einstein ring radius (related to the
Schwarzschild or gravitational radius of the lens $r_g=2GM_L/c^2$):
\begin{eqnarray}
\label{equ:einstein}
R_E & = &\left [ \frac{4GM_L}{c^2} \frac{(D_S-D_L)D_L}{D_S} \right ]^{1/2} 
				\nonumber\\
    & = & 8.1\,\left (\frac{M_L}{M_\odot} \right)^{1/2}\,
	\left ( \frac{D_S}{8\ {\rm kpc}} \right )^{1/2}\,
	\left [ (1-d)d \right ] ^{1/2} \ \ \ {\rm AU}
\end{eqnarray}
where $M_L$ is the mass of the lensing object, $D_L$ and $D_S$ are the
distances to the lens and to the source, with $d=D_L/D_S$.  The
parameterization in mass and distance gives the Einstein ring radius in
the lens plane. The Einstein angle is $R_E$ expressed in angular
coordinates:
\begin{equation}
\label{equ:angular}
\theta_E = R_E/D_L =
1.0\, \left ( \frac{M_L}{M_\odot} \right ) ^{1/2}\,
	\left ( \frac{D_L}{8\ {\rm kpc}} \right )^{-1/2}\,
	(1-d)^{1/2} \ \ \ {\rm mas}
\end{equation}
The microlensing magnification as a function of time is:
\begin{equation}
\label{equ:magnification}
A(t)= \frac{u^2(t)+2}{u(t)\,[u^2(t)+4]^{1/2}}
\end{equation}
where $u(t)$ is the projected distance between lens and source in
units of the Einstein radius. For a typical transverse velocity of
$v=200\times v_{200}$\,km\,s$^{-1}$ the time-scale of a microlensed
event is:
\begin{equation}
\label{equ:timescale}
t_E = 69.9\, \left ( \frac{M_L}{M_\odot} \right )^{1/2}\,
	\left ( \frac{D_S}{8\ {\rm kpc}} \right)^{1/2}\, 
        [(1-d)d]^{1/2} \ v^{-1}_{200}\ \ \ {\rm days}
\end{equation} 

Precise alignment is required for a detectable brightening.  The
chance of substantial microlensing magnification is extremely small,
$\sim10^{-6}$ for background stars in the Galactic bulge, nearby
Magellanic Clouds, or nearby spiral galaxy M31, even if all the unseen
Galactic dark matter is composed of objects capable of lensing
(\cite{pet81}; \cite{got81}; \cite{pac86b}).  However, microlensing
events can be distinguished from peculiar intrinsic source variability
by their achromatic behaviour.  Only since 1993, when massive
observational programmes capable of surveying millions of stars were
underway, has photometric microlensing been observed by the EROS
(Exp\'erience de Recherche d'Objets Sombres, \cite{abb+93}), OGLE
(Optical Gravitational Microlensing, \cite{usk+93}), MACHO (Massive
Compact Halo Objects, \cite{aaa+93}), and DUO (Disk Unseen Objects,
\cite{ala96}) projects. Several hundred events have now been observed,
and the first (albeit contested) events attributed to planets have
been announced. Microlensing is providing new information on the
amount of matter (dark and luminous) along these lines of sight, and a
number of recent reviews are devoted to this rapidly growing subject
(\cite{pac96}; \cite{gou96}; \cite{sac99}). Events with durations
ranging from hours to months are now reliably detected and reported
while still in progress, via the www, allowing concerted follow-up
observations by teams such as GMAN (\cite{paa+96}), PLANET (Probing
Lensing Anomalies Network, \cite{abc+96}), MPS (Microlensing Planet
Search, \cite{rbb+99}), and EXPORT (Extra-Solar Planet Observational
Research Team), which can gather detailed photometric information
about lensing events sifted from the vast survey data streams. These
are revealing microlensing `anomalies', a small subset of the already
rare microlensing events, which depart from the predicted (and
normally observed) smooth symmetric light curves (\cite{pac86b};
\cite{gri91}), and for which detailed structure in the light curves is
permitting study of the lens kinematics, the frequency and nature of
binary systems, stellar atmospheres, and the presence of planetary
systems.

\cite*{mp91} and \cite*{gl92} investigated lensing when one or more
planets orbit the primary lens, finding that detectable fine structure
in the photometric signature of the background object occurs
relatively frequently, even for low-mass planets. \cite*{gl92} found
that the probability of detecting such fine-structure is about 17\%
for a Jupiter-like planet (i.e.\ at about 5~AU from the central star),
and 3\% for a Saturn-like system; these relatively high probabilities
occur specifically when the planet lies in the `lensing zone', between
about 0.6--1.6\,$R_E$ (\cite{wam97}). Planets in this distance range,
which fortuitously corresponds to star/planet distances of a few~AU
and hence orbital radii corresponding to the habitable zone, produce
caustics (points in the source plane where the magnification is infinite)
inside the Einstein ring of the star.  Subsequent theoretical
work has included determination of detection probabilities
(\cite{bf94}), extension to Earth-mass planets including the effects
of finite source size (\cite{br96}; \cite{wam97}), determination of
physical parameters (\cite{gg97}), the calculation of the number of
detections for realistic observational programmes (\cite{pea97a};
\cite{sah97}; \cite{sac99}; \cite{gs00}), distinguishing between
binary source and planetary perturbations (\cite{gau98}),
consideration of high-magnification events (\cite{gs98}), the effect
of multiple planets in high-magnification events (\cite{gns98}), 
the possibility of observing repeating events originating from a
multiple planetary system (\cite{ds99}), and caustic-crossing 
configurations (\cite{gg00}).

\begin{figure}[t]
\begin{center}
\leavevmode
\centerline{
\epsfig{file=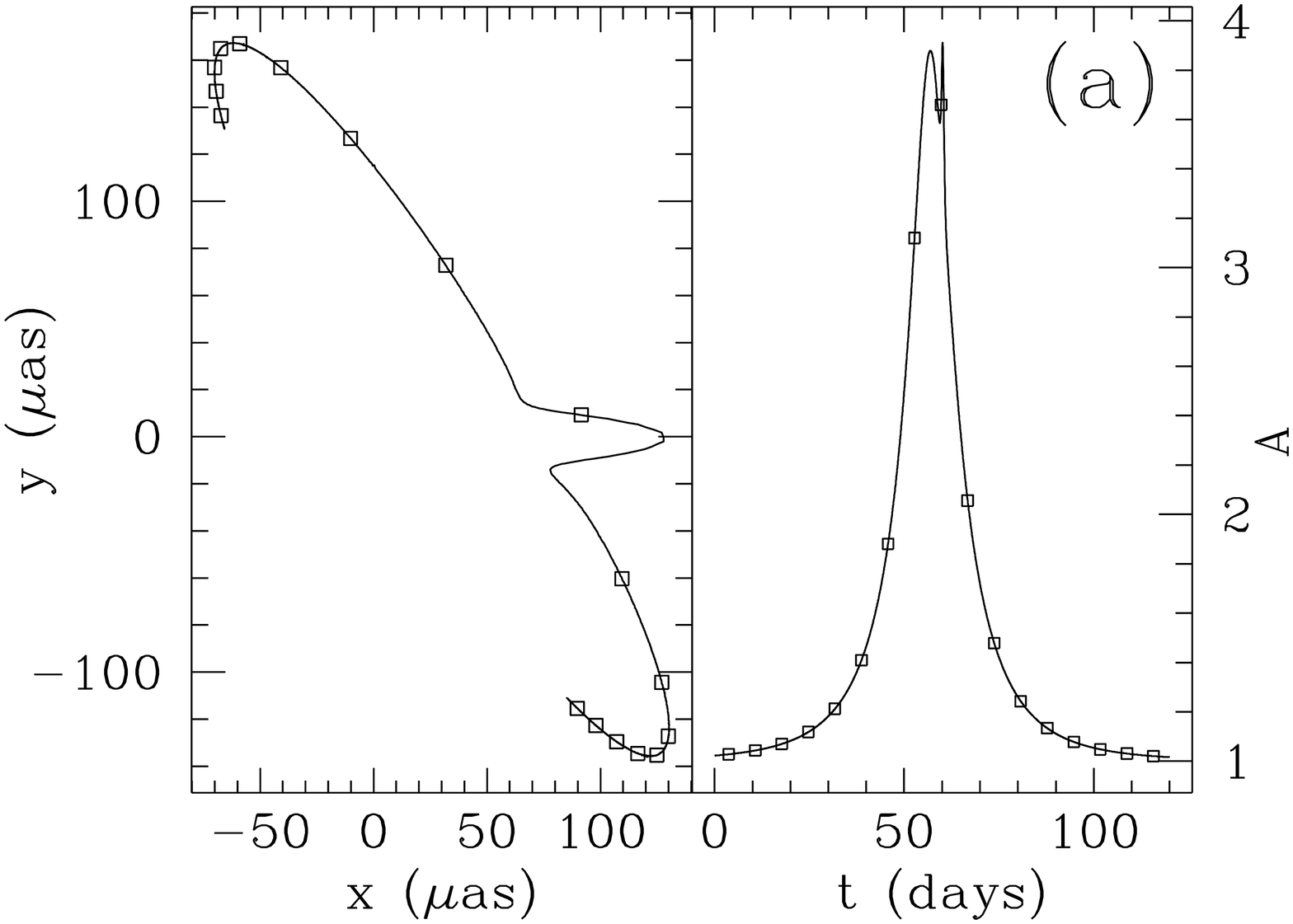,angle=270,width=5.0cm}
\hskip -20pt
\hfill
\epsfig{file=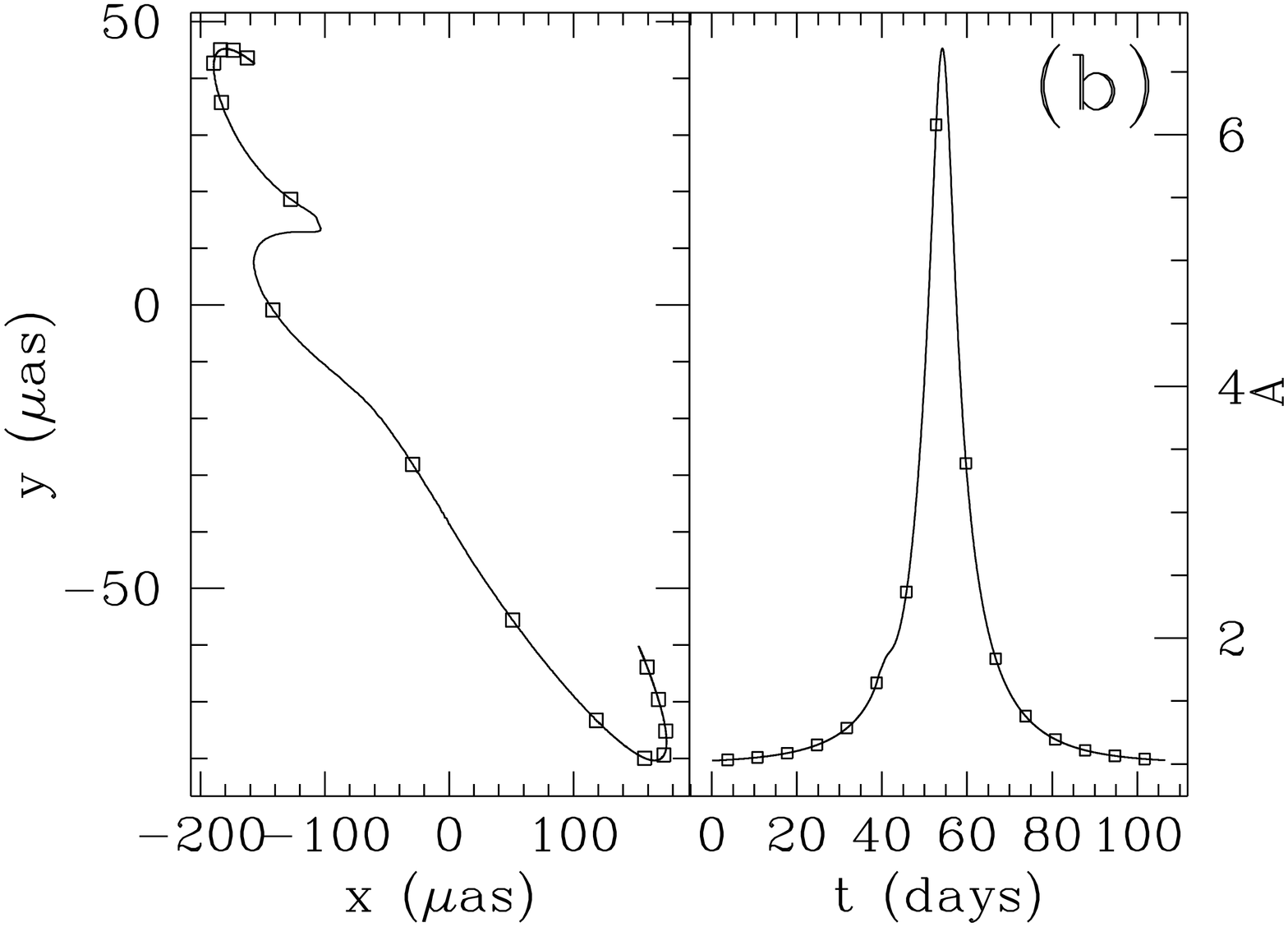,angle=270,width=5.0cm}
\hskip -20pt
\hfill
\epsfig{file=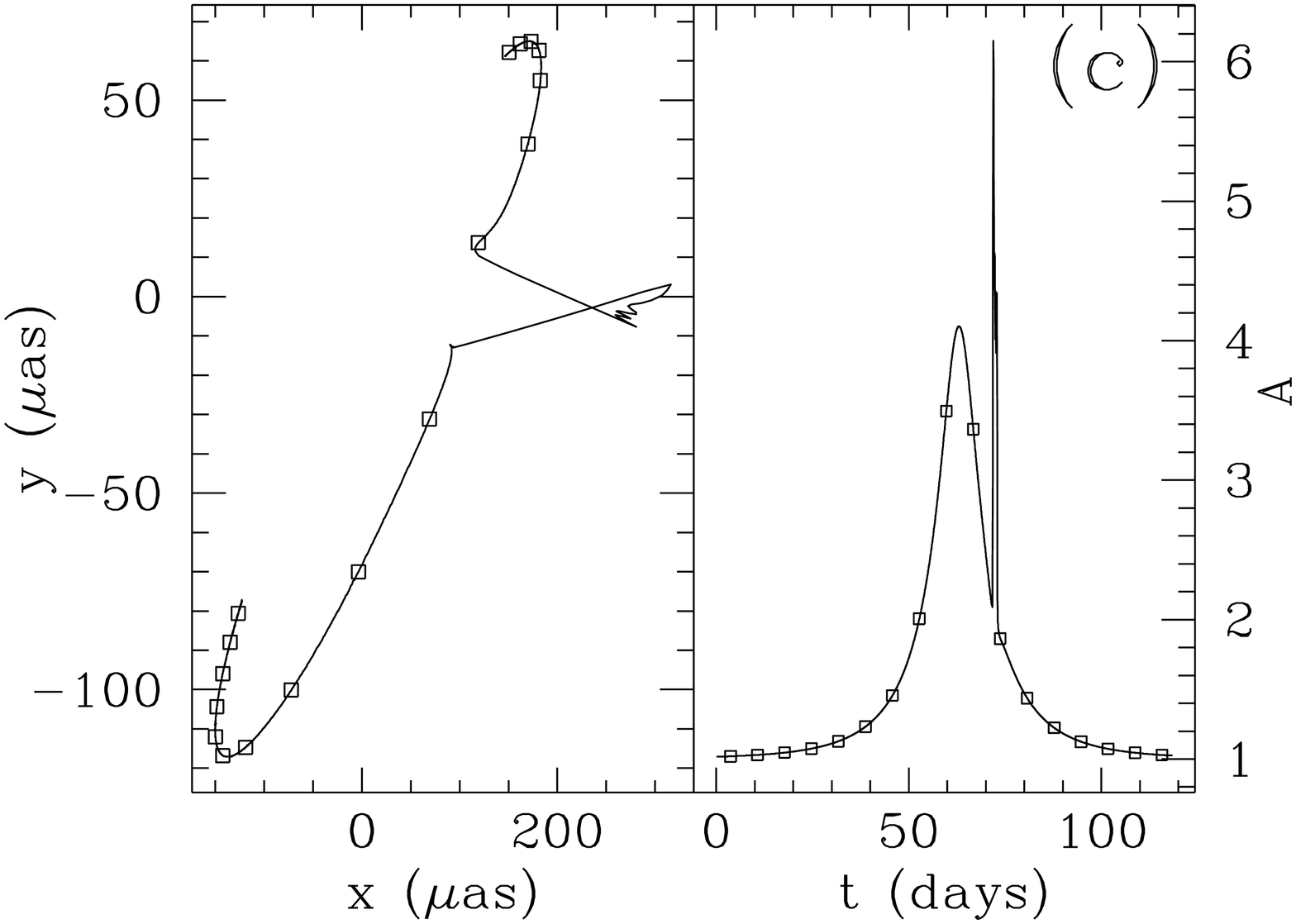,angle=270,width=5.0cm}}
\end{center}
\vspace{-10pt}
\caption{ Examples of predicted planet astrometric (leftmost plots of 
each pair) and photometric (rightmost plots) lensing 
curves (from \protect\cite{sdg99}).  All examples assume a planet to lens 
mass ratio of $q=10^{-3}$, with a primary lens
Einstein radius of 550~$\mu$as, corresponding to a Saturn mass
planet. Squares are plotted one per week: (a) $x_p=1.3$; (b)
$x_p=0.7$; (c) a caustic crossing event with $x_p=1.3$; $x_p$ is the
projected planet-lens separation in units of the Einstein radius
(courtesy of Neda Safizadeh).}
\label{fig:safizadeh}
\end{figure}

For a point-mass lens and the observer-lens-source precisely
co-aligned, the image becomes a ring of radius $R_E$, and the
magnification theoretically becomes infinite (in
equation~\ref{equ:magnification}, $A\rightarrow\infty$ as
$u\rightarrow0$).  For a single lens the caustic is the single point behind
the lens, and the critical curve (the positions of the images of these
caustics) is the Einstein ring. For single-point lenses,
high-magnification events occur when the source comes near the
caustic, with the peak magnification occurring at the projected
distance of closest approach.  When the lens consists of two
point-like objects, specifically a star and an orbiting planet, the
caustic positions and shapes depend on the planet-to-lens mass ratio
$q=M_p/M_L$ and the projected planet-lens separation $x_p$. With the
addition of a planet around a central star (both of which to a first
approximation are considered to be invisible), the light curve of the
background star will be very close to that of the single (stellar
mass) lens for most of its duration, but with fine structure
comprising additional sharp peaks, with a typical duration of a few
hours to a few days, compared with a typical primary lens event
duration of 40~days. As the planet mass decreases, the signals become
rarer and briefer: for Earth-mass planets detection probabilities are
$\sim2$\% and typical time scales are 3--5~hours (\cite{br96}).
Example magnification maps and light curves are given, for example, by
\cite*{gl92}; \cite*{wam97}; and \cite*{gs00}. Relative motion of the
source/lens/observer, including the Earth's motion around the Sun (see
also \cite{bg96}), and the effects of blending (in which the object
serving as the lens is itself luminous) introduce further 
complications.

In addition to the photometric manifestation of microlensing, the
changing alignment also results in a tiny motion of the photocentre of
the background object, an effect referred to as astrometric
microlensing.  Since a microlensed source has two (generally
unresolved) images, their centroid makes a small excursion (typically
by a fraction of a milliarcsec) around the trajectory of the source as
a result of varying magnification and image positions during lensing
(e.g.\ \cite{hnp95}; \cite{bsb98}).  The astrometric manifestation has
two advantages over photometric microlensing. First, the astrometric
cross-section is substantially larger than the photometric, and
second, the degeneracy with regard to the mass of the lens is removed
(\cite{gg97}) (ordinarily, measurements give only the planet/star mass
ratio and their projected separation in units of the Einstein radius
of the lens).  Astrometric signatures have not yet been measured,
being too small to be accessible to present ground-based observations,
but future (narrow-field) astrometric interferometers and
microarcsec-class space experiments will be able to measure these
effects.

Astrometric microlensing related specifically to planet detection has
been investigated by a number of authors (\cite{mp91}; \cite{sdg99};
\cite{hc99}; \cite{ds00}).  The planet's effect is short, unlike
parallactic and blending effects which are important over the entire
duration of the event. However, the magnitude of the planetary
perturbation can still be large for Jovian-like planets with the time
spent above 10~$\mu$as being typically of the order of days.
Figure~\ref{fig:safizadeh} shows some predicted cases from
\cite*{sdg99}.

The NASA space astrometric interferometer SIM (\cite{du99}), due for
launch in 2005, will allow the study of photometric events
alerted from ground. Scanning space astrometry missions offer less
favourable observational conditions, although GAIA
(Section~\ref{sec:astrometry}) may detect astrometric lensing motion 
independently of photometric signatures for the first time (\cite{ds00}).

\begin{figure}[tbh]
\begin{center}
\centerline{
\epsfig{file=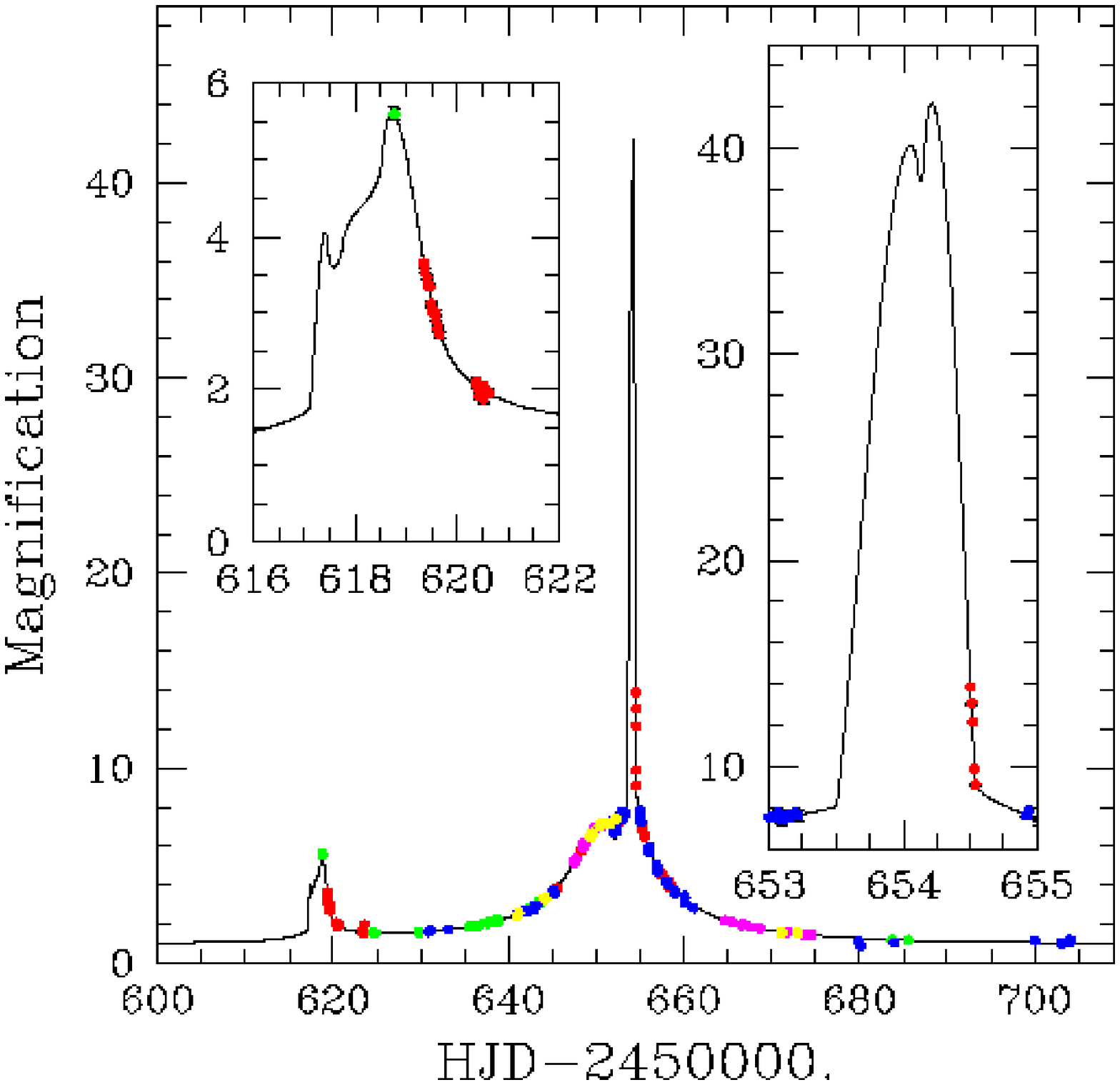,width=7.85cm,angle=0}
\hfil
\epsfig{file=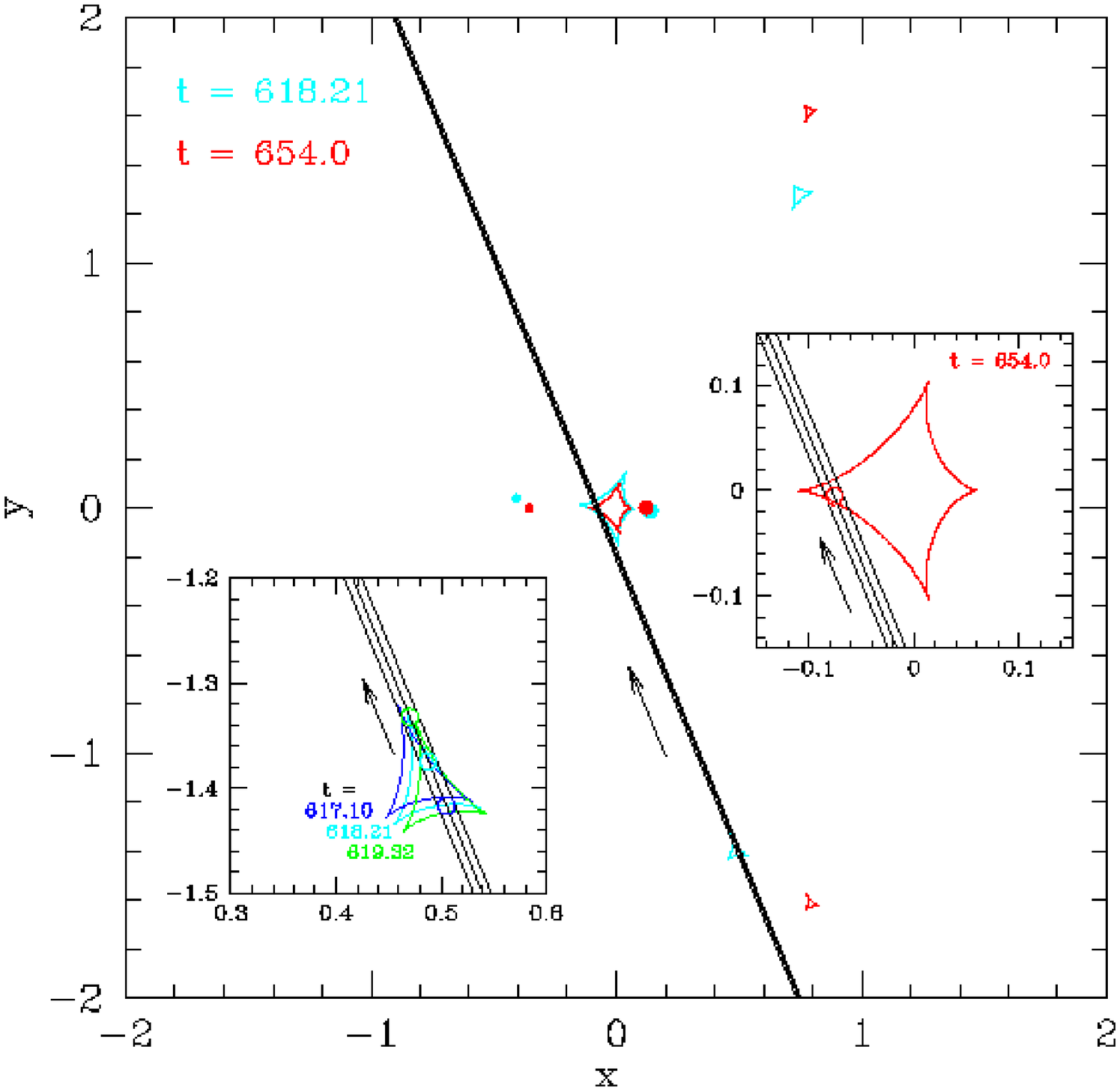,width=7.00cm,angle=0}}
\end{center}
\vspace{-20pt}
\caption{The event MACHO 97-BLG-41, from \protect\cite*{abc+00a}, 
  illustrating the fit to this complex event without invoking a
  planetary companion.  Left: magnification versus time (in
  heliocentric Julian date).  The best rotating binary model is shown
  as the solid curve, with the observational data shown as points
  (insets are zooms on the two caustic crossing regions).  Right: the
  caustic topology of the model at times near the first and second
  caustic crossings, in units of the angular Einstein radii.  The
  straight line shows the source trajectory.  The binary components
  are shown as small dots.  As time progresses, the binary rotates
  counterclockwise (causing the caustic pattern to do the same) and
  the component lenses move closer together.  The zooms show the
  source (circle) trajectory as it crosses the central caustic (right)
  and triangular caustic (left).  The triangular caustic movement
  during the entire first crossing is shown.  The source diameter is
  indicated by the distance between the parallel lines (courtesy of 
  the PLANET Collaboration).
}
\label{fig:albrow}
\end{figure}

The disadvantages of microlensing for planet detection are that
specific systems cannot be selected for study, and once they occur an
independent microlensing event is unlikely to recur for the same
system on any relevant time scale.  Advantages are the method's high
sensitivity even to low mass planetary systems, and its effectiveness
out to very large (kpc) distances, being a search technique which
requires no photons from either the planet or from the parent star (it
is the only technique identified capable of detecting interstellar
planetary mass bodies). Hundreds of microlensing events have now been
detected in the Galaxy. Results on individual objects include
observation of a caustic crossing (SMC-98-1, \cite{aaa+98a};
\cite{aaa+99}; \cite{abc+99c}; \cite{aaa+00}); 
limb darkening effects (MACHO 97-BLG-28, \cite{abc+99a};
MACHO 97-BLG-41, \cite{abc+00a};
OGLE 99-BLG-23, \cite{aab+00}); 
and limits on stellar and planetary companions (OGLE
98-BUL-14, \cite{abc+00b}). Statistical results include Large
Magellanic Cloud results for the first two years (\cite{aaa+97b});
results of the PLANET 1995 pilot campaign (\cite{abb+98}); and EROS
and MACHO combined limits on planetary mass dark matter in the
Galactic halo (\cite{aaa+98b}).

\cite*{baa+97} reported light curves from two MACHO events with
limited photometric coverage, leaving interpretation ambiguous but
providing evidence for an M-dwarf with a companion gas giant of mass
$\sim 5\,M_\Jupiter$ at $a\sim1$~AU (94-BLG-4), and for an isolated
object of mass $\sim 2\,M_\Jupiter$ or a planet more than 5--10~AU
from its parent star (95-BLG-3).  \cite*{rbf+98} reported a high
magnification event (98-BLG-35, $A\sim80$) observed towards the
Galactic centre, providing weak evidence ($4.5\,\sigma$) for a low
mass planet with mass fraction $4\times10^{-5}-2\times10^{-4}$,
corresponding to $1\,M_\oplus$ for a lens mass of $0.3\,M_\odot$.
From some 100 events monitored by the PLANET collaboration,
more than 20 have sensitivity to perturbations that would be caused 
by a Jovian mass companion to the primary lens (\cite{gs00}).  No
unambiguous signatures of such planets have been detected. These null
results indicate that Jupiter mass planets with $a=1.5-3$~AU occur in
less than one third of systems, with a similar limit applying to
planets of $3\,M_\Jupiter$ at $a=1-4\,$AU (\cite{gaa+00}).  

The event MACHO 97-BLG-41 was recently reported as the first
convincing example of planetary microlensing (\cite{brb+99}).
Although this interpretation has subsequently been disputed
(\cite{abc+00a}), the results illustrate both the modeling complexity
and the method's future potential for probing extra-solar planetary 
systems.  It was first alerted on 1997
June~18 as a possible short-duration event in the direction of the
Galactic Bulge.  From the complex light curve acquired over several
weeks, the lens system was inferred to consist of a planet of mass
$3.5\pm1.8\,M_\Jupiter$, orbiting a K-dwarf/M-dwarf binary stellar
system (\cite{brb+99}).  In this interpretation, the stars are
separated by $\sim1.8$~AU, and the planet is orbiting them at a
distance of about 7~AU. The background star is $V=19.6$~mag and likely
to be at or slightly beyond the centre of our Galaxy at 8.5~kpc.  The
probable lens distance is $\sim6.3^{+0.6}_{-1.3}$\,kpc, with a
relative motion of the lens with respect to the source of
270\,km\,s$^{-1}$.  All derived parameters are reasonable for a pair
of Galactic bulge stars.  An alternative interpretation was given by
\cite*{abc+00a}. They were able to model their (independent) 46 V-band 
and 325-I band observations, with a fit consistent with the \cite*{brb+99} 
data, with a lens consisting of a rotating binary star, having
a total lens mass $M\sim0.3\,M_\odot$, a mass ratio $q=0.34$, a lens
distance of $D_L\sim5.5$~kpc, and a binary period of $P\sim1.5$~yr,
leading to a change in binary separation of $-0.070\pm0.009\,R_E$ and
in orientation by 5\ddeg61~$\pm$~0\ddeg36 during the 35.2~days between
the separate caustic transits (Figure~\ref{fig:albrow}). While the 
interpretation for this object may remain contested, gravitational
microlensing should provide further important planetary constraints in 
the future.

\subsection{Protoplanetary disks}
\label{sec:disks}

Section~\ref{sec:planet-formation} underlines the intimate connection
between protoplanetary disks and the planets which are formed from
them.  A short discussion of disks is included here partly because of
the evidence that the observation of protoplanetary disks brings to
bear on the understanding of planetary formation, but also because the
distinction between the detection of disks and the detection of
planets is becoming smaller.

\begin{figure}[tbh]
\begin{center}
\centerline{
\hfil
\epsfig{file=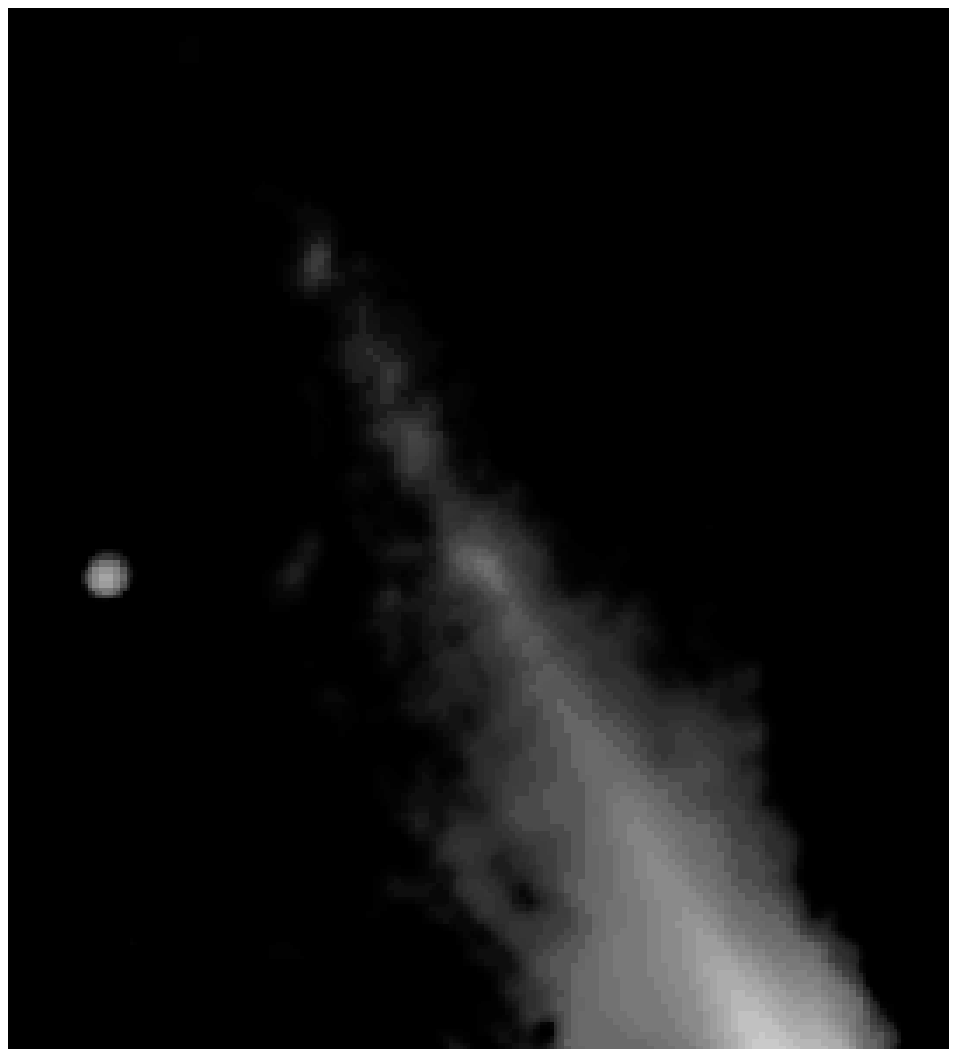,width=5.0cm,height=5.0cm,angle=0}
\hfil
\epsfig{file=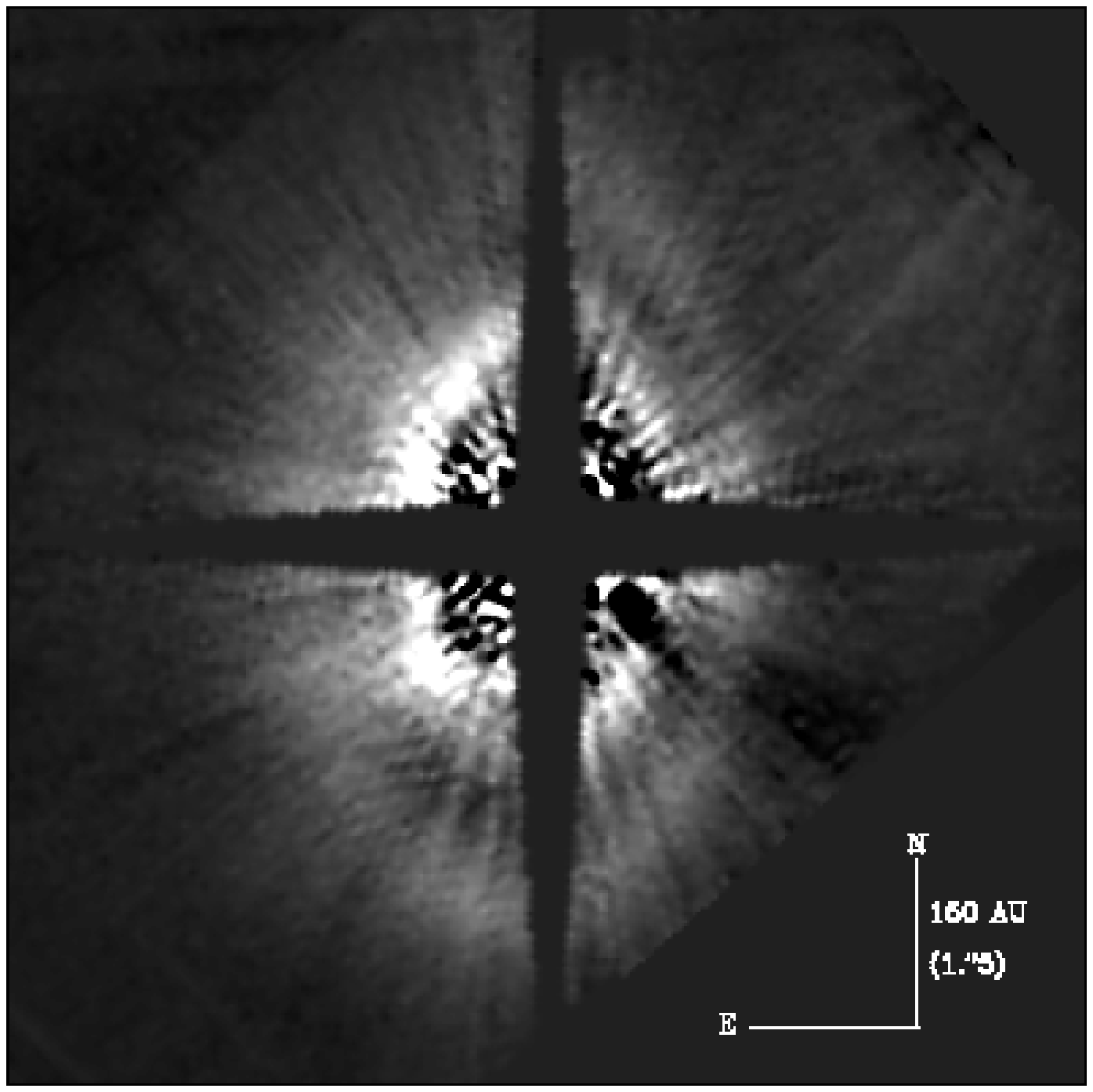,width=5.0cm,height=5.0cm,angle=0}
\hfil
\epsfig{file=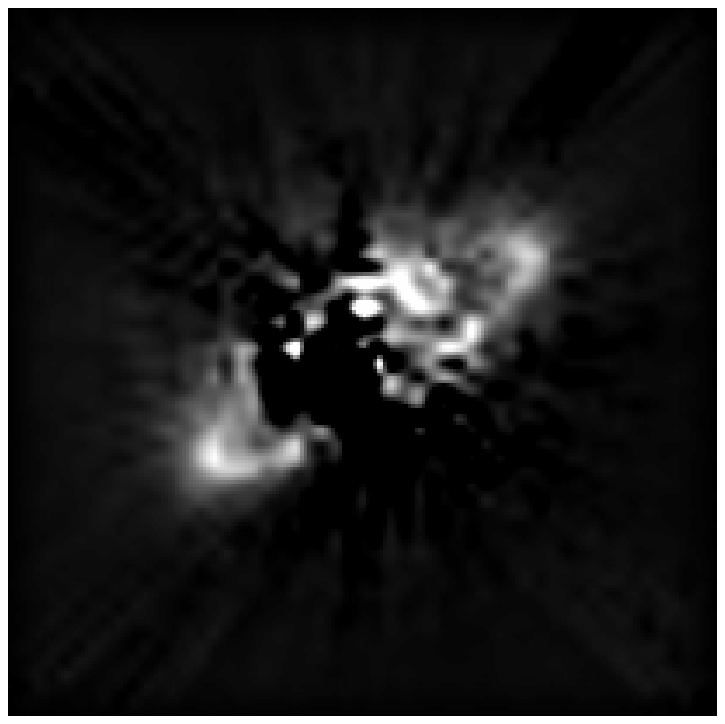,width=5.0cm,height=5.0cm,angle=0}
\hfil
}
\end{center}
\vspace{-20pt}
\caption{Images from the NASA/ESA Hubble Space Telescope. Left: part of 
  the $\beta$~Pic disk imaged by the WFPC2 instrument. Clumps of dust 
  might represent elliptical rings viewed edge-on, possibly created by 
  the gravitational force of a stellar encounter $\sim10^5$~years ago 
  (courtesy of Paul Kalas).
  Centre: the circumstellar disk of HD~141569 imaged in reflected
  light at 1.1~$\mu$m with the NICMOS instrument (from \protect\cite{wbs+99}).
  The central star and the dark vertical and horizontal bands are regions 
  obscured by a coronographic mask (courtesy of Alycia Weinberger).
  Right: the circumstellar disk around HR~4796A, also observed by 
  coronographic imaging by NICMOS (from \protect\cite{ssb+99}). 
  The ring is almost completely visible, although
  the clumpy structure arises from the coronographic system and is not real
  (courtesy of Glenn Schneider, Brad Smith, and the NICMOS IDT/EONS team). 
} 
\label{fig:hst-disk}
\end{figure}

In contrast to the observational difficulties for planets, it is now
relatively easy to observe disks (\cite{bs96x}).  Not only are they
extremely large -- a typical disk extends to of order 1000~AU from the
star -- but the surface area of the small particles which make up the
disk is many orders of magnitude larger than that of a planet.  They
emit and reflect light very well, and can be seen to relatively large
distances from the central star with modern telescopes.  Disks also
appear to be long-lived, $\sim10^6-3\times10^7$~years, and robust
against disruption by the events that commonly accompany early stellar
evolution.  They are remarkably common throughout our Galaxy, and
their properties appear quite similar to the picture of our primitive
solar nebula (Section~\ref{sec:planet-formation}).  They are
particularly evident due to their strong emission at infrared red
wavelengths, between about 2~$\mu$m and 1~mm, with a spectrum much
broader than any single-temperature black body, originating from
thermal emission over a wide variation of temperatures, from
$\sim1000$~K very close to the stars to $\sim30$~K near the outer
edges of the disks at several hundred~AU from the central star.

The star $\beta$~Pic is a prototype (see \cite{art97} for a review),
discovered from observations of the IRAS (infrared) satellite, in
which the innermost parts of the disk appear to be devoid of gas and
dust, and with the absence of diffuse material suggesting the presence
of planets (\cite{st84}).  Cometary-like bodies (\cite{bvf+94}), 
perturbations of a planet on the disk (\cite{lsr+96}), and light 
variations possibly associated with planets (\cite{lvb+97}; \cite{blc+98}
and references) have all been reported for this system.

In contrast to $\beta$~Pic, most disks occur around young stars, in 
particular around T~Tauri stars which lie close to star-forming 
clouds, and with an excess of infrared emission most likely arising 
from dusty material in orbit around, and heated by, the central star. 
Calculations imply that the disk phase of planetary formation lasts 
for only a relatively small fraction of a system's total 
lifetime, although some stars may maintain disks for a 
billion years or more, perhaps never forming any planets. 

Many disks have been detected through their infrared, sub-mm, or radio
emission, frequently showing a roughly flattened shape, with Doppler
measurements indicating rotation. Dust disks have been observed around some 
100 main-sequence stars within 50~pc of the Sun (\cite{kal98}), for example, 
around the binary BD~+31$^\circ$643 (\cite{lis97a}); $\epsilon$~Eridani
(\cite{ghm+98}); HR~4796 (\cite{krw+98}); the pre-main sequence binary
HK~Tauri (\cite{kor98}), and around HD~98800, a very young bright
planetary debris system bearing a strong similarity to the zodiacal
dust bands in our own Solar System (\cite{lhs99}).  Observations of
comets have been made in HD~100546 (\cite{gsb+97x}).  In the specific context
of the extra-solar planetary systems discovered recently are the
observations of a disk around $\rho^1$~Cnc (\cite{dlj+98};
\cite{tb98}). 

A number of disk systems have been imaged using the
Hubble Space Telescope (\cite{dwh93}), including $\beta$~Pic 
(Figure~\ref{fig:hst-disk}a; \cite{kls+00}).  In the case of HD~141569
(Figure~\ref{fig:hst-disk}b, \cite{wbs+99}), the star is at a distance
of about 100~pc, and has an age of about $10^6-10^7$~years.  Outwards
from the centre, a bright inner region is separated from a fainter
outer region by a dark band, superficially resembling the Cassini
division (the largest gap) in Saturn's rings.  The disk extends to
about 400~AU, about 13 times the diameter of Neptune's orbit, with the
gap at 250~AU.  An unseen planet may have carved out the gap, in which
case its mass can be estimated at $\sim1.3\,M_\Jupiter$, and its
orbital period as 2600~years.  If it takes $\sim300$ orbital periods
to clear such material (\cite{bcl+99}) then the gap could be opened in
$\sim8\times10^5$~years, consistent with the age of the star.  In the
case of HR~4796A (Figure~\ref{fig:hst-disk}c, 
%
%
\cite{ssb+99}) the dust
ring is confined to a width of about 17~AU. The colour of the material
(from infrared measurements at 1.1 and 1.6~$\mu$m) implies particles
with a size of a few~$\mu$m, larger than typical interstellar grains.
The implied confinement of material also argues for dynamical
constraints on the particles by one or more as yet unseen bodies.

A number of recent reviews of disk formation cover issues such as 
angular momentum transport in disks in the context
of the PSR~1257+12 pulsar planet system (\cite{rud93}); the angular
momentum evolution of young stars and disks, including the effects of
magnetic breaking, fragmentation during protostar collapse, and
viscosity (\cite{bod95}); protostars and circumstellar disks
(\cite{mcc97}); the origin of protoplanetary disks (\cite{bos98});
circumstellar disks (\cite{bec99}); and extra-solar planets and
migration (\cite{tpn00}).

\subsection{Miscellaneous signatures}
\label{sec:misc}

\cite*{ste94} has considered the direct detectability of planets during 
the short but unique epoch of giant impacts during the late stages of 
planetary formation (Section~\ref{sec:planet-formation}).  Massive 
accreting pairs, with masses of order $0.1\,M_\oplus$, would deposit 
sufficient energy to turn their surfaces 
molten and temporarily render them much more luminous at infrared 
wavelengths (ocean vaporisation or radiation to space dominate
according to mass).  This occurs even for low-velocity approaches (i.e.\ for 
co-planar, zero-eccentricity orbits at 1~AU), the gravitational 
attraction leading to impact velocities of around 10~km~s$^{-1}$. 
A luminous 1500--2500~K photosphere persisting for some $10^3$~years could be 
created for a terrestrial-mass planet (massive impacts on giant planets
would be more luminous but shorter lived) leading to an estimate of 1/250 
young stars affected.  Detection was estimated to require 1--2~nights of 
observing time per object. 

Super-flares have been observed around a number of otherwise normal
F--G main sequence stars.  These are stellar flares with energies of
$10^{33}-10^{38}$~erg, $10^2-10^7$ times more energetic than the
largest solar flares, with durations of hours to days and visible from
X-ray to optical frequencies (\cite{skd00}). \cite*{rs00} have
proposed that they are caused by magnetic reconnection between fields
of the primary star and a close-in Jovian planet, in close analogy
with flaring in RS~CVn binary systems. If the companion has a magnetic
dipole moment of adequate strength, the magnetic field lines
connecting the pair will be wrapped by orbital motion, leading to an
increase in magnetic field strength.  Interaction of specific field
loops with the passing planet will initiate reconnection events.
Super-flares on our own Sun, which would be catastrophic for life on
Earth, would not be expected since our Solar System does not have a
planet with a large magnetic dipole moment in a close orbit. Only one
super-flare star, $\kappa$~Ceti, has been searched for planets to
date, and the presence of a Saturn-mass planet is so far not excluded.
The connection between super-flares and planets is presently
conjectural.

If extra-solar planets possess a substantial magnetic field, coherent 
cyclotron radio emission could be driven by the stellar 
wind/magnetospheric interaction, as observed in our Solar System, 
perhaps episodically at the planetary rotation period (\cite{fdz99}).
Model results indicate that the most favourable candidate is presently 
$\tau$~Boo, but with an expected mean amplitude of about 2~Jy at 
28~MHz, a factor of 100 below current detectability limits. 

One terminal stage of the planet may be its spiraling in and eventual
accretion onto the central star (Section~\ref{sec:formation-revisited}).  
Modeling has been carried out for accretion onto asymptotic giant
branch stars (\cite{sl99a}) and onto solar-mass stars located on the
red giant branch (\cite{sl99b}).  In the latter case observational
signatures that accompany the engulfing of the planet include ejection
of a shell and a subsequent phase of infrared emission, increase in
the $^7$Li surface abundance, spin-up of the star because of the
deposition of orbital angular momentum, and the possible generation of
magnetic fields and the related X-ray activity caused by the
development of shears at the base of the convective envelope. Infrared
excess and high Li abundances are observed in 4--8\% of G and K stars,
and \cite*{sl99b} postulate that these signatures might originate from
the accretion of a giant planet, a brown dwarf, or a very low-mass
stars (see also Section~\ref{sec:host-properties}).

\section{Properties of extra-solar planets}
\label{sec:properties}

\begin{table}[pth]
\caption{Extra-solar planets discovered from radial velocity observations, 
ordered by increasing planet mass (of the lightest 
planet for the $\upsilon$~And system). 
The table includes discoveries to March 2000, and updated orbital parameters 
(http://exoplanets.org; for up-to-date developments see also
http://www.obspm.fr/planets):
(1) star name; 
(2) star spectral type;
(3) approximate star mass;
(4) distance (from Hipparcos);
(5) radial velocity amplitude (equation~\ref{equ:radvel1}); 
(6) planetary mass $\times\sin i$; 
(7) orbital period; 
(8) orbital semi-major axis; 
(9) orbital eccentricity; 
(10) discovery reference:
[1] \protect\cite{lms+89};
[2] \protect\cite{mq95};
[3] \protect\cite{bm96};
[4] \protect\cite{mb96};
[5] \protect\cite{bmw+97};
[6] \protect\cite{chb+97};
[7] \protect\cite{njk+97a};
[8] \protect\cite{bmv+98};
[9] \protect\cite{dfm+98};
[10] \protect\cite{mbv+98};
[11] \protect\cite{fmb+99};
[12] \protect\cite{mbv+99a};
[13] \protect\cite{bmf+99}
[14] \protect\cite{mqb+00};
[15] \protect\cite{kee+00};
[16] \protect\cite{hmb+00};
[17] \protect\cite{vmb+00};
[18] \protect\cite{smn+00};
[19] \protect\cite{qmw+00};
[20] \protect\cite{umn+00};
[21] \protect\cite{kbf+00};
[22] \protect\cite{mbv00};
[23] \protect\cite{mar00};
[24] \protect\cite{nmp+00}.
}
\label{tab:known-planets}
\lineup 
\footnotesize
\begin{tabular}{@{}llllllllll}
\br
Name&  		ST&     $M_*$&          $d$&    $K$&    $M_p\sin i$&              P&   $a$&   $e$&    Ref.\\
&               &       ($M_\odot$)&    (pc)&   (m~s$^{-1}$)&  ($M_\Jupiter$)& (days)& (AU)&   &       \\
(1)&            (2)&    (3)&            (4)&    (5)&            (6)&     (7)&   (8)&    (9)&  (10)  \\        
\mr
HD~16141        & G5IV &   0.99&    35.9&      \011  & \00.22 &  \0\075.80&   0.351& 0.28 &  [22] \\ 
HD~46375        & K1IV &   1.00&    33.4&      \035  & \00.25 & \0\0\03.024&  0.041& 0.02 &  [22] \\
HD~75289        & G0V  &   1.05&    28.9&      \054  & \00.46 & \0\0\03.508&  0.048& 0.00 &  [20]    \\ 
51~Peg          & G5V  &   0.98&    15.4&      \055  & \00.46 & \0\0\04.231&  0.052& 0.01 &  [2]     \\ 
HD~187123       & G3V  &   1.00&    47.9&      \072  & \00.54 & \0\0\03.097&  0.042& 0.01 &  [8]     \\ 
HD~209458       & G0V  &   1.03&    47.1&      \082  & \00.63 & \0\0\03.524&  0.046& 0.02 &  [16]    \\ 
$\upsilon$~And~b& F8V  &   1.10&    13.5&      \070  & \00.68 & \0\0\04.617&  0.059& 0.02 &  [5]     \\ 
$\upsilon$~And~c& --   &   -- &     -- &       \058  & \02.05 &   \0241.3&    0.828& 0.24 &  [13]     \\ 
$\upsilon$~And~d& --   &   -- &     -- &       \070  & \04.29 &    1308.5&    2.56&  0.31 &  [13]     \\ 
HD~192263       & K2V  &   0.75&    19.9&      \068  & \00.81 &  \0\024.35&   0.152& 0.22 &  [17,18] \\ 
$\rho^1$~55~Cnc & G8V  &   0.90&    12.5&      \076  & \00.93 &  \0\014.66&   0.118& 0.03 &  [5]     \\ 
$\rho$~CrB      & G2V  &   1.00&    17.4&      \061  & \00.99 &  \0\039.81&   0.224& 0.07 &  [7]     \\ 
HD~37124        & G4V  &   0.91&    33.2&      \048  & \01.13 &   \0154.8&    0.547& 0.31 &  [17]    \\ 
HD~130322       & K0III&   0.79&    29.8&       115  & \01.15 &  \0\010.72&   0.092& 0.05 &  [20]    \\ 
HD~177830       & K2IV &   1.15&    59.0&      \034  & \01.24 &   \0391.0&    1.10&  0.40 &  [17]    \\ 
HD~217107       & G7V  &   0.96&    19.7&       140  & \01.29 & \0\0\07.130&  0.072& 0.14 &  [11]    \\ 
HD~210277       & G7V  &   0.92&    21.3&      \039  & \01.29 &   \0436.6&    1.12&  0.45 &  [12]    \\ 
HD~134987       & G5V  &   1.05&    25.7&      \050  & \01.58 &   \0260.0&    0.810& 0.24 &  [17]    \\ 
16~Cyg~B        & G2.5 &   1.00&    21.6&      \050  & \01.68 &   \0796.7&    1.69&  0.68 &  [6]     \\ 
Gliese~876      & M4   &   0.32&   \04.7&       235  & \02.07 &  \0\060.90&   0.207& 0.24 &  [9,10]  \\ 
47~UMa          & G0V  &   1.03&    14.1&      \051  & \02.60 &    1084.0&    2.09&  0.13 &  [3]     \\ 
HD~12661        & K0   &   0.81&    37.2&      \091  & \02.83 &   \0264.5&    0.825& 0.33 &  [23] \\  
$\iota$~Hor     & G0V  &   1.03&    17.2&      \080  & \02.98 &   \0320.0&    0.970& 0.16 &  [15]    \\ 
HD~1237         & G6V  &   0.98&    17.6&       164  & \03.45 &   \0133.8&    0.505& 0.51 &  [24] \\  
HD~195019       & G3V  &   0.98&    37.4&       271  & \03.55 &  \0\018.20&   0.136& 0.01 &  [11]    \\ 
$\tau$~Boo      & F7V  &   1.20&    15.6&       474  & \04.14 & \0\0\03.313&  0.047& 0.02 &  [5]     \\ 
Gliese~86       & K1V  &   0.79&    10.9&       379  & \04.23 &  \0\015.80&   0.117& 0.04 &  [19]    \\ 
HD~222582       & G3V  &   1.00&    42.0&       180  & \05.18 &   \0576.0&    1.35&  0.71 &  [17]    \\ 
14~Her          & K0V  &   0.85&    18.2&      \096  & \05.44 &    1700.0&    2.84&  0.37 &  [14]    \\ 
HD~10697        & G5IV &   1.10&    32.6&       114  & \06.08 &    1074.0&    2.12&  0.11 &  [17]    \\ 
HD~89744        & F7V  &   1.40&    40.0&       257  & \07.17 &   \0256.0&    0.883& 0.70 &  [21] \\
70~Vir          & G2.5V&   1.10&    18.1&       316  & \07.42 &   \0116.7&    0.482& 0.40 &  [4]     \\ 
HD~168443       & G8IV &   0.84&    37.9&       469  & \08.13 &  \0\058.10&   0.303& 0.52 &  [12]    \\ 
HD~114762       & F7V  &   0.82&    40.6&       615  &  10.96 &  \0\084.03&   0.351& 0.33 &  [1]     \\ 
\br
\end{tabular}
\end{table}

\subsection{Masses and orbits}
\label{sec:mass-orbits}

\begin{figure}[tbh]
\begin{center}
\centerline{
\epsfig{file=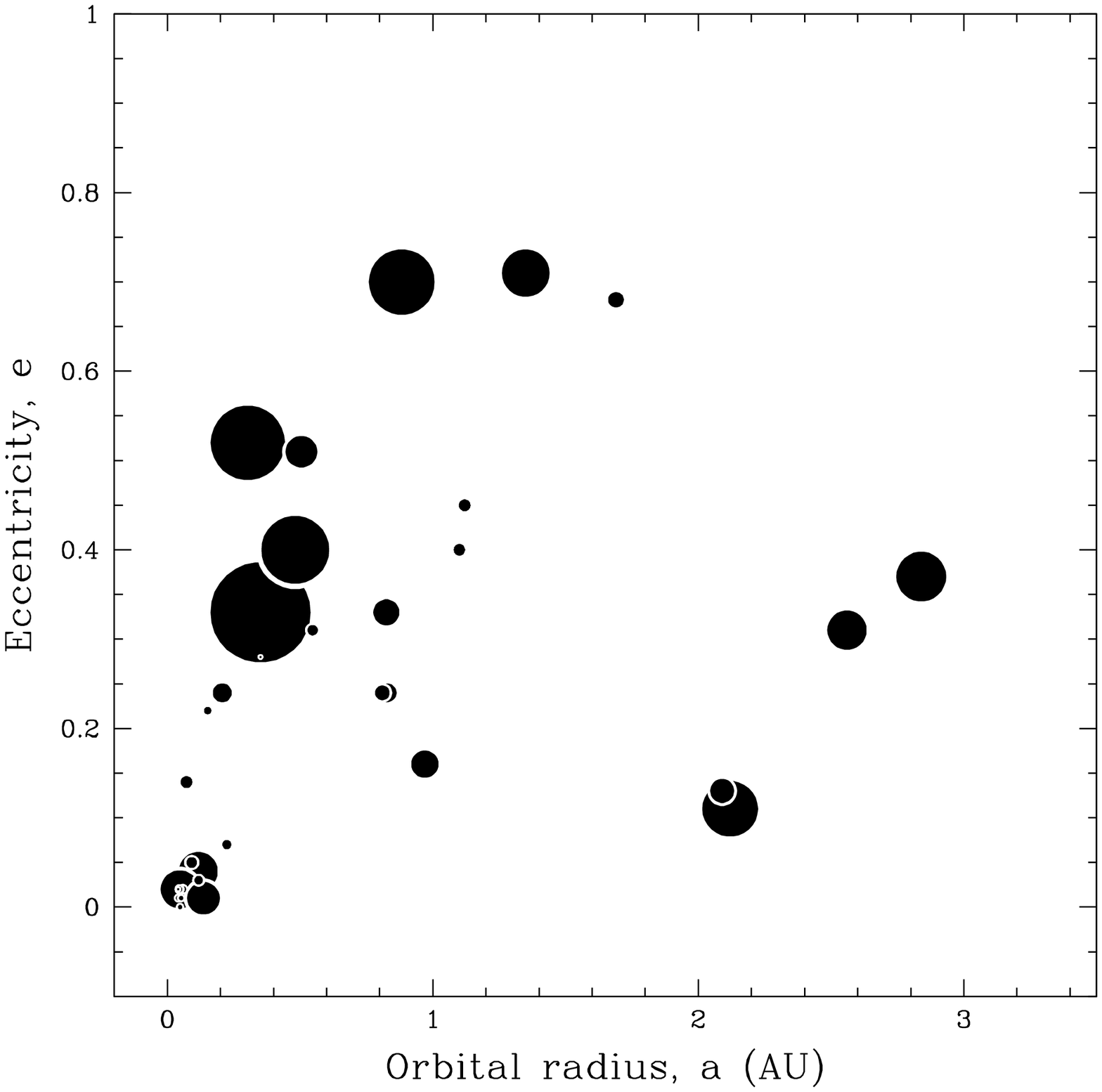,width=7.5cm,angle=0}
\hfil
\epsfig{file=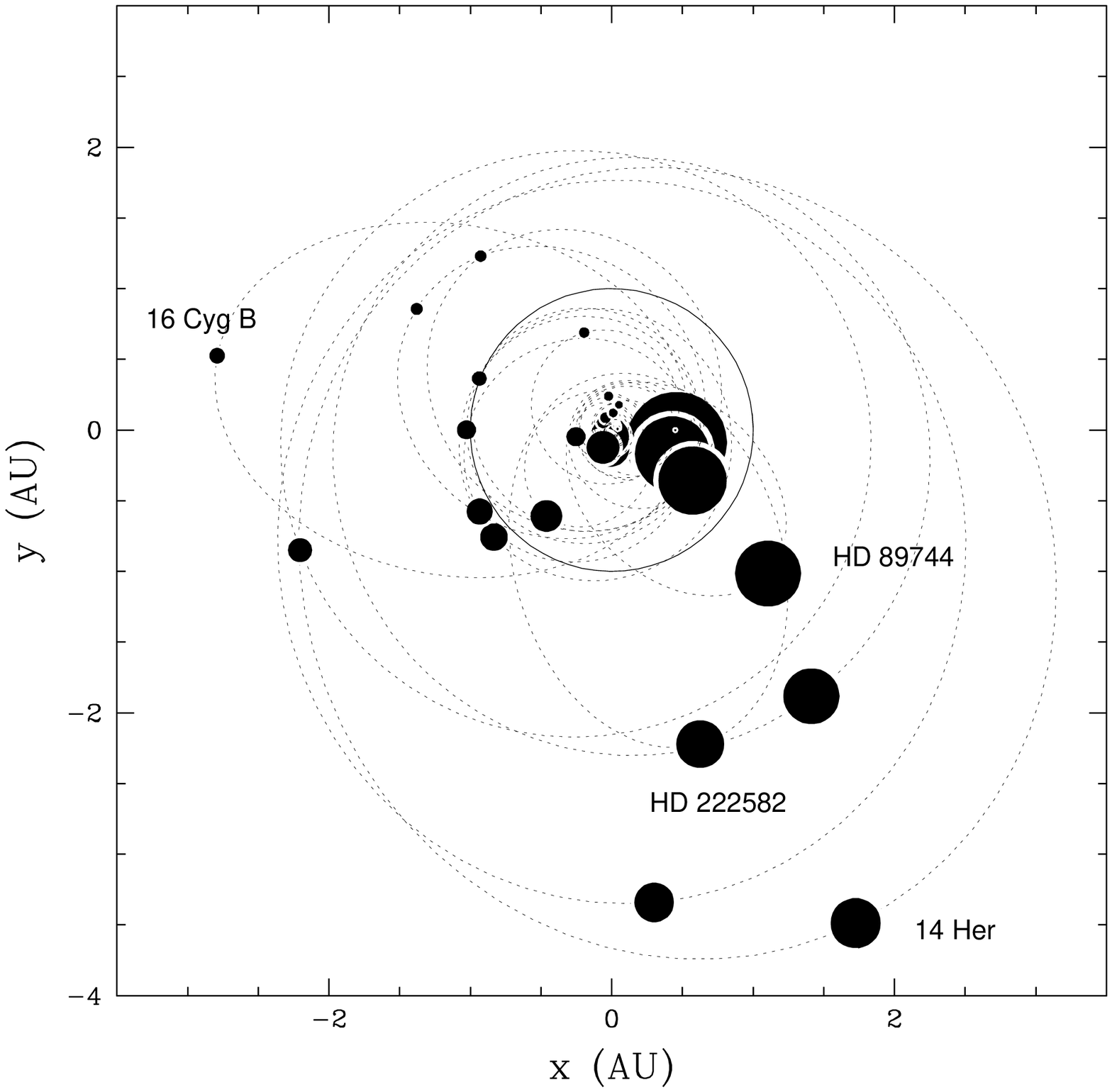,width=7.5cm,angle=0}
}
\end{center}
\vspace{-30pt}
\caption{ Left: eccentricity versus orbital radius for the known planetary 
  systems
  listed in Table~\ref{tab:known-planets}.  Diameters of the solid circles
  are proportional to the measured values of $M_p\sin i$.  
  Right: schematic of the resulting orbits.  The systems
  are shown with their relevant values of $a$ and $e$, again showing
  masses by the size of the relevant solid circle. The Earth's orbit at
  1~AU is shown as a solid line for reference.  Orbits with the largest 
  $a$ (14~Her) and largest $e$ (HD~222582, HD~89744 and 16~Cyg~B) are labeled.
}
\label{fig:mass-ecc-a}
\end{figure}

The primary characteristics of the extra-solar planets discovered to
date are given in Table~\ref{tab:known-planets}, and some 
characteristics of the extra-solar planetary orbits are
illustrated in Figure~\ref{fig:mass-ecc-a}.  

While Doppler surveys are most sensitive to small orbits, resulting in
a clear selection effect on orbital properties, what had not been
generally anticipated was the small orbital radii and large
eccentricities of many of the first systems discovered.\footnote{
Although \cite*{str52}, in considering the timeliness of radial
velocity searches, commented that {\it `It is not unreasonable that a
planet might exist at a distance of 1/50~AU... Its period around a
star of solar mass would then be about 1~day.'}.}
This trend has continued with further discoveries: 9~of the 34
known planets reside in orbits with $a<0.1$~AU, and 15 have
$a<0.3$~AU.  The correlation between $a$ (or $P$) and $e$ is 
significant (\cite{sb00}): close-in planets are nearly all in circular orbits,
while the 19 planets that orbit further out than 0.3~AU are all in
non-circular orbits with $e\ge0.1$, with 16 having $e\ge0.2$. In
contrast, most pre-discovery theories of planetary formation suggested
that extra-solar planets would be in circular orbits similar to those
in the Solar System (\cite{bos95}; \cite{lis95}).

The mass distribution rises towards lower masses, down to $M_p\sin i \simeq
0.2\,M_\Jupiter$, with companions having $M_p\sin i$
in the decade 0.5--5\,$M_\Jupiter$ outnumbering those between
5--50\,$M_\Jupiter$ by a factor~3--4 (\cite{mb98a}; \cite{mbv+99a}). 
The limited detectability of the lower mass companions implies that
the true ratio may be still higher. The underlying distribution of
$M_p$ (rather than $M_p\sin i$) cannot be derived unambiguously. For 
randomly oriented planes, $\langle M_p
\rangle = (\pi/2)\,M_p\sin i$. Thus HD~114762, 
with $M_p\sin i\sim11\,M_\Jupiter$, may well be a brown dwarf.  Similar
arguments may apply to other high mass systems, including HD~10697 for
which Hipparcos astrometry also suggests a higher mass
(\cite{zm00}). However, the possibility that the objects below 
$4\,M_\Jupiter$ are all brown dwarfs in systems seen nearly 
face on appears untenable.

\cite*{hea99} and \cite*{sb00} have investigated 
how the orbital properties compare with those of higher mass stellar
companions (brown dwarfs and low-mass stars). Solely on the basis of
the $M_p \sin i$ distribution, they find that a model comprising two different
populations is preferable, in agreement with qualitative inspection of
the histogram of $M_p \sin i$ which suggests a break in the projected 
mass distribution at around 5--7\,$M_\Jupiter$.  
In contrast, their analyses suggest that for solar-type stars,
the orbital properties of their low-mass companions in general 
(i.e.\ whether massive planets, brown dwarfs or stars) are rather
homogeneous. Thus \cite*{sb00} find that extra-solar planets and brown
dwarfs share a common probability distribution function for orbital
periods and eccentricities, a common period-eccentricity correlation,
and the same lack of a significant mass-eccentricity correlation. The
combined population displays orbital element statistics very similar
to that of compatible stellar companions.  The results are confirmed
using a much larger sample of about 400~compatible stellar companions
derived from the survey of 3347 G~dwarfs by \cite*{uml+98} (Stepinski,
private communication).  \cite*{sb00} raise the question of whether
the properties of the present sample of extra-solar planets, which
appear peculiar when viewed from the perspective of the standard model
of planet formation, could appear more natural in the context of a
population related to stellar companions (see
Section~\ref{sec:formation-revisited}). Larger samples of extra-solar
planets will be required to confirm these ideas.

\subsection{ Multiple planets and system stability}
\label{sec:stability}

Apart from the pulsar planets, only one multiple extra-solar planetary
system, $\upsilon$~And, is presently known. In addition to the
0.6\,$M_\Jupiter$ object in a 4.6-day orbit originally detected
(\cite{bmw+97}), two more distant planets were identified
from subsequent radial velocity observations, with $M_p\sin i$ of 2.0
and 4.1\,$M_\Jupiter$, $a$ of 0.82 and 2.5~AU, and large $e$ of 0.23
and 0.36 respectively (\cite{bmf+99}). Based on a specific set of 
spectroscopic orbital elements \cite*{mzt+99} derived a mass
for the outer companion of $10.1^{+4.7}_{-4.6}\,M_\Jupiter$ using
Hipparcos astrometry, compared to $M_p \sin i=4.1\,M_\Jupiter$,
implying an orbital inclination of $156^\circ$.  If the three planets
all have the same inclinations, masses of the inner two planets of
$1.8\pm0.8$ and $4.9\pm2.3~M_\Jupiter$ would follow. A large
difference between the inclinations of the outer two planets appears
to be ruled out by dynamical stability arguments (\cite{htt97};
\cite{km99x}).

With three massive planets within about 2.5~AU of each other, the 
system is well-suited to investigations of the system's dynamical stability.  
The `Hill stability' criterion requires that the planets approach no closer than
their Hill radius (or tidal radius).  The system stability therefore
depends strongly on the planetary masses, and hence orbital
inclination, since the orbits of the second and third planets brings
their relative separation close to their Hill radius, a proximity that
would make their motion chaotic.  A number of numerical studies have
been carried out to establish whether the system is stable over long
time scales (\cite{la99}; \cite{lis99}; \cite{rl00}; \cite{smb00}), confirming 
that for all except extreme values of $\sin i<0.2$ (ruled out by the
Hipparcos astrometry), most plausible orbits are stable for the 2--3~Gyr
age of the system. Numerical integration by \cite*{la99} suggest that the 
two outer planets experience extremely chaotic evolution, that the 
system tends to favour configurations in which the two outer planets 
exhibit a significant relative inclination ($i\sim15^\circ$), and that 
the eccentricity of the outer orbit is unlikely to be much larger 
than~0.3.  The considerable eccentricity of the outermost planet is 
difficult to excite through $N$-body interactions arising from a set 
of initially circular orbits.

The inner planet, with $e\sim0$, significantly exceeds the minimum 
stability criterion, suggesting little interaction with the outer two 
companions.  The outer two planets, in contrast, have high 
eccentricities, underlining the observational trend for single systems, 
and suggesting that planet-planet interactions constitute a plausible 
explanation for the orbital evolution of this system. If such 
interactions are the dominant source of orbital evolution, additional 
companion planets should ultimately be detected in most of the 
currently known systems. Alternatively, if planetary-disk interactions 
are dominant, multiple systems like $\upsilon$~And might be relatively 
rare.

The planet around 16~Cyg~B has an orbit which is particularly eccentric,
$e\sim0.7$.  16~Cyg~B is one component of a widely separated binary star
system whose eccentricity is itself very large, $e>0.54$.  The orbital
period of the binary star is very long, and therefore difficult to
measure accurately, but from astrometric observations made over about
170~years is estimated to be $>$18\,000~yr (\cite{hm99}).  It provides
an interesting test case for planetary formation theories which try to
explain these large eccentricities, perhaps through gravitational
perturbations imposed by the stellar component 16~Cyg~A.  If the
planet was originally formed in a circular orbit, with an orbital
plane inclined to that of the stellar binary by $>$45$^\circ$, then the
planet orbit oscillates chaotically between high- and low-eccentricity
states, on time scales of $10^7-10^{10}$~yr, as long as there are no
other planets with $M_p\sim M_\Jupiter$ within 30~AU (\cite{htt97};
\cite{mkr97}).  Although other highly eccentric planetary systems are
not known to be associated with stellar binaries, and therefore the
relationship between high eccentricity and stellar binarity is
unlikely to be one-to-one, perturbations caused by 16~Cyg~A remain a
plausible cause of its high eccentricity (\cite{hm99}).

\subsection{Properties of the host stars}
\label{sec:host-properties}

The host stars in Table~\ref{tab:known-planets} cover a mass range of
$M_*=0.8-1.2\,M_\odot$ (and lower for the K--M dwarfs), and a broad age 
range (\cite{fpb97}; \cite{fpb98}). Distances are well determined from the 
Hipparcos astrometry satellite measurements (\cite{esa97}). 
Detailed spectroscopic analyses of the parent stars have
been carried out (\cite{fpb97}; \cite{gon97}; \cite{fpb98}; \cite{gon98}; 
\cite{gv98}; \cite{gws99}; \cite{gim00}; \cite{gl00}).  On average, stars 
hosting planets have significantly higher metal content (elements heavier 
than He), compared to the average solar-type star in the solar
neighbourhood, although some are metal poor.  For the subset of close-in 
giants, \cite*{gl00} conclude the parent stars are all metal rich; 
values of [Fe/H]~=~0.45 for $\rho^1$~55~Cnc and 14~Her place them amongst the
most metal-rich stars in the solar neighbourhood. This suggests that their
physical parameters are affected by the process that formed the planet
(\cite{gws99}).

There are two leading hypotheses to explain the connection between
high metallicity and the presence of planets: high metallicity may
favour the formation of rocky planets (and large rocky cores of gas
giants) as a result of excess solid material in the protoplanetary
disk: essentially the metals make condensation easier.  Alternatively,
as originally formulated to explain the high Li content of certain stars
(\cite{ale67}; Li~being rapidly destroyed even at relatively low 
temperatures), the high metallicity could be due to the capture of metal-rich 
disk material by the star during its early history (\cite{lbr96}), and
possibly even to capture of the planet itself (\cite{stl+98}) as a
natural consequence of the migration scenario 
(Section~\ref{sec:formation-revisited}). A planet added to a
fully convective star would be folded into the entire stellar mass and
would lead to a negligible metallicity enhancement.  However,
main-sequence solar mass stars like the Sun have radiative cores with
relatively small outer convection zones which comprise only a few percent 
of the stellar mass. Planet capture can therefore significantly enhance the
heavy element content in the convective zone (\cite{la97}; \cite{frs99}; 
\cite{sl99b}).  A capture event may influence orbital migration of
remaining planets in the systems due to changes in angular momentum
and magnetic field (\cite{stl+98}).  If the process of migration via
the ejection of planetesimals is important (\cite{mhh+98}), many
particles can be trapped in the 3:1 or 4:1 resonances, and then pumped
into high enough eccentricites that they impact the star, leading to
metallicity enrichment and the production of evaporating bodies which
may be detectable as transient absorption lines in young stars
(\cite{qh00}).

Radial velocity searches have concentrated on F--K stars. Stars earlier 
than~F5 have fast rotation making precision radial velocity measurements 
impossible (this rotational discontinuity may be associated with planetary 
formation), while M~dwarfs have lower luminosities and relatively few have 
been surveyed.  Of these, Gliese~876 is the nearest known planetary 
system, at 4.7~pc (\cite{dfm+98}; \cite{mbv+98}).  Results 
so far suggest that planet formation is not restricted to 
more massive stars, and given that M~dwarfs outnumber 
G~dwarfs by a factor of 10, most planets in our Galaxy may orbit 
stars whose luminosity and mass are significantly lower than the Sun's.

\subsection{Physical diagnostics from transits: HD~209458}
\label{sec:hd209458}

Aside from orbital data from radial velocity measurements, specific 
planetary characteristics are presently limited. However, the recent 
detection of photometric transits for HD~209458 
(Section~\ref{sec:photometry}) provides additional important information.

The precise shape of the transit curve is determined by five
parameters (\cite{cbl+00}): the planetary and stellar radii, the
stellar mass, the orbital inclination $i$, and the limb-darkening
parameter.  For assumed values of $R_p$ and $i$ the relative flux
change can be calculated at each phase, by integrating the flux
occulted by a planet of given radius at the correct projected location
on the limb-darkened disk. Best-fit parameters yield
$R_p=1.27\pm0.02\,R_\Jupiter$ and $i=87.1\pm0.2^\circ$ which, in
combination with $M_p\sin i=0.63\,M_\Jupiter$ from the radial velocity
solution, yields $M_p=0.63\,M_\Jupiter$ ($\sin i \sim1$); similar
values are derived from the improved stellar parameters derived by
\cite*{mnt+00}. The radius is in excellent agreement with the
predictions (for a gas giant made mainly of hydrogen) of
\cite*{gbh+96}, who calculated radii for a strongly irradiated
radiative/convective extra-solar planet as a function of mass. Being
the first extra-solar planet of known radius and mass, several
important physical quantities can be derived for the first time:
\cite*{cbl+00} estimate $\rho\sim0.38$\,gm\,cm$^{-3}$, significantly
less dense than Saturn, the least dense of the Solar System gas
giants. The surface gravity is $g\sim9.7$\,m\,s$^{-2}$. Assuming an
effective temperature for the star of 6000~K, the effective
temperature of the planet is $T_p\sim1400(1-p)^{1/4}$\,K where $p$ is
the albedo.  This implies a thermal velocity for hydrogen of
$v_t\la6.0$\,km\,s$^{-1}$, a factor~7 less than the calculated escape
velocity of $v_e\sim42$\,km\,s$^{-1}$, confirming that these planets
should not be losing significant amounts of mass due to the effects of
stellar insolation.

Further physical diagnostics of this and other systems will soon
become available. In the case of HD~209458, other planets or moons in
the system could be detected from more accurate transit data, if they
exist, down to masses of $\sim1\,M_{\rm Uranus}$ at 0.2~AU, or down to
$\sim1\,M_\oplus$ from space observations.  Given the orbital
inclination, planets beyond about 0.1~AU from the star will show no
transits if the orbits are co-aligned. Detection of reflected light
would yield the planet's albedo directly since the radius is
accurately known. Observations at wavelengths longer than a few $\mu$m
may detect the secondary eclipse as the planet passes behind the star,
yielding the planet's dayside temperature, and hence constraining the
mean atmospheric absorptivity.  High-accuracy differential
spectroscopy in and out of transit may eventually lead to the
observation of absorption features added by the planetary atmosphere.

\subsection{Formation theories revisited}
\label{sec:formation-revisited}

The extra-solar main-sequence planets discovered so far 
are all more massive than Saturn, and most either orbit
very close to their stars or travel on much more eccentric paths than
any of the major planets in our Solar System.  The present small sample 
can be divided broadly into three groups (without necessarily implying
a clear distinction or different formation process between them):
Jupiter analogues, in terms of~$P$ and~$a$, and with low eccentricities
(such as 47~UMa); planets with highly eccentric orbits (such as
70~Vir); and close-in giants (or `hot Jupiters') within 0.1~AU whose
orbits are largely circular (such as 51~Peg).  While the Jupiter
analogues can be explained satisfactorily by `standard theories'
(Section~\ref{sec:planet-formation}), the other orbits provide more of 
a challenge for formation theories.  A robust formation mechanism must
explain the mass distribution, the large eccentricities, and the large 
number of orbits with $a<0.2$~AU, where both high temperature and
relatively small amount of protostellar matter available for their
agglomeration would inhibit formation {\it in situ} (\cite{bos95};
\cite{bhl00}).

If the analyses by \cite*{hea99} and \cite*{sb00} indeed point to a common
origin for the extra-solar planet and brown dwarf populations, somehow 
related to the population of stellar companions, then the standard
model of planet formation, or certain features of the standard model,
may prove to be inadequate. A formation scenario in which global
gravitational instabilities in circumstellar disks yield binaries with
a range of separations (between $R_*$ and 100~AU) and masses (perhaps 
subsequently modified by mass transfer) may then be indicated (\cite{ab92}).

Meanwhile, most theoretical efforts have so far been directed at developing 
modifications of the standard model necessary to reconcile it with the
observed system properties. Thus the existence of the close-in planets
has focussed attention on some earlier predictions that Jupiter-mass
systems (gas giants) could be formed further from the star, followed
by non-destructive migration inwards. This inward migration could be
driven by tidal interactions with the protoplanetary disk
(\cite{gt80}; \cite{lp86a}; \cite{lp86b}; \cite{wh89}), and subject to
tidal circularization in the process (\cite{tpn+98}; \cite{lpb+00};
\cite{frs99}). Radial migration is caused by inward torques between
the planet and the disk, by outward torques between the planet and the
spinning star, and by outward torques due to Roche lobe overflow and
consequent mass loss from the planet (\cite{tbg+98a}).  The inward
migration could accompany the disk material accreting onto the young
star due to viscosity within the disk, or the migrating planet could
lose angular momentum to the disk and fall inwards even if the disk
were not itself draining inwards.

The discovery of many systems with small orbital radius provides a
further complication: orbital migration time-scale is proportional to
the orbital period, which should lead to rapid orbital decay for
successively smaller orbits. This suggests that inward migration is
halted at some point until the disk evaporates.  Alternatively,
gravitational encounters between planets in a post-disk environment
could perturb a planet into an orbit with a very small periastron
distance, following which tidal interactions with the host star could
circularize the orbit (\cite{rf96}).  Recent theories and numerical
simulations have examined orbital migration, halting mechanisms, and
the origin of orbital eccentricities as part of the overall formation
scenario (\cite{lbr96}; \cite{rtl+96}; \cite{mml97}; \cite{war97a};
\cite{war97b}; \cite{mhh+98}; \cite{tbg+98a}; \cite{war98};
\cite{rud99}).

Explanations for halting the inward migration include the progressive 
raising of tidal bulges in the star that will transfer angular momentum 
from the star to the planetary orbit (\cite{lbr96}). Alternatively, holes 
in the protoplanetary disk could be cleared by the star's rotating magnetic 
field, which could entrain the hot gas, either distributing it outward, or 
channeling it onto the star along the magnetic field lines (although
such a mechanism cannot explain the planets around $\rho^1$~55~Cnc 
or $\rho$~CrB, which orbit at $a=0.1-0.2$~AU, too distant for the magnetic 
field to clear a hole).  Once the planet emerges from the inner edge 
of the disk, the inward tidal torques vanish because the surface 
density is negligible, and inward migration stops. \cite*{tbg+98a} have 
considered a mechanism involving mass loss through the planet's Roche 
lobe as the giant planet migrates inwards, leading to mass transfer 
from the planet to the star and movement outward by conservation of 
angular momentum. Stable mass transfer is achieved at a distance where 
its radius equals its Roche radius, and small stable orbits will 
result if the disk is dissipated before the planet loses all its mass. 
Our own Jupiter and the rest of the Solar System planets may be those 
bodies left stranded when the last of the protoplanetary disk cleared out.

Since orbital migration in a viscous gaseous environment is expected to
preserve circular orbits, the frequent occurrence of high orbital
eccentricity has led to a distinct class of models in which these
systems must be commonly produced.  Current attention is directed at
models involving interactions with:
(a) other orbiting planets (\cite{wm96}; \cite{rf96}; 
        \cite{li97}; \cite{lld98}); 
(b) the protoplanetary accretion disk (\cite{art93}), 
in which the proto-planet exerts
gravitational forces on the disk that launch spiral density waves, 
which in turn exert subtle forces back on the planet, pulling it away 
from pure circular motion, an effect which can build up over millions 
of years; 
(c) a companion star (\cite{htt97}; \cite{mkr97});
(d) stellar encounters in a young star cluster (\cite{ff97}; \cite{la98}).  
\cite*{mbv+99a} note the possibility that the non-circular
orbits all arise from perturbations from a bound companion star, as
proposed for 16~Cyg~B (\cite{htt97}; \cite{mkr97}), rather than from
the intrinsic dynamics of the planet formation.  


\begin{figure}[t]
\begin{center}
\centerline{\epsfig{file=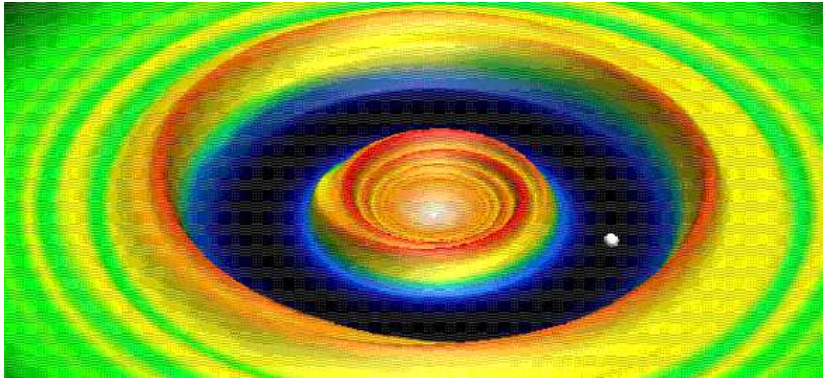,height=5.0cm,angle=0}}
\end{center}
\vspace{-20pt}
\caption{Simulation of the formation of a planetary system 
(from \protect\cite{lpb+00}).  The surface density of disk material 
has been reduced by four orders of magnitude near to the planet
of mass $M_p/M_*=10^{-3}$, and waves are clearly seen propagating 
both inward and outward away from it  (courtesy of Douglas Lin). }
\label{fig:lin-simulations}
\end{figure}

Increasingly realistic simulations of evolving planetary systems have
been made possible through improved physical models and developments
in computer speed and computational algorithms.  \cite*{lld98} have
constructed models beginning with thousands of protoplanets embedded
in a gas/dust disk, and evolved over several Gyr until orbital
stability is achieved. Planet-planet interactions typically determine
the final stable architecture of a given system, and many
of their model systems resemble, for example, the $\upsilon$~And
multiple system.  Orbit crossings and global instabilities among
planets in the disk can lead to dramatic orbit changes and large
eccentricities. \cite*{al96} and \cite*{lsa99} have modelled disk
accretion onto high-mass planets demonstrating the formation of a gap,
but with mass flow possible through it, suggesting that the opening of
a gap around a planet does not always terminate its growth.
Figure~\ref{fig:lin-simulations} shows results of simulations of
disk-planet interactions and tidal interactions with the central star
by \cite*{lpb+00}.

Current theories suggest that it is very difficult to collect all the
matter from a disk into a single planet; if the companions  to other
stars were indeed formed from disks, they are quite likely to be
members of more extensive planetary systems. Although presently there
is little information on multiple systems, and no information on
planets beyond 3~AU, this will change as radial velocity programmes
extend their temporal baseline: future results might reveal, for example, 
a population of planets residing in predominantly circular orbits which
have never experienced significant scattering or migration.  Or they
might reveal additional giant planets for all systems with giant
planets within 3~AU, as expected from dynamical evolution scenarios
that involve mutual perturbations.  Given that the Sun's planets are
all either low in mass, or orbit at larger distances, and are
therefore more difficult to discover using the Doppler technique, it
is even possible that many planetary systems will turn
out to be like our own.

\subsection{Atmospheres}
\label{sec:atmospheres}

Predicted spectral properties of extra-solar planets were made in
advance of their discovery, giving brightness versus mass and age
(\cite{bsg+95}), and including sensitivity to parameters such as
deuterium and helium abundance, rotation rate, and presence of a
rock-ice core (\cite{shb+96}).  Models have been updated subsequently,
including effects of the outer radiative zones caused by the strong
external heating which inhibits atmospheric convection. For giants at
the very small orbital distances of 51~Peg, one hundred times closer
to its primary than Jupiter, \cite*{gbh+96} calculated radii and
luminosities for a range of compositions (H/He, He, H$_2$O, and
Mg$_2$SiO$_4$). They showed that such a planet is stable to classical
Jeans evaporation, and to photodissociation and mass loss due to
extreme ultraviolet radiation, even for tidally-locked planets,
finding a mass loss rate for a gas giant at 0.05~AU of about
$10^{-16}\,M_\odot$~yr$^{-1}$.

\begin{figure}[tbh]
\begin{center}
\centerline{\epsfig{file=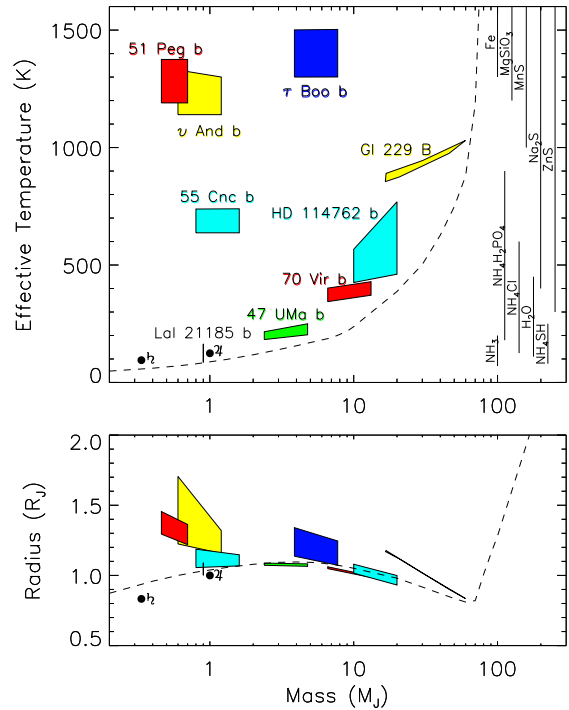,height=10.0cm,angle=0}}
\end{center}
\vspace{-30pt}
\caption{Effective temperature versus mass
  (top), and radius versus mass (bottom) for some of the early
  planets and brown dwarf candidates (from \protect\cite{gsb+97}). 
  Lalande~21185 is an unconfirmed astrometric detection 
  (see Section~\ref{sec:astrometry}) and is shown as a small vertical 
  line; Gliese~229~b (see Section~\ref{sec:imaging}) is a brown dwarf,
  and the slightly lower mass planets are possibly brown dwarfs also.
  The effective temperature and radius ranges reflect uncertainties 
  in the albedo, and the mass ranges reflect a factor of two
  uncertainty in $\sin i$.  Dotted curves correspond to models with 
  ages of $10^{10}$~years.  To the right in
  the top panel the lines depict the range of $T_{\rm eff}$ for which
  indicated species condense out near the photosphere.  Hence H$_2$O
  vapour features should be absent from the planetary spectra of
  70~Vir and 47~UMa, while Mg$_2$SiO$_4$ and Fe clouds might be in
  evidence in the spectra of $\tau$~Boo, $\upsilon$~And, and 51~Peg
  (courtesy of Tristan Guillot and Adam Burrows).
  }
\label{fig:burrows}
\end{figure}

Hydrostatic evolution calculations and atmospheric models, including
radiative effects, are being used to determine structures, radii
(e.g.\ corresponding to 10~bar), equilibrium temperatures,
luminosities (emitted and reflected), colours, and spectra of objects
with temperatures from 1300~K down to 100~K (\cite{bhl+97};
\cite{bmh+97}; \cite{gsb+97}; \cite{aah98}; \cite{bmh+98};
\cite{mar98x}; \cite{bs99}; \cite{mgs+99}).  These will help to
classify and characterise the planets as more information about them
becomes available. 
  
Evolutionary models in the luminosity/effective temperature
($L_p/T_{\rm eff}$) plane show, for example, that a few million years
after their formation, the temperatures of planets that are massive,
young, or not substantially heated by their parent star, begin to
decrease, and the planets thereafter evolve at almost constant
radius (\cite{gsb+97}).  The close-in, short-period planets are heated
by the star such that their atmosphere cannot cool substantially. The
longer period planets have a maximum radius at about 4\,$M_\Jupiter$:
for higher masses $R_p$ decreases due to electron degeneracy; lower
mass planets are smaller because their lower mass more than
compensates for their lower density. The situation changes as
stellar heating becomes significant: less massive planets are larger
because their gravity is not large enough to counter thermal
expansion.

Treatment of the close-in planets experiencing strong stellar
irradiation, using detailed radiative transfer solutions to describe
the interaction of the flux from the parent star at all relevant
depths of the planetary atmosphere, has been carried out by
\cite*{ss98}.  Compared with an isolated planet with the same
effective temperature, H$_2$ Rayleigh scattering in dust-free models
increases the planet's emission significantly shortward of Ca~H and K
at 393~nm, while inclusion of the effects of dust, formed high in
their atmospheres, increases the reflected light in the blue. 
Detailed theoretical photometric (reflected) light curves and 
polarization curves for these close-in planets are now being derived 
(\cite{sws00}).

The broad range of estimated equilibrium temperatures, from around
200~K in the case of 47~UMa to around 1500~K in the case of
$\tau$~Bootis, signifies planetary atmospheres very different from
each other, and from the planetary atmospheres in our Solar System,
with important consequences for their composition and therefore their
spectral appearance (Figure~\ref{fig:burrows}).  With increasing
temperature, chemical species likely to condense near the photosphere
range from NH$_3$, H$_2$O, and NH$_4$Cl at the lowest temperatures, up
to MnS, MgSiO$_3$, and Fe at the highest.  \cite*{sbp00} have proposed
dividing the giant planets into four albedo classes, corresponding to
four broad effective temperature ranges: a `Jovian' class with
tropospheric ammonia clouds ($T_{\rm eff}\la 150$~K); a `water cloud'
class primarily affected by condensed H$_2$O ($T_{\rm eff}\sim
250$~K); a clear class lacking cloud ($T_{\rm eff}\ga 350$~K); and a
high-temperature class for which alkali metal absorption predominates
($T_{\rm eff}\ga 900$~K). An accurate prediction of condensation in
the atmospheres is important, since the presence of clouds and
aerosols has a major influence on albedos. Planets with significant
cloud cover are expected to reflect most of the stellar light
(high albedo), hence reducing their infrared luminosity, but
increasing the amount of reflected stellar radiation.  When only rare
species condense, the albedo will be small, and the
effective temperature and infrared luminosity will be high.  


As an example of the complexities faced in the modeling, H$_2$O can
form visible clouds in the range $T_{\rm eff}=120-450$~K, which will
strongly affect the planetary spectrum and albedo (\cite{gsb+97}). At
higher temperatures, water would not be in vapour form, and would
therefore not form clouds, but would still affect the spectrum
significantly.  At lower temperatures, water would condense too deep
in the atmosphere to be detected by spectroscopy (as is the case for
Saturn, with $T_{\rm eff}=95$~K).

\section{Habitability and the search for life}
\label{sec:life}

The search for other planets is motivated by efforts to understand
their formation mechanism and, by analogy, to gain an improved
understanding of the formation of our own Solar System.  Search
accuracies will progressively improve to the point that the detection
of telluric planets in the `habitable zone' will become feasible, and
there is presently no reason to assume that such planets will not
exist in very large numbers. Improvements in spectroscopic
measurements, whether from Earth or space, and developments of
atmospheric modelling, will lead to searches for planets which are
progressively habitable, inhabited by micro-organisms, and ultimately
by intelligent life (these searches may or may not prove fruitless).  
Search strategies will be assisted by improved
understanding of the conditions required for development of life on
Earth (e.g.\ \cite{gog98}; \cite{mar98y}; \cite{mck98}) combined with
observational feasibility (e.g.\ \cite{sch94}; \cite{mlm+97};
\cite{leg98}). This will be a cross-disciplinary effort, with the
participation of astronomers, chemists and biologists.  The last few
years has seen the establishment of a number of exo-biology initiatives 
(e.g.\ \cite{bnsc99}) and numerous conferences on the search for life (e.g.\ 
\cite{cmw97}; \cite{mar97x}; \cite{wst98}), beginning to quantify
philosophical debate that has been ongoing for centuries
(\cite{cro86}; \cite{dic96}).  This section only touches upon some of 
these considerations.

Assessment of the suitability of a planet for supporting life, or
habitability, is based on our knowledge of life on Earth. With
the general consensus among biologists that carbon-based life requires
water for its self-sustaining chemical reactions (\cite{owe80}), the
search for habitable planets has therefore focused on identifying
environments in which liquid water is stable over billions of
years, as required for the development of advanced life on Earth.  
Earth's habitability over early geological time scales is complex 
and poorly understood, but its atmosphere is thought to have experienced an 
evolution in the greenhouse blanket of CO$_2$ and H$_2$O
to accommodate the 30\% increase in the Sun's luminosity over the last
4.6~billion years in order to sustain the presence of liquid water
evident from geological records (\cite{kas96}; \cite{lr98x}; 
\cite{lun99b}; \cite{lun99a}).  In the future, the Sun will increase to 
roughly three times its present luminosity by the time it leaves the 
main sequence, in about 5~Gyr. 

The habitable zone is consequently presently defined by the range of
distances from a star where liquid water can exist on the planet's
surface (\cite{har79}; \cite{kwr93}; \cite{kas96}; \cite{wkw97}).  This 
is primarily controlled by the star-planet separation, but is affected
by factors such as planet rotation combined with atmospheric
convection (models for tidally locked, synchronous rotators, are given
by \cite{jh97}). For Earth-like planets orbiting main-sequence stars,
the inner edge is bounded by water loss and the runaway greenhouse
effect, as exemplified by the CO$_2$-rich atmosphere and resulting
temperature of Venus.  The outer boundary is determined by CO$_2$
condensation and runaway glaciation, but it may be extended outwards
by factors such as internal heat sources including
long-lived radionuclides (U$^{235}$, U$^{238}$, K$^{40}$ etc.,
as on Earth, cf.\ \cite{hep78a}), tidal
heating due to gravitational interactions (as in the case of Jupiter's
moon Io), and pressure-induced far-infrared opacity of H$_2$, since
even for effective temperatures as low as 30~K, atmospheric basal
temperatures can exceed the melting point of water (\cite{ste99}). 
These considerations result, for a $1\,M_\odot$ star, in an inner
habitability boundary at about 0.7~AU and an outer boundary at 
around 1.5~AU or beyond (some authors give a more restricted
range). The habitable zone evolves outwards with time because of the
increasing Sun's luminosity with age, resulting in a narrower width of
the continuously habitable zone over $\sim4$~Gyr of around
0.95--1.15~AU.  Positive feedback due to the greenhouse effect and
planetary albedo variations, and negative feedback due to the link
between atmospheric CO$_2$ level and surface temperature may limit
these boundaries further (\cite{kwr93}; \cite{kas96}). Migration of the 
habitable zone to much larger distances, 5--50~AU, during the short
period of post-main-sequence evolution corresponding to the subgiant 
and red giant phases, has been considered by \cite*{lpm+00}.

Within the $\sim1$~AU habitability zone, Earth `class' planets can be
considered as those with masses between about 0.5--10\,$M_\oplus$ or,
equivalently, radii between 0.8--2.2\,$R_\oplus$ (\cite{bkd+97}; see
also \cite{hua60}). Planets below this mass in the habitable zone are
likely to lose their life-supporting atmospheres because of their low
gravity and lack of plate tectonics, while more massive systems are
unlikely to be habitable because they can attract a hydrogen-helium
atmosphere and become gas giants.

Habitability is also likely to be governed by the range of stellar
types for which life has enough time to evolve, i.e.\ stars not more
massive than spectral type~A.  However, even F~stars have narrower
continuously habitable zones because they evolve more rapidly, while
late K and M stars may not host habitable planets because they can
become trapped in synchronous rotation due to tidal damping. Mid- to
early-K and G stars may therefore be optimal for the development of
life (\cite{kwr93}).  Simulations of the formation of habitable
systems, within the framework of models of planet formation in general
and our Solar System in particular, are given for example by
\cite*{wet96b} and \cite*{lis97b}. 

Other effects complicate considerations of habitability: in the
absence of our Moon, thought to be an accident of accretion, an
Earth-like planet would undergo large-amplitude chaotic fluctuations
due to secular resonances (spin-axis/orbit axis or spin-axis
precession/perihelion precession) on time scales of order 10~Myr,
depending also on the planet's land-sea distribution. These motions
would result in large local temperature excursions (\cite{war74};
\cite{lr93}; \cite{ljr93}; \cite{wk97}) although perhaps not
precluding the accompanying migration of life.

\cite*{owe80} argued that large-scale biological activity on a
telluric planet necessarily produces a large quantity of O$_2$.
Photosynthesis builds organic molecules from CO$_2$, with the help of
H$^+$ ions which can be provided from different sources. In the case
of oxygenic bacteria on Earth, H$^+$ ions are provided by the
photodissociation of H$_2$O, in which case oxygen is produced as a
by-product. However, this is not the case for anoxygenic bacteria, and
thus O$_2$ is to be considered as a possible but not a necessary
by-product of life. Indeed, Earth's atmosphere was O$_2$-free until about 
2~billion years ago, suppressed for more than 1.5~billion years after 
life originated (\cite{kas96}). \cite*{owe80} noted the possibility,
quantified by \cite*{sch94} based on transit measurements, of using
the 760-nm band of oxygen as a spectroscopic tracer of life on another
planet since, being highly reactive with reducing rocks and volcanic
gases, it would disappear in a short time in the absence of a
continuous production mechanism. Plate tectonics and volcanic activity
provide a sink for free O$_2$, and are the result of internal planet
heating by radioactive uranium and of silicate fluidity, both of which
are expected to be generic whenever the mass of the planet is
sufficient and when liquid water is present. For small enough planet
masses, volcanic activity disappears some time after planet formation,
as do the associated oxygen sinks.

\cite*{acw86} showed that O$_3$ is itself a tracer of O$_2$ and, with
a prominent spectral signature at 9.6~$\mu$m in the infrared where the
planet/star contrast is significantly stronger than in the optical,
should be easier to detect than the visible wavelength lines. These
considerations are motivating the development of infrared space
interferometers (Section~\ref{sec:imaging}) for the study of lines
such as H$_2$O at 6--8~$\mu$m, CH$_4$ at 7.7~$\mu$m, O$_3$ at
9.6~$\mu$m and CO$_2$ at 15~$\mu$m (\cite{sag97}; \cite{wa98}).
Higher resolution studies might reveal the presence of CH$_4$, its
presence on Earth resulting from a balance between anaerobic
decomposition of organic matter and its interaction with atmospheric
oxygen; its highly disequilibrium co-existence with O$_2$ could be
strong evidence for the existence of life (\cite{kas96}).

The possibility that O$_3$ is not an unambiguous identification of
Earth-like biology but rather a result of abiotic processes (\cite{owe80}; 
\cite{kas96}; \cite{nrc+97}; \cite{sch99}) was considered in detail
by \cite*{loa+99}.  They considered various production processes such
as abiotic photodissociation of CO$_2$ and H$_2$O followed by the
preferential escape of hydrogen from the atmosphere.  In addition,
cometary bombardment could bring O$_2$ and O$_3$ sputtered from H$_2$O
by energetic particles (cf.\ the ultraviolet spectral signature of
O$_3$ in the satellites of Saturn, Rhea and Dione, \cite{nrc+97})
according to the temperature, greenhouse blanketing, and presence of
volcanic activity.  They concluded that a simultaneous detection of
significant amounts of H$_2$O and O$_3$ in the atmosphere of a planet
in the habitable zone presently stands as a criterion for large-scale
photosynthetic activity on the planet. Whether these would correspond to a
true signature of a biological process is less clear. An absence of
O$_2$, on the contrary, would not necessarily imply the absence of
life.

Habitability may be further confined within a narrow
range of [Fe/H] of the parent star (\cite{gon99b}; see
Section~\ref{sec:host-properties}). If the occurrence of gas giants
decreases at lower metallicities, their shielding of inner planets in
the habitable zone from frequent cometary impacts, as occurs in our
Solar System, would also be diminished (\cite{wet94}). At higher
metallicity, asteroid and cometary debris left over from planetary
formation may be more plentiful, enhancing impact probabilities.

\cite*{gon99a} has investigated whether the anomalously small motion
of the Sun with respect to the local standard of rest, both in terms
of its pseudo-elliptical component within the Galactic plane, and its
vertical excursion with respect to the mid-plane, may be explicable in
anthropic terms.  Such an orbit could provide effective shielding from
high-energy ionising photons and cosmic rays from nearby supernovae,
from the X-ray background by neutral hydrogen in the Galactic plane,
and from temporary increases in the perturbed Oort comet impact rate.
If chiral asymmetry is also a prerequisite for life (e.g.\ \cite{bai99})
habitability may further depend on the polarisation environment of the star
forming region. 

Such complex considerations may imply that the fraction of habitable 
planets is small. The search for extra-terrestrial intelligence
(SETI), nevertheless motivated by the belief that life is almost bound to
emerge under conditions resembling those on the early Earth (e.g.,
\cite{cm59}; \cite{dra61}; \cite{tow97}; \cite{lh97}; \cite{bha00}) is
not considered in this review.  Evaluation of the probability that
intelligent civilizations exist elsewhere in our Galaxy, for example
through consideration of the Drake equation (e.g.\ \cite{chy97}), or
resolution of the `Fermi paradox' (if other advanced civilizations
exist, where are they?) may lie far in the future.  Success or
persistent failure will be of considerable significance.  Concerns
such as `who will speak for Earth' when or if contact is made
(\cite{gol88}), are far from today's scientific mainstream, but
perhaps not as far as they were 10~years ago.


\section{Summary}
\label{sec:summary}

Precise radial velocity measurements have discovered 34 extrasolar
giant planets within about 50~pc over the last 5~years.  With many
radial velocity monitoring programmes underway, and an occurrence rate
of some 5\% in systems measured to date, we may expect that some
100~such planets will be discovered over the next 5~years, some 10--20
of which may be close-in systems for which transit measurements may be
possible. A number of other experimental techniques capable of
detecting extra-solar planets are under development.  Global space
astrometric measurements at the 10~microarcsec level may furnish a
list of 10--20\,000 giant planets out to 100--200~pc by the year 2020,
while accurate photometric monitoring from space should lead to the
detection and characterisation of significantly lower mass planets.
Together, such data sets will provide a vast statistical description
of planetary masses, orbits, and eccentricities, allowing important
constraints to be placed on the complex processes believed to be
involved in planetary formation.  Improved knowledge of individual
systems, in particular from transit measurements, combined with
improved atmospheric modelling and theories of habitability, will
narrow down the range of identified planets on which life may have
developed. Nearby, Earth-mass planets should exist, and would be the
natural targets for infrared space interferometers which, by 2020, may
succeed in imaging and providing evidence for life on them.  
We now know that other worlds -- large ones at least -- are common. 
Developments have been so rapid over the last few years that many 
significant developments, and many new surprises, can be predicted
with confidence.

\ack 

I am grateful to the European Southern Observatory for hospitality
while writing the major part of this review. Its preparation has
benefitted from up-to-date information on systems and www links of the
Extra-Solar Planets Encyclopaedia maintained by Dr~Jean Schneider
(http://www.obspm.fr/planets) and updated orbital parameters
maintained by Dr~Geoff Marcy (http://exoplanets.org/). On-line
journals, the NASA Astrophysics Data System (ADS), and Latex/Bibtex
have facilitated its preparation. I~am grateful to all colleagues who
readily furnished preprints and authorised the use of their figures
for this review. Figures originally published in the Astrophysical
Journal are reproduced by permission of the AAS. It is a pleasure to
thank G.~Marcy (U.C.~Berkeley and San Francisco State University),
P.D.~Sackett (University of Groningen) and F.P.~Israel and P.T.~de
Zeeuw (Sterrewacht Leiden) for comments on the manuscript. I am
particularly grateful to J.~Schneider (Observatoire de Paris-Meudon)
for numerous valuable and perceptive suggestions for improvements and
additions.

\vskip 10pt
Note added in proof: a further 8~planets were reported, from the Coralie
spectrometer observations, in a press release from the European Southern
Observatory, 2000 May~4. 

\newpage
% ...

{\footnotesize\parskip 0pt

}

\end{document}